\documentclass[12pt,letterpaper]{article}

\usepackage{amsmath}
\usepackage{amssymb}
\usepackage{epsfig}
\usepackage{graphicx}
\usepackage{cite}

\newcommand{\capdef}{}
\newcommand{\mycaption}[2][\capdef]{\renewcommand{\capdef}{#2}%
       \caption[#1]{{\footnotesize #2}}}
\makeatletter
\renewcommand{\fnum@table}{\textbf{\tablename~\thetable}}
\renewcommand{\fnum@figure}{\textbf{\figurename~\thefigure}}
\makeatother

\newcounter{myenumi}

\renewcommand{\themyenumi}{\roman{myenumi}}
{\end{list}}

\newlength{\myem}
\newcounter{mysubequation}[equation]

\makeatletter
\renewcommand{\section}{\@startsection{section}{1}{0em}{-\baselineskip}%
{\baselineskip}{\normalfont\large\bfseries}}
\renewcommand{\subsection}%
{\@startsection{subsection}{2}{0em}{-0.7\baselineskip}%
{0.7\baselineskip}{\normalfont\bfseries}}
\makeatother

\newcommand{\ie}{{\it i.e.}}

\newcommand{\eg}{{\it e.g.}}

\newcommand{\cf}{{\it cf.}}
\newcommand{\etc}{{\it etc.}}
\newcommand{\eq}{Eq.}

\newcommand{\fig}{Figure}

\newcommand{\Ref}{Ref.}
\newcommand{\Refs}{Refs.}
\newcommand{\Sec}{Section}

\newcommand{\App}{Appendix}

\newcommand{\Tab}{Table}

\newcommand{\JHFHK}{\mbox{{\sf T2HK}}$^*$}

\newcommand{\NOVA}{\mbox{\sf NO$\nu$A}}

\newcommand{\stheta}{\sin^22\theta_{13}}
\newcommand{\deltacp}{\delta_\mathrm{CP}}
\newcommand{\ldm}{\Delta m_{31}^2}
\newcommand{\sdm}{\Delta m_{21}^2}
\newcommand{\equ}[1]{\eq~(\ref{equ:#1})}
\newcommand{\figu}[1]{\fig~\ref{fig:#1}}

\newcommand{\bi}{\begin{itemize}}
\newcommand{\ei}{\end{itemize}}

\newcommand{\cpI}{DeRujula:1998hd,Dick:1999ed,Donini:1999jc,Freund:1999gy}
\newcommand{\cpII}{Arafune:1997hd,Minakata:1997td,Minakata:1998bf,Tanimoto:1998sn}
\newcommand{\cpO}{Barger:1980jm}

\begin{document}
%%%%%%%%%%%%%%%%%%%%%%%%%%%%%%%%%%%%%%%%%%%%%%%%%%%%%%%%%%%%%%%%%%%%%
%%%%                     Title-page                              %%%%
%%%%%%%%%%%%%%%%%%%%%%%%%%%%%%%%%%%%%%%%%%%%%%%%%%%%%%%%%%%%%%%%%%%%%

\begin{titlepage}

% the footnote symbols are only redefined for the title page !
\renewcommand{\thefootnote}{\alph{footnote}}

\vspace*{-3.cm}
\begin{flushright}
TUM-HEP-634/06\\
MADPH-06-1459

%hep-ph/
\end{flushright}

\vspace*{0.5cm}

\renewcommand{\thefootnote}{\fnsymbol{footnote}}
\setcounter{footnote}{-1}

{\begin{center}
{\Large\bf Optimization of a neutrino factory oscillation experiment}
\end{center}}
\renewcommand{\thefootnote}{\alph{footnote}}

\vspace*{.8cm}
%\vspace*{.3cm}
{\begin{center} {\large{\sc
                P.~Huber\footnote[1]{\makebox[1.cm]{Email:}
                phuber@physics.wisc.edu},~
                M.~Lindner\footnote[2]{\makebox[1.cm]{Email:}
                lindner@ph.tum.de},~
                M.~Rolinec\footnote[3]{\makebox[1.cm]{Email:}
                rolinec@ph.tum.de},~
                W.~Winter\footnote[4]{\makebox[1.cm]{Email:}
                winter@ias.edu}
                }}
\end{center}}
\vspace*{0cm}
{\it
\begin{center}

\footnotemark[1]%${}^,$\footnotemark[2]%
       Department of Physics, University of Wisconsin, \\
       1150 University Avenue, Madison, WI 53706, USA

\vspace*{1mm}

\footnotemark[2]${}^,$\footnotemark[3]%
       Physik--Department, Technische Universit\"at M\"unchen, \\
       James--Franck--Strasse, 85748 Garching, Germany

\vspace*{1mm}

\footnotemark[2]%${}^,$\footnotemark[3]%
       Max--Planck--Institut f\"ur Kernphysik, \\
       Postfach 10 39 80, 69029 Heidelberg, Germany

\vspace*{1mm}

\footnotemark[4]%
       School of Natural Sciences, Institute for Advanced Study, \\
       Einstein Drive, Princeton, NJ 08540, USA

\vspace*{1cm}

\today
\end{center}}

\vspace*{0.3cm}

\begin{abstract}

  We discuss the optimization of a neutrino factory
  experiment for neutrino oscillation physics in terms of muon energy,
  baselines, and oscillation channels (gold, silver, platinum). In
  addition, we study the impact and requirements for detector
  technology improvements, and we compare the results to beta beams.
  We find that the optimized neutrino factory has two baselines, one
  at about $3 \, 000$ to $5 \, 000 \, \mathrm{km}$, the other at about
  $7 \, 500 \, \mathrm{km}$ (``magic'' baseline). The threshold and
  energy resolution of the golden channel detector have the most
  promising optimization potential. This, in turn, could be used to
  lower the muon energy from about $50 \, \mathrm{GeV}$ to about $20
  \, \mathrm{GeV}$.  Furthermore, the inclusion of electron neutrino
  appearance with charge identification (platinum channel) could help
  for large values of $\sin^2 2 \theta_{13}$. Though tau neutrino
  appearance with charge identification (silver channel) helps, in principle,
  to resolve degeneracies for intermediate $\sin^2 2 \theta_{13}$, we find
  that alternative strategies may be more feasible in this 
  parameter range. As far as matter density
  uncertainties are concerned, we demonstrate that their impact can be
  reduced by the combination of different baselines and channels.
  Finally, in comparison to beta beams and other alternative
  technologies, we clearly can establish a superior performance for a
  neutrino factory in the case $\sin^2 2 \theta_{13} \lesssim 0.01$.
  
\end{abstract}

\vspace*{.5cm}

\end{titlepage}

\newpage

\renewcommand{\thefootnote}{\arabic{footnote}}
\setcounter{footnote}{0}

%%%%%%%%%%%%%%%%%%%%%%%%%%%%%%%%%%%%%%%%%%%%%%%%%%%%%%%%%%%%%%%%%%%%%
%                     Introduction                                  %
%%%%%%%%%%%%%%%%%%%%%%%%%%%%%%%%%%%%%%%%%%%%%%%%%%%%%%%%%%%%%%%%%%%%%

\section{Introduction}

Neutrino oscillation physics has entered the precision age by the
measurements of the leading atmospheric and solar oscillation
parameters. The status of current data and parameter estimates can
be found, \eg, in \Refs~\cite{Fogli:2005cq,Schwetz:2006dh}. The common
framework typically used for neutrino oscillations includes three active flavors, which
can accommodate all data besides the LSND evidence~\cite{Aguilar:2001ty}. 
The rough picture is that there are
two mass splittings which are different by a factor of $\sim30$, as well
as one possibly maximal mixing angle, and one large, but certainly not maximal, mixing
angle. The third mixing angle is known to be not larger than the
Cabibbo angle. That leaves a number of open questions even within the
standard framework: The value of the small mixing angle $\theta_{13}$,
the ordering of the mass eigenstates, the value of the Dirac-type CP
phase, and whether there is maximal mixing in the neutrino sector.  A
number of experiments in the near future are targeted towards
providing answers to some of those questions. They all have in common
that they only will succeed for values of $\stheta$ not too much below
the current bound.  Examples are future reactor experiments with near
and far
detectors~\cite{Minakata:2002jv,Huber:2003pm,Anderson:2004pk,Ardellier:2004ui}
and neutrino beams~\cite{Ables:1995wq,Itow:2001ee,Ayres:2004js}.
Since none of them will be able to provide more than indications and
hints towards the mass hierarchy and leptonic CP violation, there will be
the need for an advanced neutrino oscillation facility. It will
allow to make firm statements and precise measurements in order to
finally reveal the underlying theoretical structures. There is a large
number of contenders for this advanced neutrino facility, but it seems
there is consensus that the most capable of these are neutrino
factories~\cite{Geer:1998iz,Albright:2000xi,Apollonio:2002en} and
higher gamma beta beams~\cite{Burguet-Castell:2003vv,Huber:2005jk}.
They will allow for high precision measurements for large $\stheta$,
or will have excellent discovery reaches for small values of
$\stheta$.

In this study, we focus on the optimization and physics reach of a
neutrino factory. Earlier studies discussing the potential and optimization
of such an experiment include
\Refs~\cite{Barger:1999fs,Cervera:2000kp,Burguet-Castell:2001ez,Freund:2001ui}.
We extend those results by including the full parameter correlations,
degenerate solutions, the matter density uncertainty as well as
detector effects, backgrounds and systematical errors in the analysis.
We investigate the continuous dependence of the performance on
$L$ and $E_\mu$ as well. In addition, we improve on the use of disappearance
information by using a data sample without charge identification, we
test the combination of baselines, and we discuss different channels
and improvements to the detection system. The objective of this work
is to identify the ``optimal'' setup given current possibilities
\emph{and} to determine the factors which have the greatest
potential for improvement. Based on that, we will formulate the
requirements which upgrades of  a neutrino factory
should fulfill in order to yield a certain level of performance gain.
These requirements will allow to focus R\&D for further optimization
in the coming years.

In order to optimize the neutrino factory, we consider the sensitivities to 
$\stheta$, the mass hierarchy, and CP
violation, as well as we discuss measurement of the leading
atmospheric parameters. For the underlying oscillation theory, we
assume unitary three-flavor neutrino oscillations without significant
``new physics'' effects, since there is so far no convincing evidence
for such effects. However, note that the physics motivation for a
neutrino factory does include many more measurements than discussed in
this study, some of which may lead to a very different optimization.  
Already from the point of view of oscillation physics, an
important application, which is not discussed here, are precision
measurements of $\stheta$ and $\deltacp$, as soon as $\stheta > 0$ is
established (see, \eg, \Refs~\cite{Huber:2002mx,Huber:2004gg}). Furthermore,
the issue of whether $\theta_{23}$, if it is not maximal, is smaller
or larger than $\pi/4$ (octant) is an important question~\cite{Donini:2005db}.
Furthermore, using a very long baseline $L \gtrsim 5 \, 000 \,
\mathrm{km}$, one can verify the MSW effect at high
significance~\cite{Cervera:2000kp,Freund:2000ti} even for $\stheta =
0$ by the solar appearance term~\cite{Winter:2004mt}.  In addition, a
baseline $L \simeq 6 \,000 \, \mathrm{km}$ could be required for the
mass hierarchy determination for $\stheta = 0$ using disappearance
information~\cite{deGouvea:2005hk,deGouvea:2005mi}, and the
possible application of a matter density measurement may have some
impact on the baseline choice as well~\cite{Winter:2005we}. Except from
standard three-flavor oscillation, another important category are
``new physics'' tests, where a large volume of literature exists, see,
\eg,
\Refs~\cite{Ota:2001pw,Huber:2002bi,Huber:2001de,Campanelli:2002cc,Garbutt:2003ih,Barger:2004db,Blennow:2005yk,Blennow:2005qj,Ota:2005et,Kitazawa:2006iq,Klinkhamer:2006ae}.
For a discussion of present bounds on non-standard neutrino interactions,
see \Ref~\cite{Davidson:2003ha}. In addition, the neutrino factory
front-end can be used for high statistics muon and neutrino physics
experiments~\cite{Mangano:2001mj}. Therefore, the physics motivation
for a neutrino factory does not only depend on the measurements
discussed in this study, but is based on a much wider range of
physics. Obviously, the ``optimal'' neutrino factory for testing new 
physics effects may be completely different from the one found as optimal
with respect to, for instance, CP violation. This should be kept in mind in
interpreting our results.

This study is organized as follows: In \Sec~\ref{sec:sim} we describe
the neutrino oscillation phenomenology in terms of oscillation
probabilities, we discuss the impact of using different channels, and
we describe our simulation methods. Next in \Sec~\ref{sec:optstd}, we
introduce and optimize our ``standard'' neutrino factory, which
corresponds to the one from earlier studies with some slight
improvements. This neutrino factory represents common knowledge on
detector technology and does not include additional channels, as well
as we focus on a single baseline optimization in that section. As far
as the detector optimization is concerned, we introduce possible
improvements to the golden channel detector in \Sec~\ref{sec:det}, and
illustrate where to concentrate the R\&D. As a completely independent
discussion, we introduce more oscillation channels
(silver, platinum) in \Sec~\ref{sec:channels}, we demonstrate how
these affect the energy and baseline optimization, and we discuss ways
of improving these channels. In the last \Sec~\ref{sec:comp},
we compare different optimized setups from earlier sections, include
the evaluation of two simultaneously operated baselines, and show the
results for the combination of different optimization strategies.
Finally, we summarize our results in \Sec~\ref{sec:summary}. Note that
a more detailed summary on ``where to concentrate the efforts'' in
terms of categories can be found in \Sec~\ref{sec:efforts}.

%%%%%%%%%%%%%%%%%%%%%%%%%%%%%%%%%%%%%%%%%%%%%%%%%%%%%%%%%%%%%%%%%%%%%%%%%%%%%%%%
\section{Phenomenology and simulation methods}
\label{sec:sim}

In this section, we describe the theoretical framework for much
of this work. Following that, performance indicators and standard
values will be defined setting the stage for the actual results.

It was recognized quite early that the most favorable transition to
study genuine three flavor effects and the influence of matter on
neutrino propagation in an accelerator experiment is the so called
``golden channel''
$\nu_e\rightarrow\nu_\mu$~\cite{\cpO,\cpI,\cpII,Cervera:2000kp}.  It
is sensitive to $\stheta$, $\deltacp$, and sign($\Delta m^2_{31}$).
However, a measurement using the golden channel suffers from
correlations and degeneracies.  This, already, can be seen from the
oscillation probability. In the following we will use an expansion of
$P_{e \mu}$ in the small parameters $\sin 2\theta_{13}$ and
$\alpha\equiv\frac{\Delta m^2_{21}}{\Delta m^2_{31}}\sim \pm 0.04$ up
to second order~\cite{Freund:2001pn,Cervera:2000kp}
\begin{equation}
P_{e \mu}\ \simeq\ \sin^2 2\theta_{13}\ T_1\ +\ \alpha\ \sin 2\theta_{13}\ T_2
+\ \alpha\ \sin 2\theta_{13}\ T_3\ +\ \alpha^2\ T_4,
\end{equation}
where the individual terms are of the form
\begin{eqnarray}
T_1 & = & \sin^2\theta_{23}\ \frac{\sin^2[( 1-\hat{A})\Delta]}{(1-\hat{A})^2} \, , \label{equ:t1} \\
T_2 & = & \sin\deltacp\ \sin2\theta_{12}\ \sin2\theta_{23}\ 
\sin\Delta \frac{\sin(\hat{A}\Delta)}{\hat{A}} \frac{\sin[( 1-\hat{A})\Delta]}{(1-\hat{A})} \, , \label{equ:t2} \\
T_3 & = & \cos\deltacp\ \sin2\theta_{12}\ \sin2\theta_{23}\ 
\cos\Delta\ \frac{\sin(\hat{A}\Delta)}{\hat{A}} \frac{\sin[( 1-\hat{A})\Delta]}{(1-\hat{A})} \, , \label{equ:t3} \\ 
T_4 & = & \cos^2 2\theta_{23}\ \sin^2 2\theta_{12}\
\frac{\sin^2(\hat{A}\Delta)}{\hat{A}^2}  \, , \label{equ:t4} 
\end{eqnarray}
with $\Delta\equiv\frac{\Delta m^2_{31}\ L}{4\ E_{\nu}}$ and
$\hat{A}\equiv\frac{2 \sqrt{2} G_F n_e E}{\Delta m^2_{31}}$.
Obviously this expression depends on all of the oscillation
parameters: the solar parameters $\Delta m^2_{21}$ and $\theta_{12}$,
the leading atmospheric parameters $\Delta m^2_{31}$ and
$\theta_{23}$, as well as $\theta_{13}$ and $\deltacp$. Thus, if one
wants to extract information on only one parameter, correlations with
all the other parameters, which are not exactly known, will
deteriorate the achievable precision or limit. For example, one can
only easily extract a continuous set of combinations between $\stheta$
and $\deltacp$ from this channel. The usual strategy in order to
resolve this correlation between $\stheta$ and $\deltacp$ is the
inclusion of antineutrino running with\footnote{See
  \Ref~\cite{Akhmedov:2004ny} for the replacement rules in the
  probabilities using different channels.}
\begin{eqnarray}
P_{\bar{e}\bar{\mu}} & = &
P_{e\mu}(\deltacp\rightarrow-\deltacp,\hat{A}\rightarrow-\hat{A})\ = \\
 & \simeq & \sin^2 2\theta_{13}\ T_1(\hat{A}\rightarrow-\hat{A})\ -\ \alpha\ \sin
 2\theta_{13}\ T_2(\hat{A}\rightarrow-\hat{A}) \nonumber\\ 
 & + &  \alpha\ \sin 2\theta_{13}\ T_3(\hat{A}\rightarrow-\hat{A})\ +\
 \alpha^2\ T_4(\hat{A}\rightarrow-\hat{A}) 
\end{eqnarray} 
In combination with the golden channel from neutrino running only, one
intersection besides the original solution in $(\theta_{13},\deltacp)$ remains. 

Even with neutrino and antineutrino running, there is a set of eight discrete degeneracies affecting the performance~\cite{Barger:2001yr}:
\begin{itemize}
\item Intrinsic ($\theta_{13}$,$\deltacp$) 
  degeneracy with ($\theta_{13}$,$\deltacp$) $\rightarrow$ 
  ($\theta_{13}'$,$\deltacp'$)~\cite{Burguet-Castell:2001ez}.
\item Sign-degeneracy with $\Delta m^2_{31}\rightarrow -\Delta m^2_{31}$~\cite{Minakata:2001qm}.
\item Octant-degeneracy with $\theta_{23}\rightarrow\pi/2-\theta_{23}$~\cite{Fogli:1996pv}.
\end{itemize}
The octant-degeneracy does not influence our results and discussions,
since the true value for the atmospheric mixing angle is set to the
current best-fit value $\sin^22\theta_{23}=1$, which means that the degenerate
solution is the same as the original solution. The sign-degeneracy is
only exact in the vacuum case where $\hat{A}=0$ and the sign change of
$\Delta m^2_{31}$ can be exactly compensated by the additional
transformation $\deltacp\rightarrow -\deltacp$. In matter,
however, as for a neutrino factory baseline, this
degeneracy is not exact, and the second degenerate solution appears at a different 
value of $\deltacp$ and may in some cases affect the sensitivity to
CP violation. For the true value
$\deltacp=3\pi/2$, the sign-degenerate solution can move as function of
$\stheta$ and be located  at fit
values for $\deltacp$ near the CP conserving values. Therefore, CP
conservation cannot be excluded even if CP is maximally violated. This
effect is also called ``$\pi$-transit''~\cite{Huber:2002mx}.  
Similarly, the intrinsic or mixed degeneracies can destroy the CP violation
sensitivity if one of the degeneracies cannot be distinguished from
CP conservation.

Using neutrino and antineutrino channels, as well as spectral data,
one can resolve the intrinsic degeneracy
completely~\cite{Freund:2001ui}, and one can resolve the
sign-degeneracy (and thus determine the mass hierarchy) if the true
value of $\sin^22\theta_{13}$ is not too small. Nevertheless,
correlations and degeneracies cannot be resolved in all areas of the
parameter space. Thus, it is worthwhile to discuss additional
strategies to resolve these problems. For example, the correlation
with the phase $\deltacp$ can be completely resolved if the golden
channel measurement is performed at the so-called ``magic baseline''
at approximately 7500~km. At this baseline, by definition the
condition $\sin(\hat{A}\Delta)=0$ is fulfilled and only the term $T_1$
remains, while $T_2$, $T_3$, and $T_4$ vanish (\cf, \equ{t2} to
\equ{t4}). Therefore, a clean measurement of $\stheta$ becomes
possible~\cite{Huber:2003ak}. Note that one looses the sensitivity to
the CP phase at the magic baseline, and thus needs an additional
detector at a different baseline for a measurements of $\deltacp$.

The two additional appearance channels, the silver
$\nu_e\rightarrow\nu_\tau$ channel~\cite{Donini:2002rm} and the platinum
$\nu_\mu\rightarrow\nu_e$ channel~\cite{Bueno:2001jd} in principle allow to reduce the
effects of correlations and degeneracies since there the dependence on
the oscillation parameters is slightly different. The probability for
the silver channel is given by the expression
  \begin{eqnarray}
P_{e \tau} & = &
P_{e\mu}(s^2_{23}\leftrightarrow c^2_{23},\sin2\theta_{23}\rightarrow-\sin2\theta_{23})\ = \\
 & \simeq & \sin^2 2\theta_{13}\ T_1\ -\ \alpha\ \sin
 2\theta_{13}\ T_2 \nonumber\\ 
 & - &  \alpha\ \sin 2\theta_{13}\ T_3\ +\ \alpha^2\ T_4 \, . \nonumber
\end{eqnarray}  
Note, that the transformation $s^2_{23}\leftrightarrow c^2_{23}$ does
not affect the probability since $s^2_{23} = c^2_{23}$ holds for the
assumption of maximal mixing. The dependency on the CP phase is
different than for the other channels, since here the sign of the
CP-odd (containing $T_2$) and CP-even (containing $T_3$) terms is
changed relative to the other two terms, while only the sign of the
CP-odd term is changed by switching to the golden antineutrino channel
or the platinum channel. 
The platinum channel appearance probability
is given by 
\begin{eqnarray}
P_{\mu e} & = &
P_{e\mu}(\deltacp\rightarrow-\deltacp)\ = \\
 & \simeq & \sin^2 2\theta_{13}\ T_1\ -\ \alpha\ \sin
 2\theta_{13}\ T_2 \nonumber\\ 
 & + &  \alpha\ \sin 2\theta_{13}\ T_3\ +\ \alpha^2\ T_4, \nonumber
\end{eqnarray} 
which indicates that the platinum channel is comparable to
anti-neutrino running without the change due to the matter effect
(T-conjugated channel).
This could support measurements of $\deltacp$ for large values 
of $\stheta$, which are usually restricted 
due to the matter density uncertainty along this baseline.

A neutrino factory experiment, however, will \emph{not} measure
probabilities, but event rates, which are a convolution of the
oscillation probabilities, cross sections, detector efficiencies, {\it
  asf}.  Thus the relative merit of each option only can be assessed
by performing a simulation of event rates and a subsequent statistical
analysis of those simulated data. In this study this is done with the
GLoBES software~\cite{Huber:2004ka}. As input or so called true
values, we use, unless stated otherwise (taken from~\cite{Maltoni:2004ei})
\begin{eqnarray}
\Delta m^2_{31}=2.2^{+1.1}_{-0.8}\cdot10^{-3}\,\mathrm{eV}^2\quad\sin^2\theta_{23}=0.5^{+0.18}_{-0.16} \, , \nonumber \\
\Delta m^2_{21}=8.1^{+1.0}_{-0.9}\cdot10^{-5}\,\mathrm{eV}^2\quad\sin^2\theta_{12}=0.3^{+0.08}_{-0.07} \, , \nonumber\\
\sin^2\theta_{13}=0^{+0.047}_{-0}\quad \deltacp=0^{+\pi}_{-\pi} \, 
\label{equ:params}
\end{eqnarray}
where the ranges represent the current $3\sigma$ allowed ranges (see
also \Refs~\cite{Fogli:2003th,Bahcall:2004ut,Bandyopadhyay:2004da}).
In addition, we assume a 5\% external measurement for $\sdm$ and
$\theta_{12}$ from solar experiments at that time (see, \eg,
\Ref~\cite{Bahcall:2004ut}), and include matter density uncertainties
of the order of 5\%~\cite{Geller:2001ix,Ohlsson:2003ip} uncorrelated
between different baselines. We include the
$\mathrm{sgn}(\ldm)$-degeneracy and the
$(\theta_{13},\deltacp)$-degeneracy, whereas the octant degeneracy
does not appear for maximal mixing.

In order to allow for a concise comparison of the various options, we introduce
performance indicators. They are mainly chosen for their ability to
condense the information about the performance of a given setup
into a very small set of numbers. We are aware that this implies a
certain loss of information and detail. In the cases where this data
compression results in a bias towards or against a certain setup, we
provide more details in the text and/or additional figures.
As the performance indicators for the purpose of optimization, we
choose the $\stheta$, maximal CP violation, and mass hierarchy
sensitivities. We define the $\stheta$ sensitivity as the largest fit
value of $\stheta$ which fits the true $\stheta=0$. As illustrated in
\App~C of \Ref~\cite{Huber:2004ug}, this definition does not depend on
$\deltacp$ and the mass hierarchy if correlations and degeneracies are
taken into account. Compared to the $\stheta$ discovery potential, it
corresponds to the hypothesis of no signal. For the sensitivity to
maximal CP violation $\deltacp=\pi/2$ or $3 \pi/2$ (simulated value),
we test if one can exclude CP conservation $\deltacp=0$ {\em and}
$\deltacp=\pi/2$ (fit values) at the chosen confidence level.
Including correlations and degeneracies, this implies that any
degenerate solution fitting one of these two values destroys the CP
violation sensitivity. In addition, we define to have sensitivity to a
chosen mass hierarchy (normal or inverted simulated hierarchy) if we
can exclude {\em any} solution with the wrong hierarchy at the chosen
confidence level. Note that in all cases the unused oscillation
parameters are marginalized over (effect of correlations).
At some points, we will also use the $\stheta$ discovery potential,
which tests the hypothesis of nonzero $\stheta$ compared to the fit
$\stheta=0$. This performance indicator depends on the (simulated)
$\deltacp$ and the mass hierarchy. Furthermore, we will use the
``Fraction of (true) $\deltacp$'' as performance indicator for the
$\stheta$ discovery potential and the CP violation and mass hierarchy
sensitivities. Since the performance of all of these indicators
depends on the simulated/true $\deltacp$, the ``Fraction of (true)
$\deltacp$'' quantifies for what fraction of all possible values the
respective quantity can be discovered. 
Note that we will in most cases only discuss a normal simulated mass hierarchy,
since we know from earlier studies that the results do not look qualitatively
very different for the inverted hierarchy (see, \eg, \Ref~\cite{Huber:2005jk}).
The reason is the symmetric operation in the neutrino and antineutrino modes,
which means that the different hierarchy only means slightly adjusted statistics
due to different neutrino and antineutrino cross sections, \etc .\footnote{For the case of
the inverted hierarchy, the antineutrino appearance is matter enhanced instead of
the neutrino appearance, which means that statistics between the neutrino and
antineutrino rates becomes somewhat more
balanced; see, \eg,  Fig.~11 in \Ref~\cite{Huber:2005jk}. Therefore, the assumption
of the normal hierarchy may actually be the more conservative choice.}

%%%%%%%%%%%%%%%%%%%%%%%%%%%%%%%%%%%%%%%%%%%%%%%%%%%%%%%%%%%%%%%%%%%%%%

\section{Optimization of our ``standard'' neutrino factory}
\label{sec:optstd}

In this section, we define and optimize our standard neutrino factory
for specific performance indicators in a self-consistent matter. We
will discuss deviations and possible improvements from this definition
and their consequences in the following sections.

\subsection{Definition of our ``standard neutrino factory''}

As our ``standard neutrino factory'', we use the definition NuFact-II
from \Ref~\cite{Huber:2002mx} with some modifications that we will
discuss below.  This setup uses $1.06 \cdot 10^{21}$ useful muon
decays per year and a total running time of four years in each
polarity (corresponding to $5.3 \cdot 10^{20}$ useful muon decays per
year and polarity for a simultaneous operation with both polarities).
The detector is a magnetized iron detector with a fiducial mass of $50
\, \mathrm{kt}$ located in a distance $L$ from the source. We allow
the baseline $L$ and the muon energy $E_{\mu}$ to vary within a
reasonable range. In the standard setup, we only include the
$\nu_{\mu}$ appearance and disappearance channels (for neutrinos and
antineutrinos), where we assume that the best information on the
leading atmospheric parameters is determined from the experiment's own
disappearance channels.

\begin{figure}[t]
\begin{center}
\includegraphics[width=\textwidth]{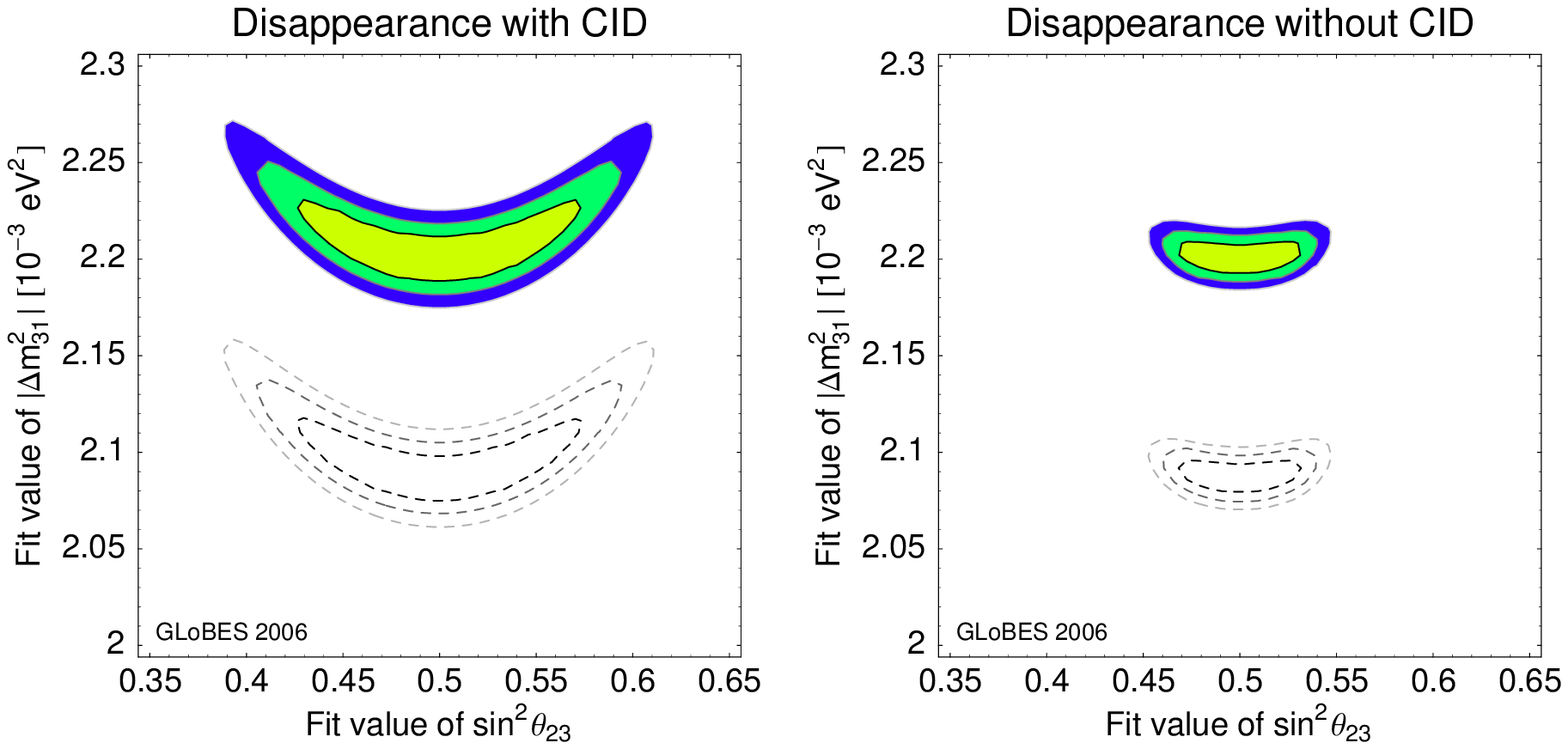}
\end{center}
\mycaption{\label{fig:cid} Comparison of
  $\ldm$-$\theta_{23}$-precision between CID (left) and no CID (right)
  in the disappearance channel including all correlations ($1\sigma$,
  $2 \sigma$, $3\sigma$, 2 d.o.f., $\stheta=0$). The appearance
  information is added as usual with CID. Dashed curves correspond to
  the inverted hierarchy solution.}
\end{figure}

Compared to the NuFact-II setup from \Ref~\cite{Huber:2002mx}, we
split the raw dataset into two samples: One with charge identification
(CID), the other without, where the dataset with charge identification
is used for the appearance channel and modeled according to
\Ref~\cite{Huber:2002mx}. As it can be seen from \figu{cid}, it turns
out to be useful {\em not} to use the CID information for the
disappearance channels (\cf, \Ref~\cite{deGouvea:2005mi}). This allows
to use also the low energy bins with full efficiency, which maximizes
the oscillatory signal. The price
one has to pay for that is that the neutrino and antineutrino rates
have to be added in this case, which is not a major problem for the
disappearance channel~\cite{Freund:2000ti}. However, as pointed out in
\Ref~\cite{deGouvea:2005mi}, the higher event rates at low energies
may lead to relatively fast oscillations especially for long
baselines, which can lead to problems for large muon energies and
small bin numbers. Therefore, we change the binning and use $43$ bins
in total.\footnote{We use $43$ variable bins from $1 \, \mathrm{GeV}$
  to $E_{\mu}$: 18 bins of $\xi \times 500 \, \mathrm{MeV}$, 10 bins
  of $\xi \times 1 \, \mathrm{GeV}$, and 15 bins of $\xi \times 2 \,
  \mathrm{GeV}$ from the lowest to the highest energy, where $\xi =
  (E_{\mu}-1)/49$ is an overall scale factor ($\xi=1$ correspond to
  the ``canonical'' $50 \, \mathrm{GeV}$ neutrino factory).} In
addition, we use the filter feature from GLoBES in order to average
any fast oscillations already on the probability level over a width of
$150 \, \mathrm{MeV}$.\footnote{We use the energy resolution type~2 to
  compensate for this additional energy smearing; \cf, GLoBES
  manual~\cite{Huber:2004ka}.} We have tested that this choice
appropriately describes the low energy range where the first
significant events enter in order to allow muon energies up to about
$100 \, \mathrm{GeV}$ in combination with baselines up to about $9 \,
000 \, \mathrm{km}$. Since we do not use CID in the disappearance
channel, we use the MINOS efficiencies and threshold from
\Refs~\cite{Ables:1995wq,Huber:2004ug} in this channel. Note that we
now have two different energy threshold functions. The
fact that there are almost no events below about $4 \, \mathrm{GeV}$
in the appearance channel is appropriately modeled.\footnote{For
  details on the shape of the appearance channel threshold function,
  the efficiencies, and model of the energy resolution, see \App~B.2
  of \Ref~\cite{Huber:2002mx}.} Finally, we choose 2.5\% for the
signal normalization errors, 20\% for the background normalization
errors, and $\sigma_E =0.15 \, E_\nu$ for the energy resolution.

\subsection{Optimized reach in $\boldsymbol{\stheta}$ as function of energy and baseline}

We now discuss the optimization of $L$ and $E_{\mu}$ for the $\stheta$
sensitivity, CP violation sensitivity, and mass hierarchy sensitivity,
assuming that we have a single ``standard'' neutrino factory
experiment. In this section, we are mainly interested in the reach in
$\stheta$, \ie, the smallest values of $\stheta$ for which a given
performance indicator can be probed.

\begin{figure}[t!]
\begin{center}
\includegraphics[width=10cm]{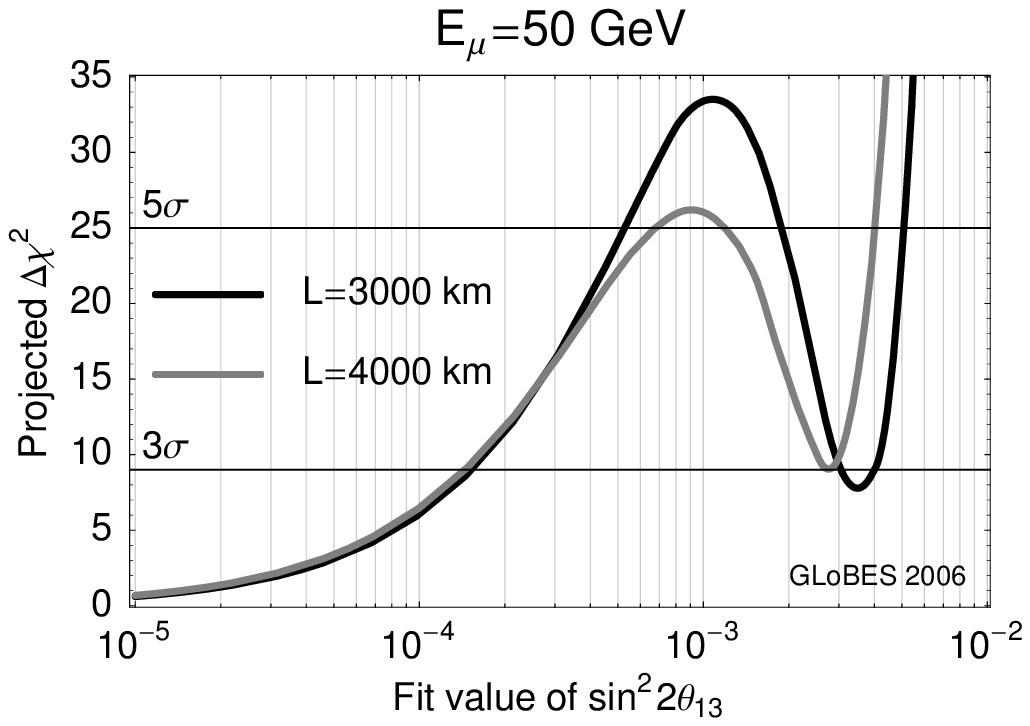}
\end{center}
\mycaption{\label{fig:degdid} Projected $\Delta \chi^2$ for the
  $\stheta$ sensitivity as function of the fit value of $\stheta$ for
  $E_\mu = 50 \, \mathrm{GeV}$ and two different baselines as given in
  the plot legend (includes degeneracies).}
\end{figure}

For the $\stheta$ sensitivity, one has to be carful to interpret the
results with respect to the
$(\theta_{13},\deltacp)$-degeneracy~\cite{Burguet-Castell:2001ez}. We
illustrate this challenge in \figu{degdid}. Since we define the
$\stheta$ sensitivity as the largest $\stheta$ which fits $\stheta=0$,
any degenerate solution will destroy the $\stheta$ sensitivity. The
interpretation in terms of the $(\theta_{13},\deltacp)$-degeneracy is
then as follows: If there is no signal (hypothesis $\stheta=0$), this
degeneracy (see bumps in the right-hand side of the figure) will
destroy the $\stheta$ sensitivity, \ie, a fake solution with a
relatively large $\stheta$ will still be consistent with $\stheta=0$.
Therefore, one will not be able to exclude that $\stheta$ could be
rather large. Since we want the results to be robust with respect to
this definition, we choose $\Delta \chi^2 =25$ (corresponding to
$5\sigma$) for all $\stheta$ sensitivity plots. As one can read off
from \figu{degdid}, choosing the $\Delta \chi^2=9$ (corresponding to
$3 \sigma$) would imply that very small changes in luminosity and
configuration could, depending on the baseline, lead to jumps of the
$\stheta$ sensitivity by an order of magnitude. For example, if one
was not able to achieve the originally anticipated luminosity by 10\%
for $\Delta \chi^2 =9$, the results would look qualitatively
different. However, for $\Delta \chi^2=25$, the degeneracy will always
be visible in the two different baseline cases in \figu{degdid}, and
these two cases will be interpreted as qualitatively similar (which
they in fact are).

\begin{figure}[t!]
\begin{center}
\includegraphics[width=0.9\textwidth]{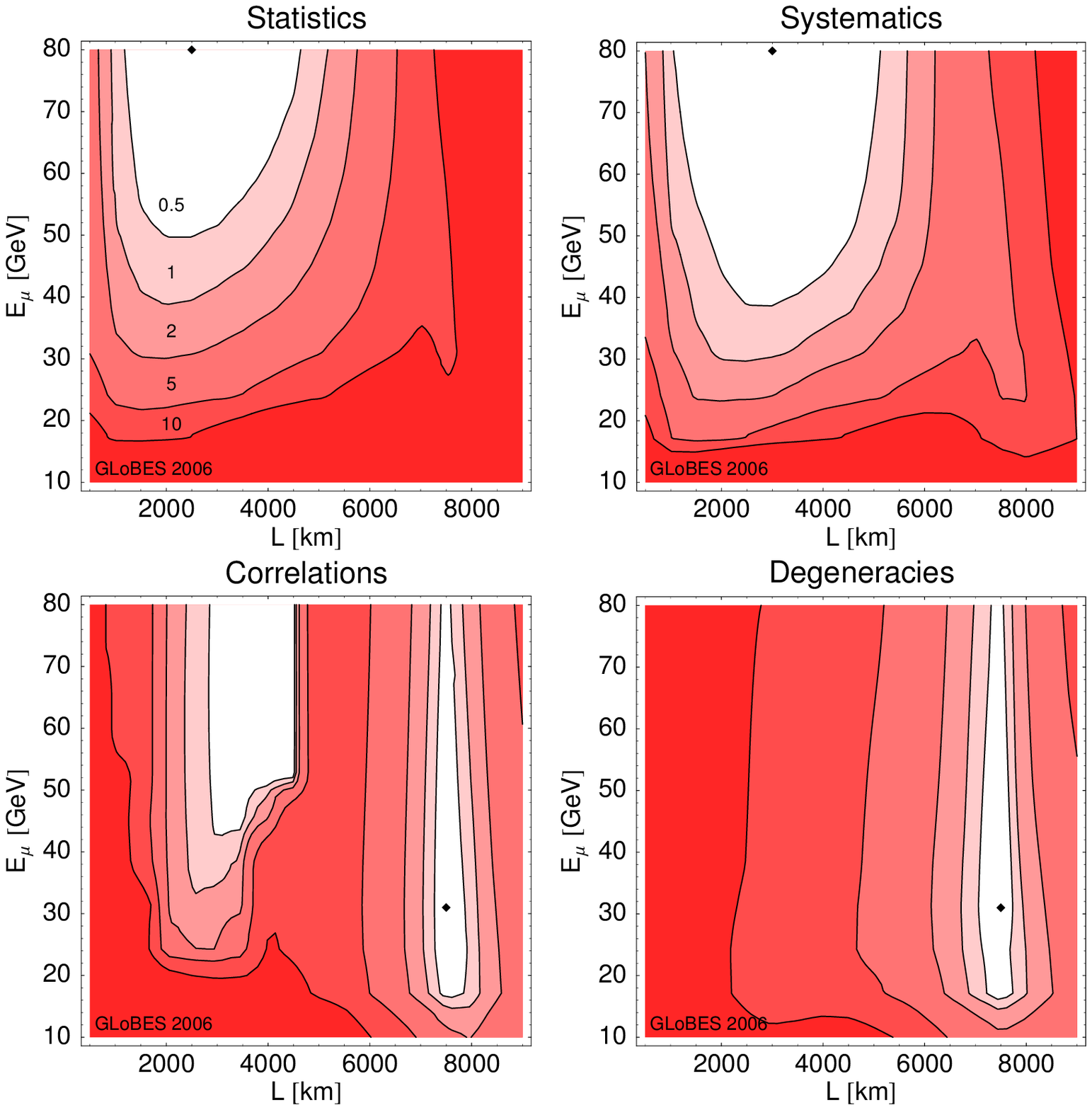}
\end{center}
\mycaption{\label{fig:t13sens} Sensitivity to $\stheta$ ($5 \sigma$)
  relative to the optimum (white) within each plot.  The different
  panels correspond to successively taking into account statistics,
  systematics, correlations, and degeneracies. The different contours
  represent the region within a factor of 0.5, 1, 2, 5, and 10 above
  the optimal sensitivity in each plot. The optimal sensitivities are
  $\stheta < 1.4 \cdot 10^{-5}$ (statistics), $2.8 \cdot 10^{-5}$
  (systematics), $2.4 \cdot 10^{-4}$ (correlations), and $5.0 \cdot
  10^{-4}$ (degeneracies), obtained at the energies and baselines
  marked by the diamonds.}
\end{figure}

In \figu{t13sens}, we show the $\stheta$ sensitivity ($5 \sigma$)
relative to the optimum (white) within each plot, \ie, the absolute
minima in the different plots are different. Not surprisingly, for
the systematics and correlations limits, baselines from $1 \, 000$ to
$ 4 \, 000 \, \mathrm{km}$ with as much muon energy as
possible give the best sensitivities. However, including correlations
and degeneracies, the ``magic baseline''~\cite{Huber:2003ak} at about
$7 \, 500 \, \mathrm{km}$ becomes more emphasized, where a
correlation- and degeneracy-free measurement of $\stheta$ is possible.
Most importantly, the optimal muon energies do not need to be higher
than about $40 \, \mathrm{GeV}$, even $30 \, \mathrm{GeV}$ are
absolutely sufficient for the long baseline. The reason for this is
that the $\stheta$-term in the appearance probability does not drop as
function of baseline at the mantle matter resonance energy. Therefore,
matter effects prefer lower energies, whereas higher muon energies
imply higher event rates and a relative decrease of events at the
mantle resonance. The optimum is determined by a balance between these
two factors. We have compared these results with the $\stheta$
discovery reach (systematics only). We find that qualitatively the
$\stheta$ discovery reach for a CP fraction of $0$ (best case of
$\deltacp$) is very similar to the upper row of \figu{t13sens}, but
the $\stheta$ discovery reach for a CP fraction of $100\%$
(conservative $\deltacp$) corresponds more to the lower row of
\figu{t13sens}. This result is not very surprising, since the
$\stheta$ sensitivity basically corresponds to the conservative, {\it
  i.e.} worst case true value of $\deltacp$, discovery reach.

\begin{figure}[t!]
\begin{center}
\includegraphics[width=10cm]{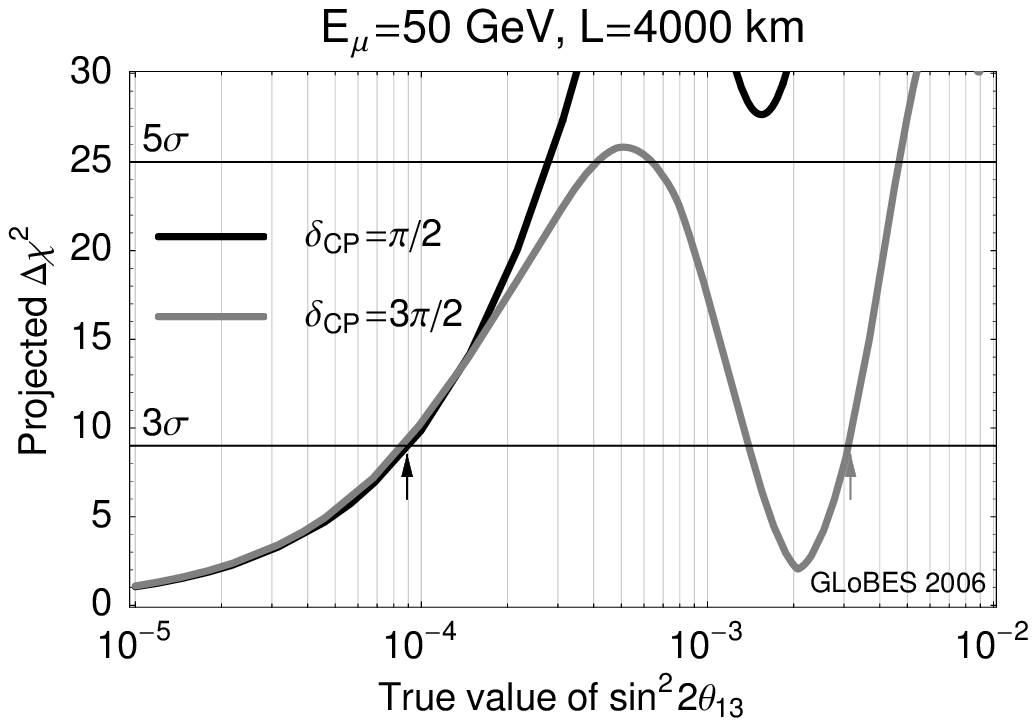}
\end{center}
\mycaption{\label{fig:cpdid} Projected (marginalized) $\Delta \chi^2$
  for the sensitivities to maximal CP violation $\deltacp=\pi/2$ and
  $\deltacp=3 \pi/2$ as function of the true value of $\stheta$ for
  $E_\mu = 50 \, \mathrm{GeV}$ and $L=4 \, 000 \, \mathrm{km}$.
  Sensitivity to maximal CP violation is given where $\Delta \chi^2>9$
  ($3 \sigma$). The arrow represents the ``conservative'' reaches for
  $\deltacp=\pi/2$ (gray arrow) and $\deltacp=3 \pi/2$ (black arrow)
  used for \figu{cpel}, \ie, thresholds in $\stheta$ above which CP
  violation can be found for any value of $\stheta$.}
\end{figure}

In order to discuss the CP violation sensitivity in terms of the
$L$-$E$-optimization, we have to sufficiently condense the
information. Since we are interested in the $\stheta$ reach, \ie, the
smallest values of $\stheta$ for which one can measure leptonic CP
violation, we have to define how to deal with ``gaps'' in the
$\stheta$ direction. This is illustrated in \figu{cpdid}, where CP
violation sensitivity is given for all shown true values of $\stheta$
when the function is above the chosen confidence level line.
Obviously, for maximal CP violation $\deltacp=3 \pi/2$, there is a gap
independent of the choice of $\Delta \chi^2=9$ or $25$, which is not
present for $\deltacp=\pi/2$. Therefore, we choose $\Delta \chi^2=9$
for all mass hierarchy and CP violation measurements because the
qualitative interpretation hardly depends on the confidence level. In
order to determine the $\stheta$ reach, we choose the rightmost
intersection with the chosen CL line, as illustrated by the arrows for
the two different curves (``conservative reach'').  In order to
illustrate the details for CP violation and mass hierarchy we will
show both figures with compressed information as well as we will later
show all regions where these measurements are possible. Note that the
interpretation is very different from the $\stheta$ sensitivity: Since
we show the CP violation sensitivity as function of the true
$\stheta$, we are, in principle, interested in all regions of the
parameter space where we can measure leptonic CP violation. This means
that one can measure CP violation if nature has chosen a value in the
sensitive regions. If there is only a small gap, not finding CP
violation in some sense would be ``bad luck''. On the other hand it is
a real risk for the experiment to fail. For the $\stheta$ sensitivity,
however, such a gap in the fitted value of $\stheta$ would mean that
we could not establish a small exclusion limit.

\begin{figure}[t!]
\begin{center}
\includegraphics[width=0.9\textwidth]{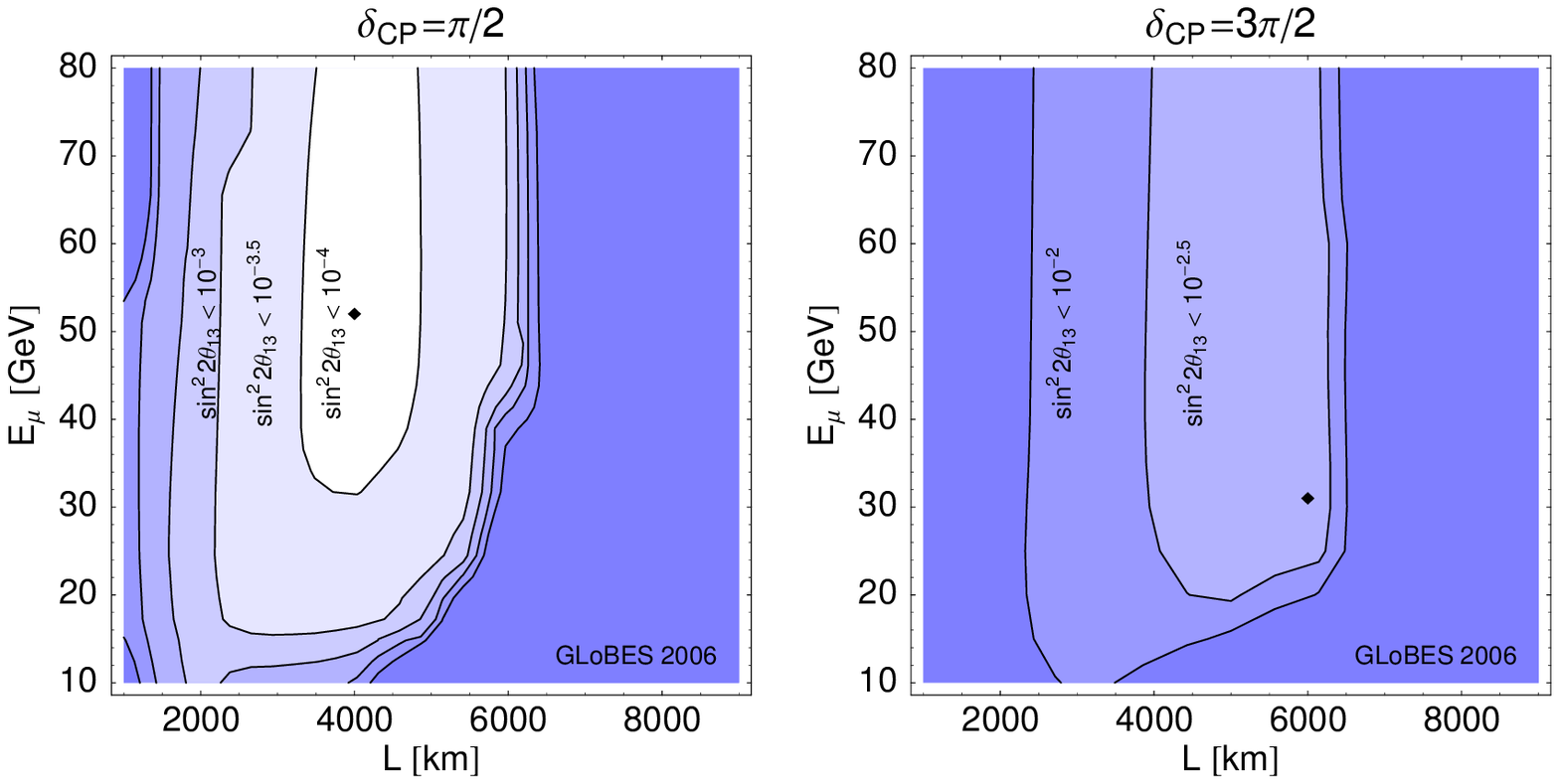}
\end{center}
\mycaption{\label{fig:cpel} Sensitivity to maximal CP violation
  ($\deltacp=\pi/2$ or $3 \pi/2$) for a normal mass hierarchy as
  function of $L$ and $E$. The sensitivity is given as absolute reach
  in $\stheta$ at the $3 \sigma$ confidence level including all
  parameter correlations and degeneracies. The minima are at $\stheta
  = 8.8 \cdot 10^{-5}$ (left plot) and $\stheta = 1.3 \cdot 10^{-3}$
  (right plot) and marked by the diamonds. See text for more
  explanations and definition of the $\stheta$ reach.}
\end{figure}

We show in \figu{cpel} the sensitivity to maximal CP violation for the
two different choices of $\deltacp$. For $\deltacp=\pi/2$, we find the
optimal performance at about $3 \, 000 - 5 \, 000 \, \mathrm{km}$ for
$E_{\mu} \gtrsim 30 \, \mathrm{GeV}$, where large energies are not
necessary. For the case $\deltacp = 3\pi/2$, the absolute $\stheta$
reach is rather poor, where we again have in this case shown the most
conservative value of $\stheta$ above which CP violation can be
determined. In this case, degeneracies affect the CP violation
performance. As it has been demonstrated in \Ref~\cite{Huber:2004gg},
the ``magic baseline'' can be used to resolve these degeneracies in
the third and fourth quadrants of $\deltacp$. Therefore, in order to
have optimal performance, a second baseline is necessary if $\deltacp$
turned out to be in this region.

\begin{figure}[t!]
\begin{center}
\includegraphics[width=\textwidth]{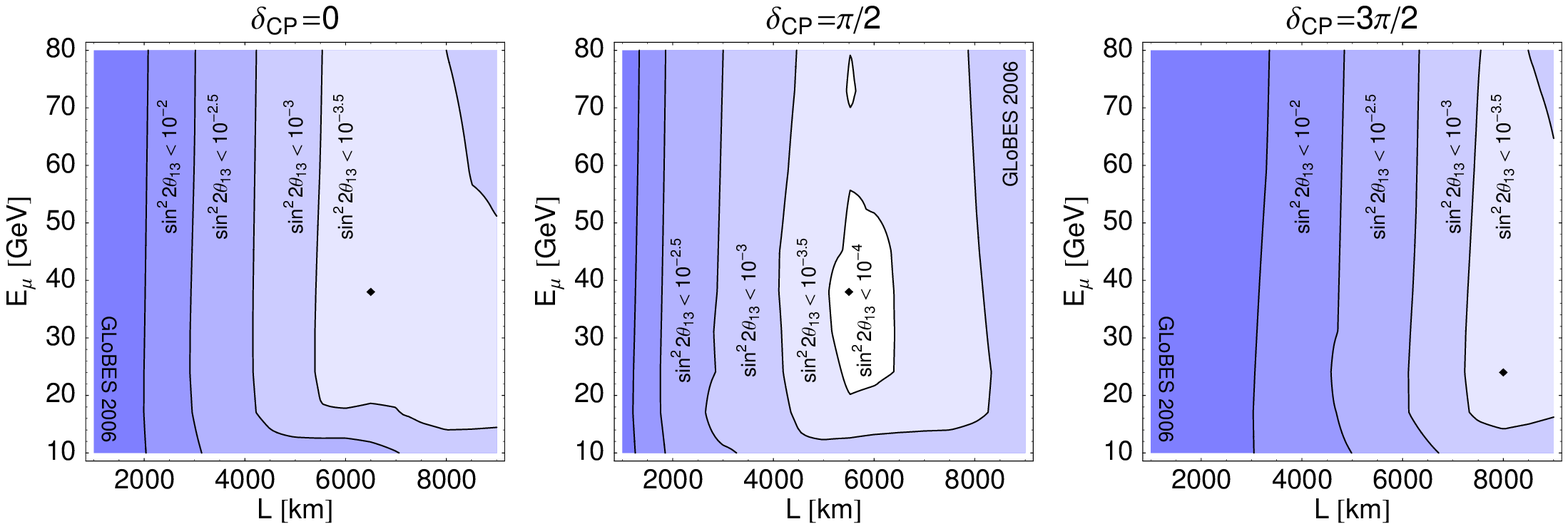}
\end{center}
\mycaption{\label{fig:signel} Sensitivity to a normal mass hierarchy
  for different values of $\deltacp$ (plot labels) as function of $L$
  and $E$. The sensitivity is given as absolute reach in $\stheta$ at
  the $3 \sigma$ confidence level including all parameter correlations
  and degeneracies. The minima are at $\stheta = 1.8 \cdot 10^{-4}$
  (left plot), $\stheta = 6.7 \cdot 10^{-5}$ (middle plot), and
  $\stheta = 1.6 \cdot 10^{-4}$ (right plot) and marked by the
  diamonds. See text for more explanations and definition of the
  $\stheta$ reach.}
\end{figure}

The normal mass hierarchy sensitivity reach in $\stheta$ is shown in
\figu{signel} for different values of $\deltacp$. 
As one can read off this figure, the mass hierarchy
sensitivity essentially increases with the baseline because of matter
effects. This means that for very small true values of $\stheta$, one
will need a very long baseline. The muon energy is of secondary
interest, as long as it is larger than about $20 \, \mathrm{GeV}$. In
fact, for $\deltacp=\pi/2$ or very long baselines $L > 8\, 000 \,
\mathrm{km}$, having a muon energy larger than $50 \, \mathrm{GeV}$ is
unfavorable because of the matter resonance at lower energies. In all
cases, the ``magic baseline'' $L \simeq 7\, 500 \, \mathrm{km}$ is
near the optimum.
There is, however, one feature which is not shown in \figu{signel}:
For certain values of $\deltacp$, there are gaps in the $\stheta$
direction (similar to \figu{cpdid}). In \figu{signel}, such gaps occur
for $\deltacp = 3\pi/2$, and we have chosen to show the most
conservative value of $\stheta$ above which mass hierarchy sensitivity
can be achieved for all values of $\stheta$. Therefore, \figu{signel},
right, actually shows the ranges for the ``gap-less'' determination of
the mass hierarchy. Thus, for very long baselines $L \gtrsim 7\, 500
\, \mathrm{km}$, the mass hierarchy can be determined in the full
shown range of $\stheta$. Note that in this case such a baseline
itself allows to resolve the degeneracies.

As far as the dependence on the true $\ldm$ is concerned, we have
tested somewhat larger values of $\ldm$, which could be suggested by
the latest MINOS results~\cite{Tagg:2006sx}, for the $\stheta$ and CP
violation sensitivities.  For the $\stheta$ sensitivity, the ``magic
baseline'' choice does not depend on $\ldm$. However, for $\ldm$
somewhat larger than the current best-fit value, the rates at both the
short and long baseline choices increase, and so does the absolute
performance at both baselines. However, it turns out that the relative
improvement at the magic baseline is even stronger, \ie, this baseline
choice becomes even more emphasized. For CP violation, the effect of a
larger $\ldm$ is essentially an improvement of the absolute reach
without baseline re-optimization (but slightly larger values of $E_\mu
\gtrsim 40 \, \mathrm{GeV}$ preferred). In addition, the baseline
window where one can measure CP violation becomes slightly broader.
For the mass hierarchy sensitivity, the absolute baseline length
determines the $\stheta$ reach, which means that the optimization
should hardly depend on $\ldm$.

\subsection{Optimized precision of the leading atmospheric parameters}

Except from any suppressed three-flavor effects, a neutrino factory
might be useful for the precision measurement of the leading
atmospheric parameters $\ldm$ and $\theta_{23}$. For simplicity, we
discuss the case of the true $\stheta=0$ in this section, because
$\stheta>0$ yields complicated correlations in the disappearance
channel (\cf, \eq~(33) in \Ref~\cite{Akhmedov:2004ny}). In addition,
we do not include degeneracies for the $\ldm$ precision.\footnote{The
  solution of the inverted hierarchy is, depending on the definition
  of the large mass squared splitting, always somewhat off the
  original solution. However, there is no qualitative difference to
  the best-fit solution for $\stheta=0$.} For $\theta_{23}$, we are
mainly interested in deviations from maximal mixing, which turns out
to be a useful indicator for neutrino mass
models~\cite{Antusch:2004yx}. Of course, this indicator is only useful
if $\theta_{23}$ is consistent with maximal mixing before the neutrino
factory operation.  However, the precision of $\theta_{23}$ behaves
very similar.

\begin{figure}[t!]
\begin{center}
\includegraphics[width=0.9\textwidth]{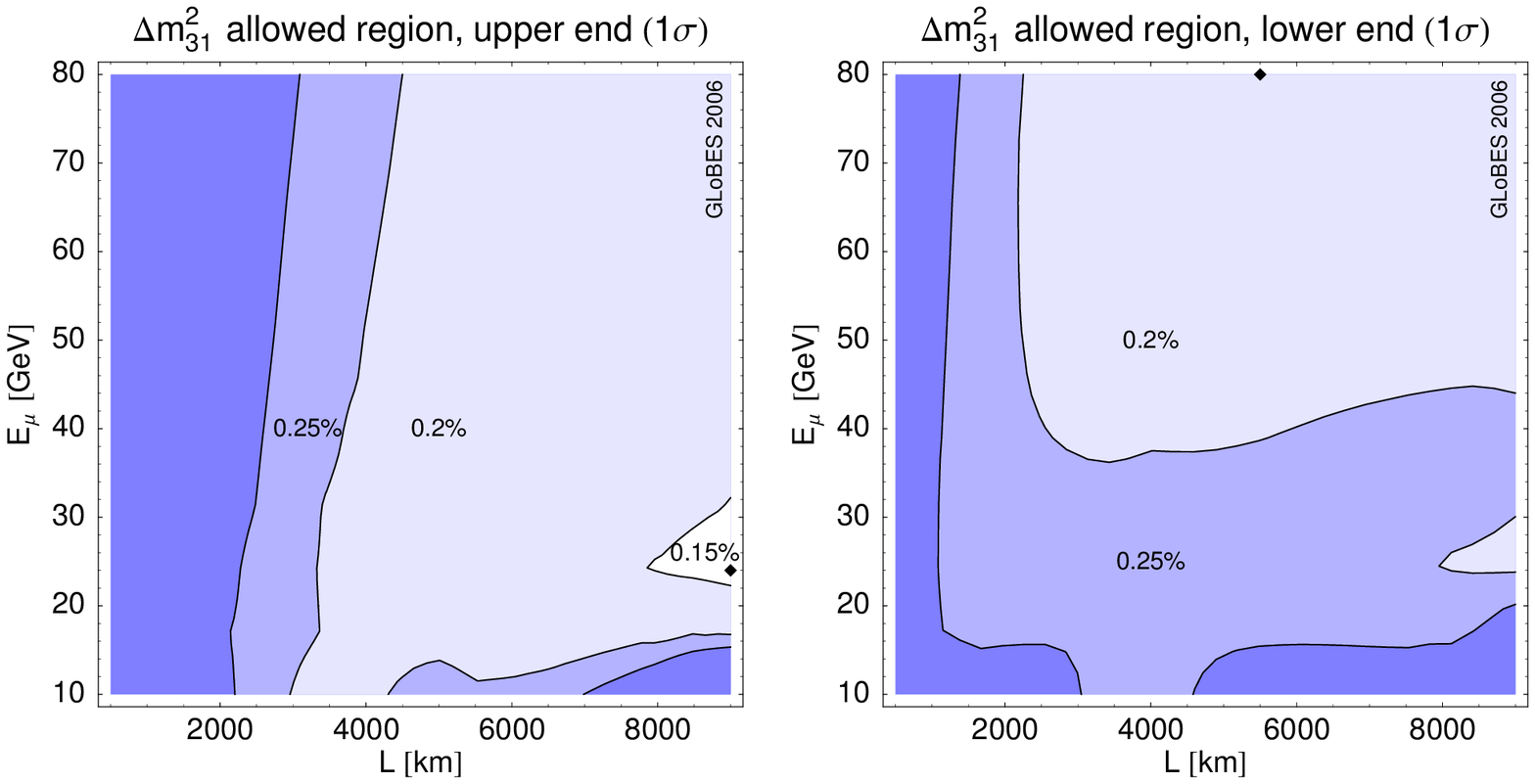}
\end{center}
\mycaption{\label{fig:dmel} Precision of $\ldm$ for a normal mass
  hierarchy and $\stheta=0$ as function of $L$ and $E$. The precision
  is given as relative precision $\ldm$ in per cent at the $1 \sigma$
  confidence level including all parameter correlations. The upper end
  (left panel) and lower end (right panel) of the allowed region are
  given separately because the $\Delta \chi^2$ is quite asymmetric.
  The minima occur at 0.14\% (left) and 0.18\% (right) and are marked
  by the diamonds.  }
\end{figure}

We show in \figu{dmel} the precision of $\ldm$ for a normal mass
hierarchy and $\stheta=0$ as function of $L$ and $E$. The precision is
given as relative precision $\ldm$ in per cent at the $1 \sigma$
confidence level including all parameter correlations. The upper end
(left panel) and lower end (right panel) of the allowed region are
given separately, because the $\Delta \chi^2$ is quite asymmetric in
many cases.  As the most important result, the separate analysis of
the dataset without CID yields an extremely good relative precision of
$\ldm$ of the order of $0.2 \%$ for $L \gtrsim 3 \, 000 \,
\mathrm{km}$ and $E_{\mu} \gtrsim 40 \, \mathrm{GeV}$. This extremely
high precision comes, compared to \Ref~\cite{Freund:2001ui}, from the
ability to resolve the oscillation maximum at low energies for long
enough baselines and good enough statistics because of the improved
threshold function without CID. In addition, the overall efficiency of
the disappearance channel is higher without CID. Though the total rate
decreases for longer baselines, more oscillation maxima can be
resolved. Note that we have included sufficiently many bins at low
energies to incorporate these effects. In general, the first
oscillation maximum can be found at
\begin{equation}
\frac{L_{\mathrm{max}}}{\mathrm{km}} \sim 564 \, \frac{E}{\mathrm{GeV}} \, ,
\label{equ:lmax}
\end{equation}
which more or less determines the optimal configuration.  If $L \ll
L_{\mathrm{max}}$, the $\sin$-term in the oscillation probability can
be expanded and $\theta_{23}$ and $\ldm$ are highly correlated.  This
means that $L \gtrsim 1 \, 000 \, \mathrm{km}$ is a necessary
condition to be able to disentangle $\theta_{23}$ from $\ldm$ because
of the energies where the first significant events enter ($\sim 2 \,
\mathrm{GeV}$). In addition, this formula explains the optimum for
$E_{\mu} \sim 10 \, \mathrm{GeV}$ at about $3 \ 500 \, \mathrm{km}$ if
one takes into account that the mean energy is somewhat below
$E_{\mu}$. In summary, a neutrino factory optimized for $\ldm$ has an
very good precision compared to all other available technologies.

\begin{figure}[t!]
\begin{center}
\includegraphics[width=0.4\textwidth]{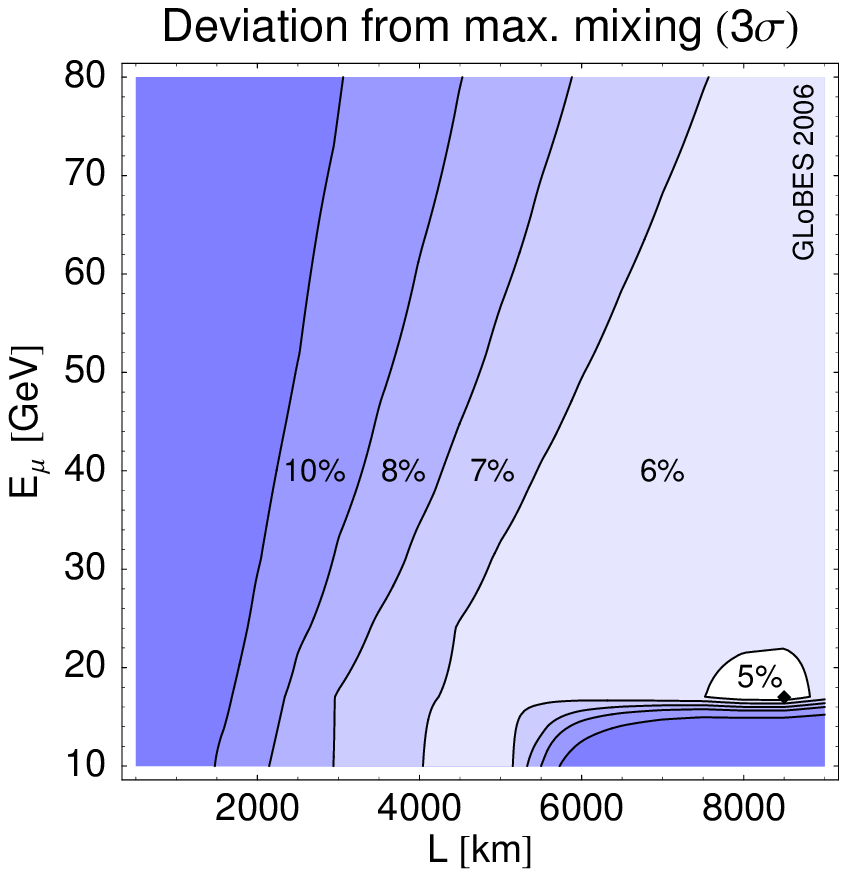}
\end{center}
\mycaption{\label{fig:devmaxel} Sensitivity to deviations from maximal
  mixing for a normal mass hierarchy and $\stheta=0$ as function of
  $L$ and $E$. The sensitivity is given as relative deviation of
  $\sin^2 \theta_{23}$ from $0.5$ in per cent at the $3 \sigma$
  confidence level including all parameter correlations, where only
  the upper branch $\sin^2 \theta_{23} >0.5$ is taken into account.
  The minimum as at 4.2\% and marked by the diamond.}
\end{figure}

For $\theta_{23}$, we show in \figu{devmaxel} the sensitivity to
deviations from maximal mixing for a normal mass hierarchy and
$\stheta=0$ as function of $L$ and $E$. The sensitivity is given as
relative deviation of $\sin^2 \theta_{23}$ from $0.5$ in per cent at
the $3 \sigma$ confidence level including all parameter correlations.
Note that only the upper branch $\sin^2 \theta_{23} >0.5$ is taken
into account, because there is hardly any sensitivity to the
$(\theta_{23}, \pi/2 - \theta_{23})$ ambiguity~\cite{Fogli:1996pv} and
the problem is very symmetric around $\theta_{23} = \pi/4$. We find a
very similar qualitative and quantitative behavior compared to
\Ref~\cite{Freund:2001ui}. However, the low energy performance for
very long baselines $L \gtrsim 6 \, 000 \, \mathrm{km}$ is
significantly improved because the efficiencies at lower energies are
better without CID. Most importantly, it is very hard to improve the
sensitivity to deviations from maximal mixing with the given setup,
probably because of the rather large normalization uncertainties. In
particular, T2HK could achieve a similar quantitative
performance~\cite{Antusch:2004yx}.

\subsection{Optimization for large $\boldsymbol{\stheta}$}

\begin{figure}[t!]
\begin{center}
\includegraphics[width=0.9\textwidth]{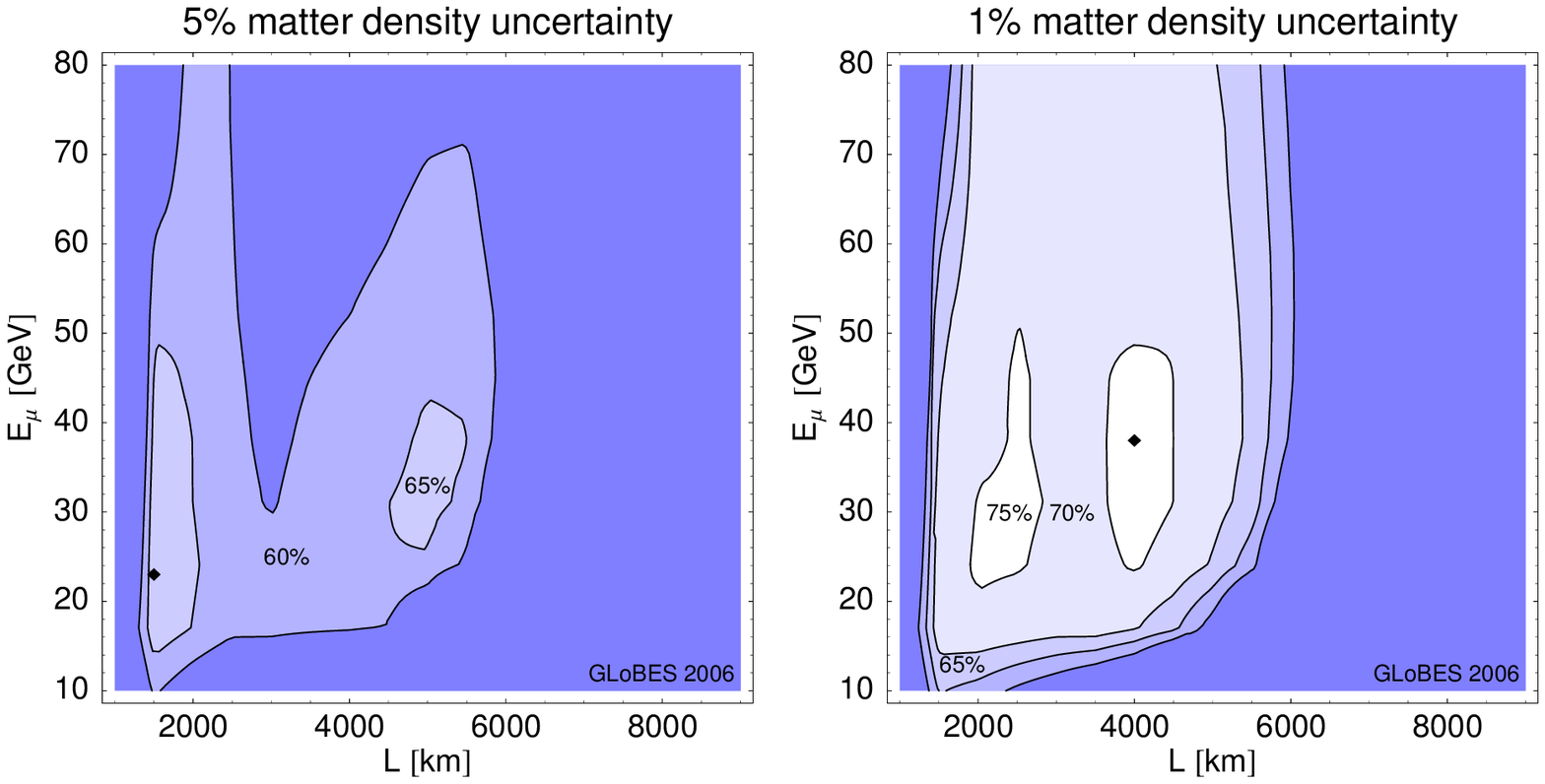}
\end{center}
\mycaption{\label{fig:lthetael} Fraction of (true) $\deltacp$ as
  function of $L$ and $E$ for the measurement of CP violation for
  $\stheta=0.1$ and a normal mass hierarchy ($3\sigma$, including all
  parameter correlations and degeneracies). The left panel corresponds
  to a matter density uncertainty of $5\%$, and the right panel to a
  matter density uncertainty of $1\%$. The optima are at 68\% (left
  plot) and 77\% (right plot) and are marked by the diamonds.}
\end{figure}

Let us now assume that $\stheta$ is large, such as $\stheta \simeq
0.1$, which means that it will be quickly found by the next generation
of superbeam experiments.  However, it is well known that for large
$\stheta$ matter density uncertainties affect the precision
measurements of $\stheta$ and $\deltacp$ (see, \eg,
\Refs~\cite{Huber:2002mx,Ohlsson:2003ip}).  Therefore, it is an
interesting question if the optimization changes for large $\stheta$,
and if one can exceed the performance of conventional techniques.
Since maximal CP violation measurements and mass hierarchy
measurements should for most parameter values not be a problem in this
case, but usually depend on the true $\deltacp$, we choose as a
performance indicator the fraction of all values of $\deltacp$ for
which CP violation or the mass hierarchy can be established at the $3
\sigma$ confidence level.

For the mass hierarchy, we find that the optimization is hardly
affected by the matter density uncertainty. As a general rule of
thumb, one can measure it for all values of $\deltacp$ for $L \gtrsim
1\, 000 \, \mathrm{km}$ almost independent of the muon energy. Later,
we will also demonstrate that the discovery of $\stheta$ is possible
independent of $\stheta$. This means that the only relevant question
for large $\stheta$ is the CP violation sensitivity, which we show in
\figu{lthetael} as the CP fraction for $\stheta=0.1$ and a normal mass
hierarchy as function of $L$ and $E$.  The left panel corresponds to a
matter density uncertainty of $5\%$, and the right panel to a matter
density uncertainty of $1\%$. As a first result, the maximum
achievable CP fraction depends on the matter density uncertainty, and
is only marginally affected by a different baseline choice in the
window between $1 \, 500$ and $5 \, 500 \, \mathrm{km}$ for small
matter density uncertainties.  Comparing \figu{lthetael}, right, with
\figu{cpel}, left, also illustrates that for small values of the
matter density uncertainty, the ``usual'' optimization for CP
violation is qualitatively recovered. However, from \figu{lthetael},
left, we can read off a very different result for large matter density
uncertainties which are more realistic for current
knowledge~\cite{Geller:2001ix,Ohlsson:2003ip,Pana}.  As a very
important result, the often used combination $L=3 \, 000 \,
\mathrm{km}$, $E_{\mu}=50 \, \mathrm{GeV}$ performs especially bad. It
is not trivial to explain this minimum: First of all, small muon
energies are preferred because matter density uncertainties hardly
affect the the $\stheta$-term in \equ{t1} (which is acting as
background to the $\deltacp$ measurement) close to the matter
resonance (\cf, \fig~3 of \Ref~\cite{Ohlsson:2003ip}).  Second, small
baselines are preferred because there the matter effects are small in
general, and therefore also the absolute impact of matter density
uncertainties is small.  Third, there is a second optimum for $L
\simeq 5 \, 000 \, \mathrm{km}$, where the CP asymmetric term is
enhanced for $E \sim 10 \, \mathrm{GeV}$ (\cf, \equ{lmax}; remember
that the mean energy of the spectrum is considerably below the muon
energy). These factors together cause the structure in
\figu{lthetael}, left. From \figu{lthetael}, right, we can read off
that the optimal performance for small matter density uncertainties is
reached in a wide range of $L$ and $E_{\mu}$.

\begin{table}[t]
\begin{center}
\begin{tabular}{lrrr}
\hline
Performance indicator & $L$ [km] & $E_{\mu}$ [GeV] \\
\hline
{\bf Three-flavor effects:} \\
$\stheta$ sensitivity & $\sim 7 \, 500 $ (``magic baseline'') & 20-50 \\
Mass hierarchy sensitivity & $\gtrsim 6 \, 000 $ & 20-50 \\
Max. CP violation sensitivity & $\sim 3 \, 000 - 5 \, 000$ & $>$ 30 \\
\hline
{\bf Leading atmospheric parameters:} & \\
$\ldm$ precision & $ \gtrsim  3 \, 000$ & $\gtrsim 40$\\
Deviation from maximal mixing ($\theta_{23}$) & $\gtrsim 3 \, 500 + 50 \cdot E_{\mu} /\mathrm{GeV}$ & $\gtrsim 20$ \\
\hline
{\bf Optimization for large $\boldsymbol{\stheta}$:} \\
Mass hierarchy sensitivity & $> 1 \, 000$ & $>10$\\
CP violation sensitivity ($\Delta \rho = 1\% \, \bar{\rho}$)  &  $\sim 1 \, 500 - 5 \, 500$ & 20-50  \\
CP violation sensitivity ($\Delta \rho = 5\% \, \bar{\rho}$) &  $\sim 1 \, 500 - 2 \, 000$ & 20-50  \\
 & $\sim 4 \, 500 - 5 \, 500$ & 20-40  \\
\hline
\end{tabular}
\end{center}
\mycaption{\label{tab:requirements} Requirements for the near-optimal performance of our ``standard neutrino factory'' (one individual experiment) for $\ldm = 0.0022 \, \mathrm{eV}^2$ for different performance indicators.}
\end{table}

\vspace*{0.3cm}

We summarize in \Tab~\ref{tab:requirements} the requirements for the
optimization of our standard neutrino factory. There are two very
important results. For the baselines, we need two different baselines
for the optimal performance: A ``shorter'' baseline $L \sim 3 \, 000 -
5 \, 000 \, \mathrm{km}$ for CP violation and leading atmospheric
parameter measurements, and a longer baseline $L \simeq 7 \, 500 \,
\mathrm{km}$ for the sensitivity to $\stheta$, mass hierarchy
sensitivity reach, and the disentanglement of degeneracies for CP
violation measurements. For the muon energies, we find that $E_{\mu}
\gtrsim 20 \, \mathrm{GeV}$ is sufficient for most applications, and
$E_{\mu} \sim 40 \, \mathrm{GeV}$ should be on the safe side.
Therefore, we find that the main challenge for a neutrino factory will
be the baseline, which can affect the physics potential much more than
a muon energy lower than previously assumed. For the rest of this
work, we will consider two baselines for our standard neutrino
factory, one at $4 \, 000 \, \mathrm{km}$ right at the optimal
$\stheta$ reach for CP violation (and close to the optimum for large
$\stheta$), and one at $7 \, 500 \, \mathrm{km}$ (magic baseline). For
the muon energy, we will use $50 \, \mathrm{GeV}$, unless stated
otherwise.

%%%%%%%%%%%%%%%%%%%%%%%%%%%%%%%%%%%%%%%%%%%%%%%%%%%%%%%%%%%%%%%%%%%%%%%%%%%%%%%%%%%%
\section{Detector requirements}
\label{sec:det}

A neutrino factory requires a large investment into accelerator R\&D
and infrastructure. Therefore, it is worth to consider an increased
effort on the detector side of the experiment. The aspect of joint
optimization of both accelerator and detector has so far been
neglected, where the main problem is the lack of reliable performance
predictions for large magnetic detectors.  The goal of this section is
{\em not} to prove the feasibility of certain detector properties or
parameters, but to demonstrate the possible gain in physics reach if
certain properties can be achieved.  Therefore, the following
statements or assumptions about the detector performance are not to be
mistaken as a claim of feasibility, but should be understood as
desirable improvements to be determined by extensive R\&D.
Nevertheless, we have tried to choose our assumptions not too far away
from what seems to be possible~\cite{ISSdetectorWG}. We will, however,
discuss how variations of our assumptions affect the physics results
in some cases. Thus, the results may serve as guideline where to focus
efforts in detector R\&D, and will be indicative of the expected
improvements as well. They should be interpreted as ``optimization
potential of the detector'' rather than as ``optimized detector''.

\subsection{Improved detector assumptions}
\label{sec:requirements}

The main limitation of a neutrino factory compared to other advanced
neutrino facilities comes from the fact the standard detector has a
relatively high neutrino energy threshold (necessary for charge
identification), which makes the first oscillation maximum basically
inaccessible (\cf, \Ref~\cite{Cervera:2000vy}). All measurements have
therefore to be performed in the high energy tail of the oscillation
probability off the oscillation maximum. In different words, a
neutrino factory is optimized for high statistics in the appearance
channel, not for operation at the oscillation maximum. This is the
reason why it seems to be the experiment most affected by the
eightfold degeneracy~\cite{Barger:2001yr,Huber:2002mx}.  A number of
solutions to this degeneracies problem has been proposed, amongst them
it has been studied what a better detector in terms of a better
neutrino energy threshold could achieve~\cite{Huber:2002mx}. We will
pick up this starting point and discuss improvements in the detection
threshold and energy resolution in this section.

The high neutrino energy threshold in \Ref~\cite{Cervera:2000vy} is
the result from optimizing for the purest possible sample of wrong
sign muons, which clearly puts the emphasis on events with a high
energy muon. The lower the muon energy is, the higher the likelihood
to mis-identify the muon charge or the nature of the event (CC vs NC)
becomes.  Thus the background increases with decreasing neutrino
energy, since the average muon energy will decrease with the neutrino
energy. The background fraction scales with the neutrino energy such
as a power law with a spectral index around $-2$. Our background model
assumes that whatever happens with the threshold will only affect
events below the threshold, but not events above, \ie , there is only
down-feeding of background but no up-feeding. The reason behind this
assumption is that a mis-identified NC event always should have a
reconstructed energy which is lower than the true energy, since there
is missing energy in every NC event. In order to roughly match the
total background obtained in \Ref~\cite{Cervera:2000vy}, we use a
background fraction $\beta E_\nu^{-2}$ with $\beta=10^{-3}$.
Integrating this background fraction from $4\,\mathrm{GeV}$ to
$50\,\mathrm{GeV}$ yields an average background fraction of $5\cdot
10^{-6}$. We assume this background fraction separately for the
background from neutral currents and wrong sign muons.

Achieving a lower threshold probably requires a finer granularity of
the detector, \ie , a higher sampling density in the calorimeter. This
should at the same time improve the energy resolution of the detector.
We use a parameterization $\sigma_E \, [\mathrm{GeV}]=\sigma \,
\sqrt{E_\nu \, [\mathrm{GeV}]}+0.085\,\mathrm{GeV}$ with $\sigma =
0.15$ for the energy resolution (as compared to $\sigma_E=0.15 \,
E_\nu$ before, corresponding to $\sigma \simeq 0.5$), where the
constant part models a lower limit from Fermi motion.\footnote{For the
  neutrino factory, this lower limit turns out to be of secondary
  importance because there are practically no events in the relevant
  energy range.} For definiteness, we take the neutrino energy
threshold to be $1\,\mathrm{GeV}$, and the efficiency to be constant
$0.5$ for all appearance neutrino events above threshold.  This setup
of combined lower threshold, increasing background fraction, and
better energy resolution will be called ``optimal appearance''.
Similar numbers are quoted for the \NOVA\ 
detector~\cite{Ambats:2004js}, which is a totally active
calorimeter\footnote{Using an air coil system similar to the one in
  ATLAS, it should be possible to magnetize a detector like this.}.

In order to illustrate the sensitivity of the results to
these numbers, we will use the following setups:
\begin{enumerate}
\item
 {\bf Standard} detector, as from the last section.
\item {\bf Optimal appearance:} $\sigma=15\%$, $\beta=10^{-3}$, full
  efficiency of 50\% already reached at $1 \, \mathrm{GeV}$.
\item
 {\bf Better threshold:} Same as 2), but $\sigma=50\%$ (similar to 1).
\item
 {\bf Better energy resolution:} Same as 2), but old threshold from 1).
\end{enumerate}

As before, we we assume that the systematical background uncertainty
is $20\%$ and the corresponding error $s$ for the signal is $s=2.5\%$
for all these setups.

\subsection{Impact on physics reach}

\begin{figure}[t]
\begin{center}
\includegraphics[width=0.45\textwidth]{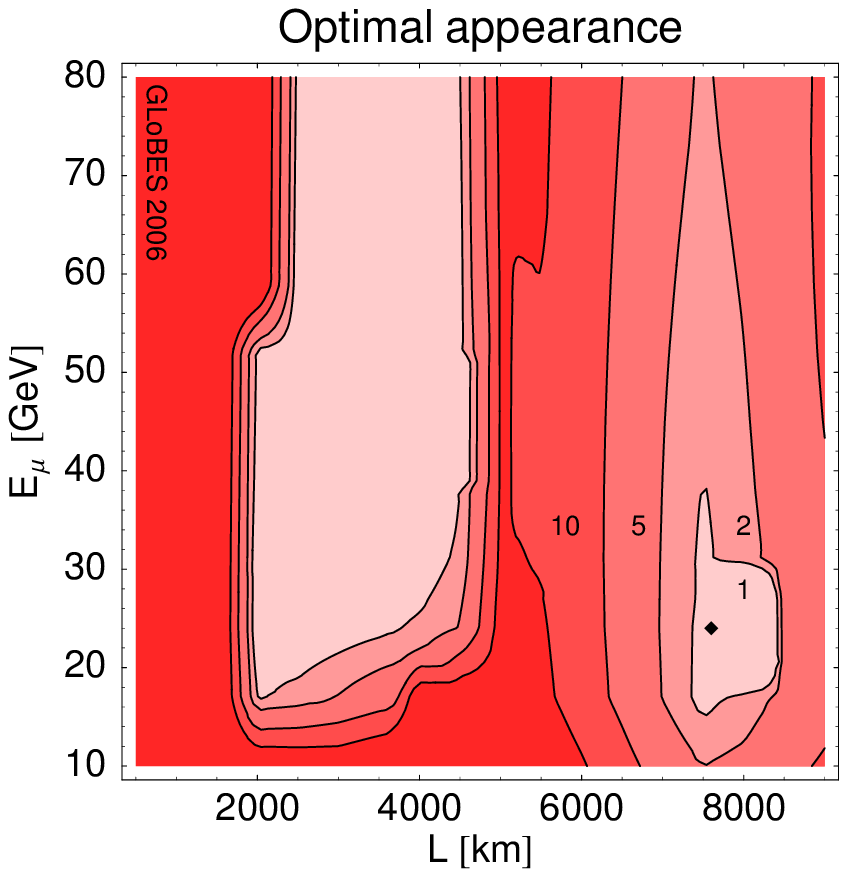}\hspace*{0.4cm}%
\raisebox{-0.5cm}{\includegraphics[width=0.51\textwidth]{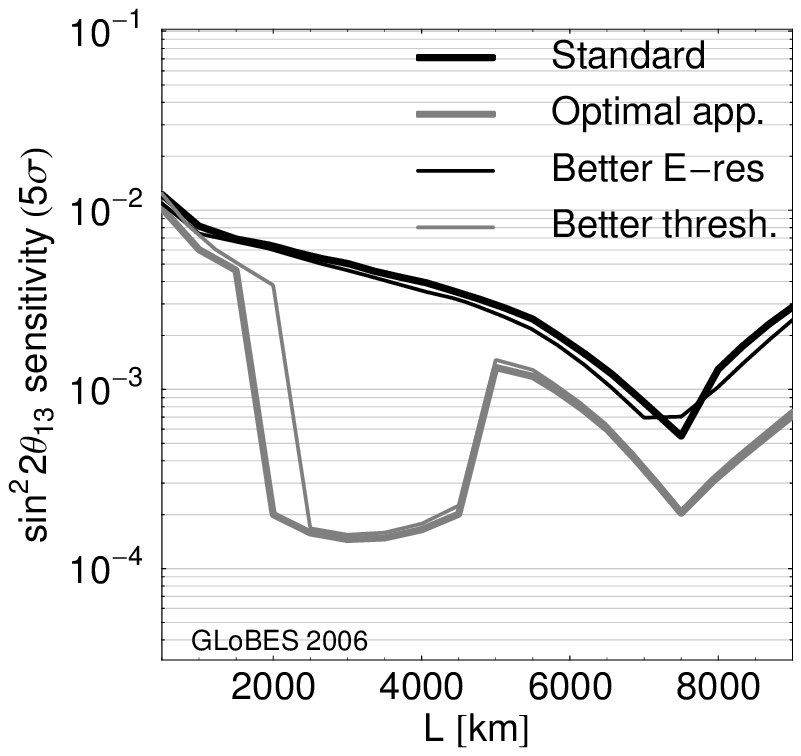}}
\end{center}
\mycaption{\label{fig:th13le} $\stheta$ sensitivity ($5\sigma$) for
  several improved detector options.  The left hand panel shows the
  $\stheta$ sensitivity as function of baseline and muon energy
  relative to the minimum for the `optimal appearance' detector
  including degeneracies similar to \figu{t13sens} (lower right). The
  minimum appears at $\stheta=1.1 \cdot 10^{-4}$ marked by the
  diamond. The right hand panel shows the $\stheta$ sensitivity as
  function of baseline for different detector options (see plot
  legend) and $E_\mu=50 \, \mathrm{GeV}$ fixed. Note that the better
  energy resolution option uses a different background model, which
  leads to the crossing with the ``standard'' curve at $L \sim 7 \,500
  \, \mathrm{km}$.}
\end{figure}

Changing the detector threshold by a large amount certainly should
impact the choice of the optimal baseline and muon energy. In the left
panel of \fig~\ref{fig:th13le}, the sensitivity to $\stheta$ at
$5\,\sigma$ is shown for the optimal detector as a function of the
baseline and muon energy including degeneracies.  The optimum is
marked by the diamond and has a value of $\stheta = 1.1 \cdot
10^{-4}$, it is located at around $7\,500\,\mathrm{km}$ and
$E_\mu=24\,\mathrm{GeV}$ similar to \figu{t13sens} (lower right).
Compared to \figu{t13sens}, the second optimum at shorter baselines is
still present including degeneracies, and the allowed muon energies
tend to be rather low. Even energies as low as $20\,\mathrm{GeV}$ now
work reasonably well for both baselines.
Next is is interesting to see whether the improvements are mainly due
to the lower threshold or energy resolution. This is illustrated in
the right hand panel of \fig~\ref{fig:th13le}, where different
combinations of better threshold or energy resolution are compared
with the standard setup with respect to their $\stheta$ sensitivity
(in this figure, $E_\mu$ is fixed to $50 \, \mathrm{GeV}$).  The main
effect for the $\stheta$ sensitivity improvement clearly comes from
lower energy threshold, the better energy resolution only plays a very
minor role. Note that the optimum in this figure occurs at around $3
\, 000 \, \mathrm{km}$ for the optimal detector because we have fixed
the muon energy. A comparison to \figu{th13le}, left, illustrates that
this is not the global minimum in $L$-$E_\mu$-space.

\begin{figure}[t]
\begin{center}
\includegraphics[width=\textwidth]{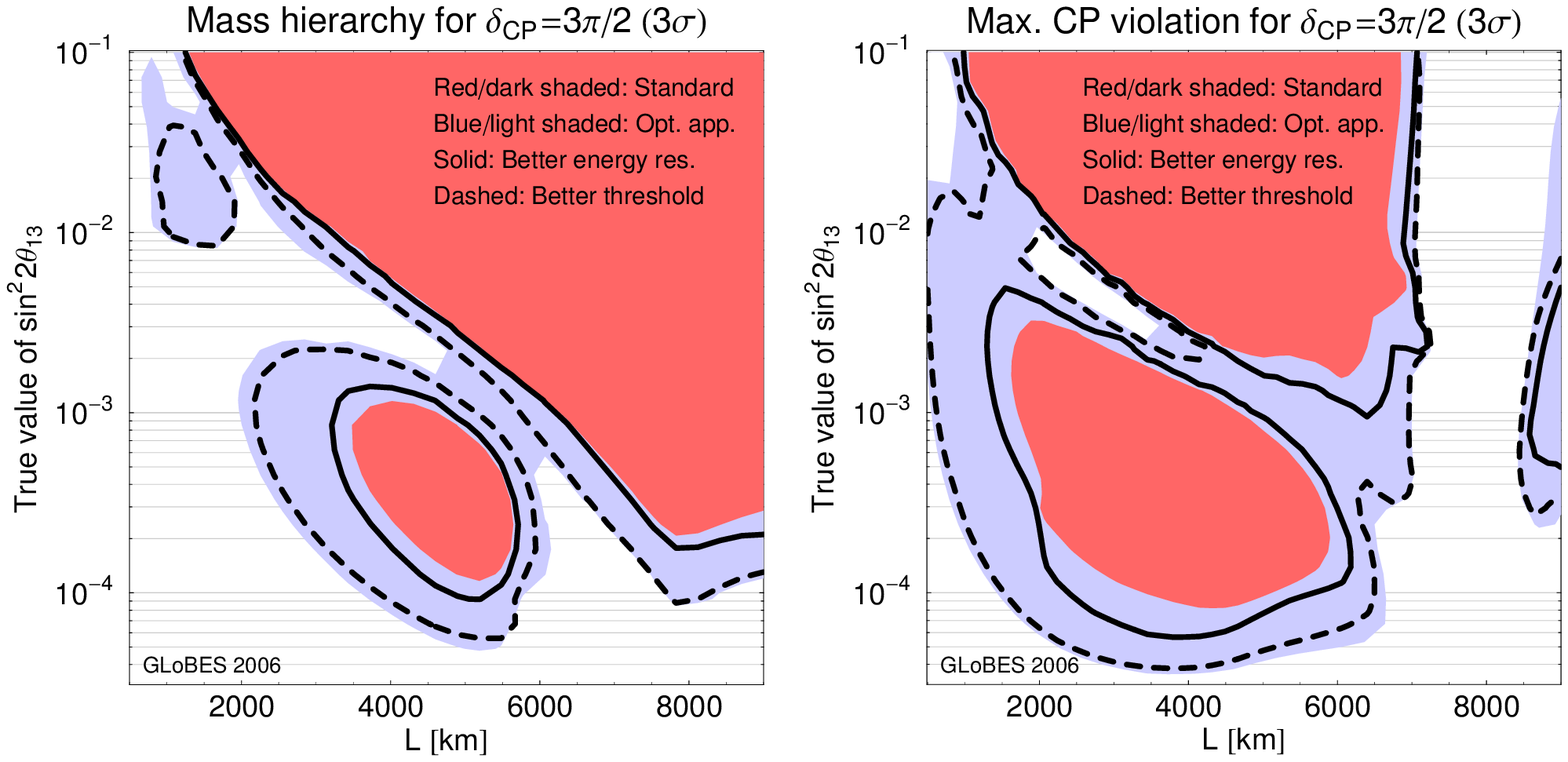}
\end{center}
\mycaption{\label{fig:mhcpoptdet} The normal mass hierarchy (left) and
  CP violation (right) sensitivities ($3 \sigma$) as function of
  baseline and true $\stheta$ for a normal hierarchy and $\deltacp=3
  \pi/2$, different detector options (see legend), and $E_\mu=50 \,
  \mathrm{GeV}$.  Sensitivity is given in the shaded/enclosed
  regions.}
\end{figure}

The behavior for the other performance indicators CP violation and
mass hierarchy is slightly different, as we discuss with
\figu{mhcpoptdet}.  In this figure, $\deltacp=3 \pi/2$ was chosen
since for this specific value degeneracies have a large impact
(compared to $\deltacp = \pi/2$) and any improvements are most obvious
there.
The left hand panel shows the sensitivity to the mass hierarchy at
$3\,\sigma$, where sensitivity is given within the shaded/ marked
areas.  The red (dark) shaded regions shows the result for the
standard detector whereas the blue (light) shaded region shows the
result for the optimal setup. Clearly, the accessible range in
$\stheta$ improves as well as the constraints on the baseline become
somewhat weaker for the better detector. The difference between having
only a better threshold (dashed line) and only a better energy
resolution (solid line) is quite large. Therefore, for the mass
hierarchy the main improvement comes from to the lower threshold as
well.
For CP violation in the right panel of \figu{mhcpoptdet}, the detailed
picture looks different but the conclusion is the same: Large
improvements come from a lower threshold, and there is only minor
influence of the energy resolution. The choice of the optimal $L$ and
$E$ seems to be basically unaffected by a better detector.

\begin{figure}[t]
\begin{center}
\includegraphics[width=\textwidth]{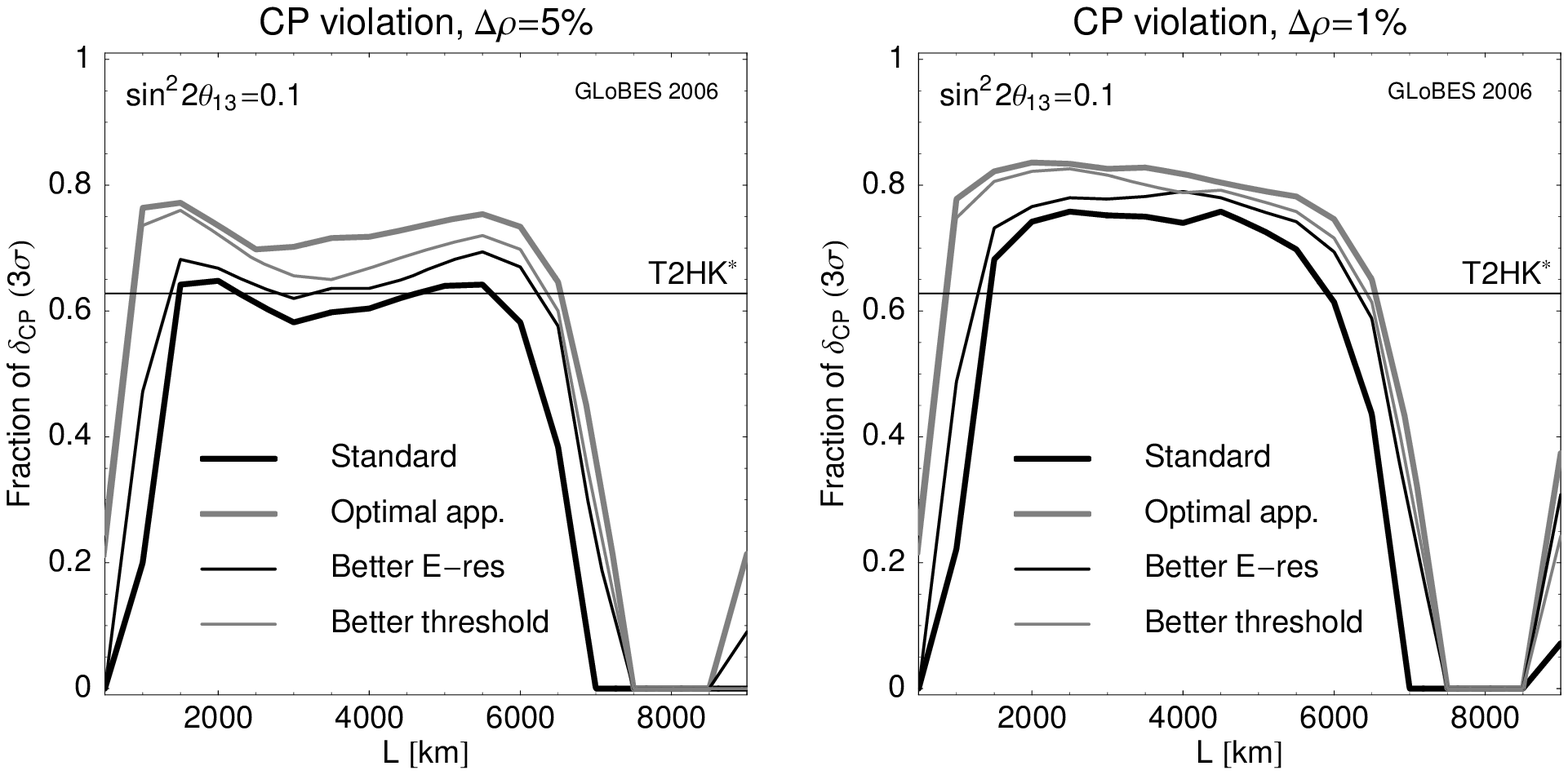}
\end{center}
\mycaption{\label{fig:CPfracoptdet} The CP fraction for the
  sensitivity to CP violation ($3 \sigma$) for a normal hierarchy as
  function of baseline for different detector options (see legend) and
  $E_\mu=50 \, \mathrm{GeV}$. The left plot correspond to $5\%$ matter
  density uncertainty, and the right plot to $1\%$ matter density
  uncertainty.}
\end{figure}

One important issue in this context is the performance of a neutrino
factory if $\stheta$ turns out to large, such as around $0.1$. There
will be information regarding this case from reactor experiments by
around 2010, such as from Double
Chooz~\cite{Ardellier:2004ui,Huber:2006vr}.  Note that we have stated
earlier that $\stheta$ discovery and mass hierarchy measurements are
unproblematic for large values of $\stheta$, which means that the
optimization is determined by $\deltacp$ measurements.  We show in
\figu{CPfracoptdet} the fraction of $\deltacp$ for the sensitivity to
CP violation as a function of the baseline for $\stheta=0.1$ and
different combinations of experimental setup and matter density
errors. In the case of large $\stheta$, alternative technologies, such
as superbeams, can be very competitive in their physics reach.
Therefore, we show for comparison as the grey line the CP fraction for
which \JHFHK\ would be sensitive to CP violation.\footnote{\JHFHK\ is
  the off-axis T2K upgrade as defined as in \Ref~\cite{Huber:2002mx},
  but uses a $m_{\mathrm{Det}} = 500 \, \mathrm{kt}$ water Cherenkov
  detector.  It is operated two years in the neutrino running mode and
  six years in the antineutrino running mode with a target power of $4
  \, \mathrm{MW}$. The baseline is $L=295 \, \mathrm{km}$.} In the
left hand panel the results are shown for the canonical value for the
matter density uncertainty of $5\%$. Clearly the standard neutrino
factory setup does not perform better than the superbeam. The
situation changes once better detectors are considered.  The optimal
setup defined previously would yield a significant improvement over
the superbeam for nearly all choices of the baseline above
$1500\,\mathrm{km}$.  It also can be seen that the improvement comes
from both the lower threshold and better energy resolution. In this
scenario, the detector performance is crucial in making the case for a
neutrino factory.
The right hand panel shows the result if the matter density
uncertainty could be reduced down to $1\%$. Quite obviously this would
further improve the performance of neutrino factory, as well as it
affects the baseline somewhat.
We have checked that these results for the optimal detector hold for a
lower muon energies around $20\,\mathrm{GeV}$ as well, \ie, though $50
\, \mathrm{GeV}$ do not harm, $20 \, \mathrm{GeV}$ are sufficient in
this case. Thus, for the case of large $\stheta$, we conclude that
improving the detector energy resolution and energy threshold would
allow to choose a shorter baseline of about $1500\,\mathrm{km}$ and a
muon energy of $20\,\mathrm{GeV}$, while the option $4000 \,
\mathrm{GeV}$ at $50 \, \mathrm{km}$ does not mean a significant loss
in sensitivity (the loss is, depending on the matter density
uncertainty, about $0.05$ to $0.08$ in the CP fraction between the
optimum and this point).  Furthermore, for one neutrino factory
baseline only, it can be concluded that lower threshold, better energy
resolution, and lower matter density uncertainty would equally help to
improve the performance.

\subsection{Systematics impact and disappearance channel}

Above we have defined a background model and assumed a certain
systematical uncertainty on the signal. Here we show how our results
for the measurement of CP violation change if we modify the input
values for the background fraction $\beta$ and the signal
normalization error $s$ (as defined in \Sec~\ref{sec:requirements}).
In addition, we discuss the impact of energy resolution and
systematics on the disappearance channel.

\begin{figure}[t]
\begin{center}
\includegraphics[width=8cm]{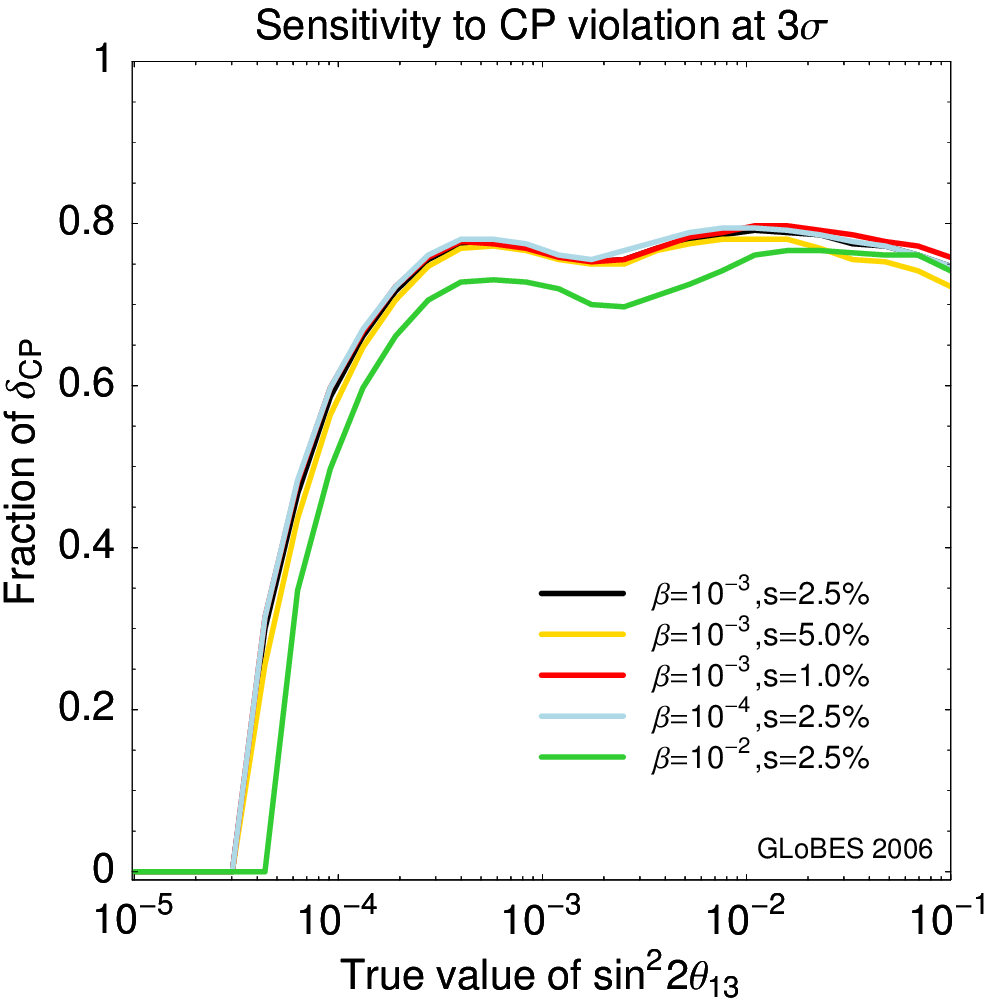}
\end{center}
\mycaption{\label{fig:cpsys} Fraction of $\deltacp$ for which CP
  violation can be established at $3\,\sigma$ as function of
  $\stheta$. The different curves are for different assumptions about
  background and systematical uncertainty of the signal as defined in
  the legend.  }
\end{figure}

\figu{cpsys} shows the impact of varying the systematics parameters
$\beta$ or $s$ on the CP violation measurement. To a very good
approximation, it is safe to say that varying $s$ from $1\%$ to $5\%$
does not change the results at all. Furthermore, \figu{cpsys} also
shows that $\beta$ is only important as far as it may not become too
large, but even a factor of 10 is not devastating.  Note, that the
error on the background is assumed to be $20\%$, which is quite
conservative compared to the numbers usually quoted for superbeams.
Certainly the impact of an increased background will be strongly
reduced by reducing this uncertainty.

The disappearance channels are mainly used to determine the 
atmospheric neutrino parameters $\ldm$ and $\sin^2\theta_{23}$. As
shown in \figu{cid}, the obtainable accuracies are very impressive
even with the standard setup. It also can be seen from that
figure that having a low threshold is important in order to properly
cover the first oscillation dip.
\begin{figure}[t]
\begin{center}
\includegraphics[width=\textwidth]{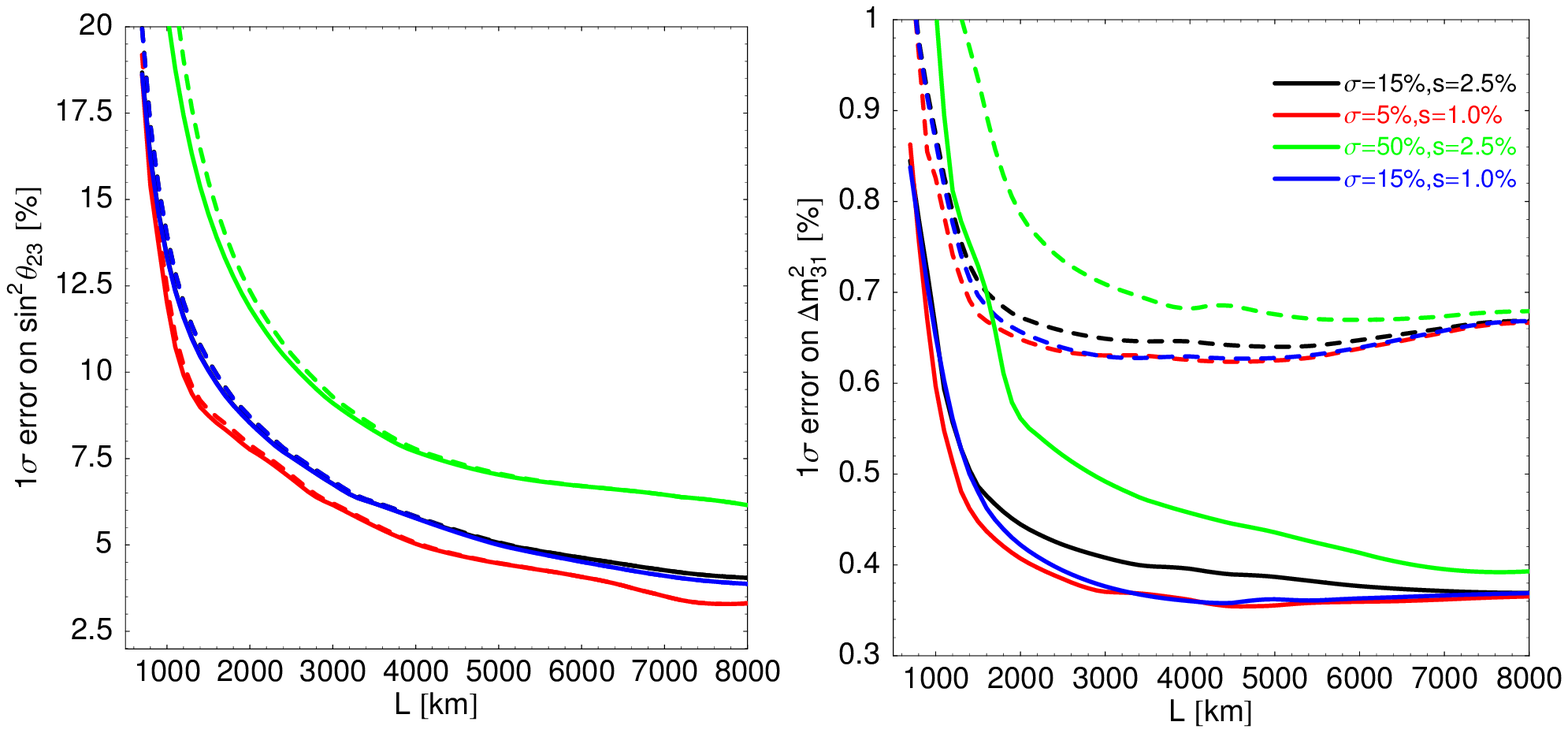}
\end{center}
\mycaption{\label{fig:atmprec} The relative $1\,\sigma$
  (full width) errors on $\sin^2\theta_{23}$ (left hand panel) and
  $\ldm$ (right hand panel) as a function of the baseline. The result
  is shown for various combinations of energy resolution $\sigma$ and
  systematic error $s$. The dashed lines assume an uncertainty on
  $\sdm$ and $\theta_{12}$ of $10\%$, whereas the solid lines are
  calculated for the default value of $5\%$. The results are computed
  for $\stheta\equiv0$.}
\end{figure}
The optimal detector considered here will not improve the threshold
for the disappearance channel (because we do not use charge identification
for that channel), but the energy resolution will be much
better. It has been demonstrated in \Ref~\cite{deGouvea:2005mi} that the energy
resolution has a large influence on the accuracy for the leading
parameters. In \figu{atmprec} the relative $1\,\sigma$
(full width) errors on $\sin^2\theta_{23}$ (left hand panel) and
$\ldm$ (right hand panel) as a function of the baseline are shown. The
different colored lines correspond to different values of the energy
resolution $\sigma$ and the normalization error of the
signal $s$. Interestingly, the signal error $s$ seems to be quite
unimportant. The energy resolution, on the other hand, has a
relatively large impact, especially at the shorter baselines. The dashed
lines show the results if the error on the solar parameters were
$10\%$ instead of $5\%$, and one can see that this would deteriorate
the results considerably. Irrespective of the error on the solar
parameters and the energy resolution, longer baselines are preferred
especially for $\sin^2\theta_{23}$.

%%%%%%%%%%%%%%%%%%%%%%%%%%%%%%%%%%%%%%%%%%%%%%%%%%%%%%%%%%%%%%%%%%%%%%%%%
\section{Addition of silver and platinum channel data}
\label{sec:channels}

So far we have discussed the $\nu_{\mu} \rightarrow
\nu_\mu$-disappearance channel for the leading atmospheric oscillation
parameters, and the ``golden'' $\nu_e \rightarrow \nu_\mu$-appearance
channel for sub-leading three-flavor effects, \ie , $\stheta$,
$\deltacp$ and the sign of $\Delta m^2_{31}$.  Besides these channels,
the neutrino flavors contained in the beam of a neutrino factory allow
for additional oscillation channels which could help to resolve
correlations and degeneracies: the $\nu_e \rightarrow
\nu_\tau$-appearance channel (``silver channel'') and the $\nu_\mu
\rightarrow \nu_e$-appearance channel (``platinum channel''); for
details on the phenomenology, see \Sec~\ref{sec:sim}.  In this
section, we first describe the definition for the silver and platinum
channels as is used throughout this work and discuss technical issues
relevant for these channels as well.  The silver channel has been
studied in great detail in the context of the OPERA experiment and
thus is very well understood in terms of the detector. For the
platinum channel the situation is slightly less favorable since no
reliable data on electron charge identification was available.  For
both additional channels, we define two setups, a standard scenario
with a conservative choice of parameters and an optimistic scenario
which certainly would require a considerable detector R\&D effort to be
realized. The idea is, to explore the possible maximal gain in physics
reach which could be obtained by the optimistic setups in order to
show whether an increased effort in R\&D is necessary. For the silver
channel, we also investigate optimization issues concerning the
placement of the second detector. For both, we formulate the
requirements to reach a certain level of improvement in the physics
performance. Then, in the last subsection, we analyze the impact of
this additional channel information for the three performance
indicators introduced in \Sec~\ref{sec:sim} (the sensitivity limit to
$\stheta$, the sensitivity to maximal CP violation and the sensitivity
to the sign of $\Delta m_{31}$), and we compare the different
additional channels performances.

\subsection{Silver channel}

For the silver channel, the tau neutrinos are detected which are
oscillating from the electron neutrinos in the beam. Since the
neutrino energies at a neutrino factory are above the tau production
threshold, tau leptons can be produced in charged-current reactions.
The detection of these tau leptons from the $\nu_e\rightarrow\nu_\tau$
oscillation is called ``silver channel'' and was already discussed in
the literature~\cite{Donini:2002rm,Autiero:2003fu}.  The observation
of the produced tau leptons is not possible at the ``golden''
detector, which means that a second Opera-like Emulsion Cloud Chamber
(ECC) detector is assumed for the measurement. This kind of detector
is capable of distinguishing the tau lepton events from other events
by the observation of the decay topology of the tau decay. Our
description of the silver channel follows \Ref~\cite{Autiero:2003fu}.
The discussed OPERA-like ECC detector is capable of observing the
decay of the charged-current produced tau leptons into muons. We
incorporate an energy dependent threshold for the decay-produced muon
identification. The evolution of this threshold was taken from
Figure~7 in \Ref~\cite{Autiero:2003fu}.  The energy resolution is
assumed to be $20\% \times E$, which is also an optimistic choice. We
assume silver channel data taking only during the $\mu^+$-stored phase.
\begin{table}[t]
\begin{center}
\begin{tabular}{lr} \hline
Background source & Rejection factor \\ \hline
Neutrino induced charm production & $10^{-8} \, \times \, (N_{CC}(\nu_e)+N_{CC}(\nu_\mu))$ \\ 
Anti-neutrino induced charm production & $3.7 \cdot 10^{-6} \, \times \, N_{CC}(\bar{\nu}_\mu)$ \\ 
$\tau^+\rightarrow\mu^+$ decays & $10^{-3} \, \times \, N_{CC}(\bar{\nu}_\tau)$ \\ 
$\mu$ matched to hadron track & $7 \cdot 10^{-9} \, \times \, N_{CC}(\bar{\nu}_\mu)$ \\ 
Decay-in-flight and punch-trough hadrons & $6.97 \cdot 10^{-7} \, \times N_{NC}\, +$ \\
& $+ \, 2.1 \cdot 10^{-8} \, \times \, N_{CC}(\nu_e)$\\ 
Large-angle muon scattering & $10^{-8} \, \times \, N_{CC}(\nu_\mu)$ \\ \hline
\end{tabular}
\end{center}
\mycaption{\label{tab:silverbckg} The background sources and rejection factors for the silver channel
measurement in the $\mu^+$-stored phase. The numbers are taken from \Ref~\cite{Autiero:2003fu}.}
\end{table}
As indicated above, we define two setups representing the current
``standard'' assumptions and the improvement potential in the spirit
of the last section for the golden detector:
\begin{itemize}
\item Standard: {\bf Silver} \\
  We assume the ECC detector to have a fiducial mass of 5~kt as in
  Ref.~\cite{Autiero:2003fu}. In addition, we apply an overall signal
  efficiency of approximately 10\%, which was chosen to reproduce the
  signal event numbers from Table~4 in \Ref~\cite{Autiero:2003fu}.
  The background rejection factors are taken from
  \Ref~\cite{Autiero:2003fu} as well, and are summarized in
  \Tab~\ref{tab:silverbckg}.
\item Optimistic: {\bf Silver$^*$} \\
  In the standard scenario, it was assumed, that only leptonic tau
  decays can be observed.  But, in principle, all the other decay
  channels of the tau lepton might be analyzed as well, this increases
  the signal by a factor of five. At the same time, we assume that
  those improvements necessary for identifying hadronic tau decays will
  allow to reduce the background somewhat and hence we take only three
  time the value of the standard setup~\cite{Migliozzi}. Furthermore, we 
  assume a fiducial detector mass of 10~kt. 
\end{itemize}

\begin{figure}[t!]
 \begin{center}
 \includegraphics[width=0.4\textwidth]{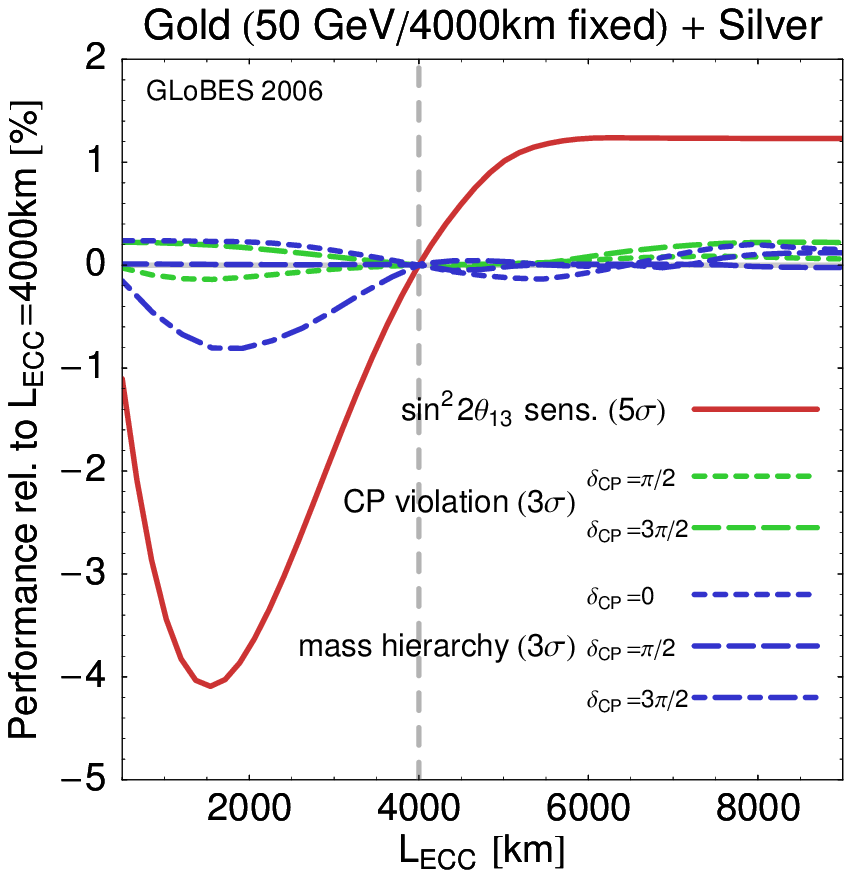} \hspace{0.1cm}
 \includegraphics[width=0.4\textwidth]{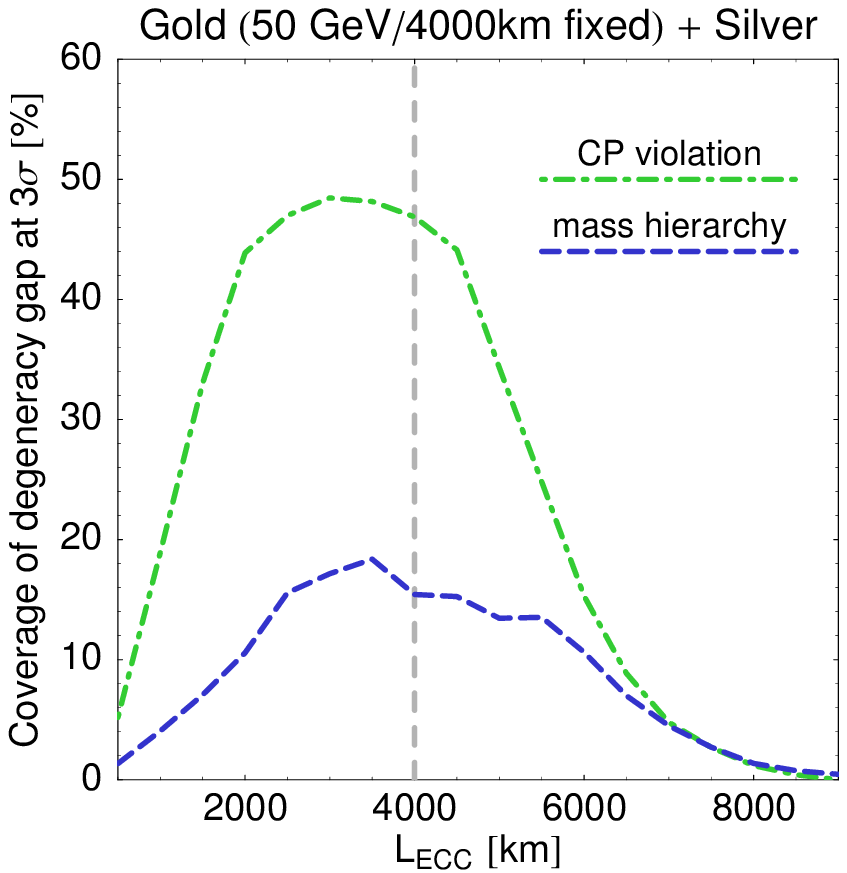}
 \end{center}
 \mycaption{\label{fig:SilverL} Optimization of the ECC detector
   baseline (standard silver channel scenario) for a fixed golden
   channel with $\mathrm{E_\mu=50\,GeV}$ and
   $\mathrm{L_{MID}=4000\,km}$. The ECC detector baseline is varied
   from 500~to~9000~km. Left-hand side (statistics dominated regime): The
   smallest $\stheta$ for which sensitivity can be found for the three
   performance indicators $\sin^22\theta_{13}$ sensitivity (solid
   red/dark curve), sensitivity to maximal CP violation (bright
   grey/green dashed curves), and sensitivity to normal mass hierarchy
   (dark grey/blue dashed curves) relative to the case of
   $\mathrm{L_{ECC}=L_{MID}=4000\,km}$ (negative numbers are better).
   Right-hand side (degeneracy resolution potential): The coverage of
   the sensitivity gap for sensitivity to maximal CP violation
   (dashed-dotted curve) and sensitivity to normal mass hierarchy
   (dashed curve), which appears for the golden channel alone for a
   medium $\sin^22\theta_{13}\sim 10^{-2.5}$ and $\delta_{CP}=3\pi/2$
   at 3$\sigma$. See text for details.  }
 \end{figure}
 The silver channel detector can be operated independently from the
 golden channel detector, and can, in principle, be located at a
 different baseline. We checked that the additional silver channel
 data does not affect the golden channel baseline optimization.  On
 the other hand, one can think of optimizing the silver baseline for a
 optimized fixed golden baseline of $\mathrm{L_{MID}=4000\,km}$, \ie,
 placing the silver channel detector at a different baseline.  In
 \figu{SilverL}, the impact of a variation of the ECC detector
 baseline for the standard silver channel scenario is shown for the
 three performance indicators sensitivity to $\stheta$, sensitivity to
 maximal CP violation, and sensitivity to normal mass hierarchy. There
 are two topics illustrated in this figure: On the left-hand side, we
 give the absolute reach in small $\stheta$, \ie, the smallest true
 value of $\stheta$ for which we still find sensitivity (statistics
 dominated regime). The absolute reach is shown relative to the
 $\mathrm{L_{ECC}=L_{MID}=4000\,km}$ case, where negative numbers
 refer to better performance. For CP violation and mass hierarchy
 sensitivity, the impact of the silver baseline variation is within
 less than 1\%. The best sensitivity limit to $\stheta$ is given at a
 ECC detector baseline of~1500~km, but also here, the effect is only
 4\% because of the low event rate for small $\stheta$. Note that this
 effect would hardly be visible on a logarithmic scale, such as in
 \figu{mhcpoptdet}. A different topic is illustrated in
 \figu{SilverL}, right (degeneracy resolution potential): As easily
 visible in \figu{mhcpoptdet}, the golden channel measurement suffers
 significantly from degeneracies for true $\delta_{CP}=3\pi/2$ at the
 $4 \, 000 \, \mathrm{km}$ baseline. Therefore, at medium true values
 of $\stheta\sim10^{-2.5}$, the sensitivities to maximal CP violation
 and the normal mass hierarchy are lost, and a sensitivity gap
 appears. On the right-hand side of \figu{SilverL} the coverage of
 this sensitivity gap is shown for the inclusion of silver channel
 data with varied ECC detector baselines. The gap is defined as the
 size of the region without sensitivity at $3\,\sigma$ in units of
 $\log \stheta$. The golden channel is again fixed to an optimized
 setup with $\mathrm{E_\mu=50\,GeV}$ and $\mathrm{L_{MID}=4000\,km}$.
 As can be seen, the optimal ECC baselines to cover the sensitivity
 gap as much as possible is found between 2500~and~5000~km.  This
 effect is visible on logarithmic scales of $\stheta$, since we define
 the coverage width of the gap on a logarithmic scale. Because of this
 effect and because it is more cost effective, we will therefore
 assume in the following that the ECC detector be located at the
 golden main detector baseline.

\subsection{Platinum channel}

\begin{table}[t]
\begin{center}
\begin{tabular}{lr} \hline
Background source & Rejection factor \\ \hline 
Muon disappearance  & $10^{-3} \, \times \, N_{CC}(\nu_\mu) \, (N_{CC}(\bar{\nu}_\mu))$ \\ 
Tau appearance & $5 \cdot 10^{-2} \, \times \, N_{CC}(\nu_\tau) \, (N_{CC}(\bar{\nu}_\tau))$ \\  
Neutral current reactions & $10^{-2} \, \times \, N_{NC}$ \\ 
Wrong sign electron/positron & $10^{-2} \, \times \, N_{CC}(\bar{\nu}_e) \, (N_{CC}(\nu_e))$ \\ \hline
\end{tabular}
\end{center}
\mycaption{\label{tab:platinumbckg} The background sources and rejection factors for the platinum channel
measurement for the $\mu^-$-stored phase, while the brackets refer to the $\mu^+$-stored phase. The numbers, besides the background from electron/positron CID, are taken from \Ref~\cite{NUMI714}.}
\end{table}

Besides the previously considered channels, the neutrino beam of a
neutrino factory allows to observe neutrino oscillations from the
$\nu_{\mu}/\bar{\nu}_{\mu}\rightarrow\nu_e/\bar{\nu}_e$ channel, which
is often called ``platinum channel''. This is the T-conjugated
oscillation channel to the golden channel, and corresponds to the
CP-conjugated golden channel with different matter effect. Therefore,
it should allow to resolve the correlations and degeneracies of the
golden channel measurements as well. Again, as for the silver channel,
we define two different scenarios, one conservative and one
optimistic. For the description of the platinum channel, we roughly
follow the $\nu_e$-appearance performance of the MINOS detector, which
has been estimated in \Ref~\cite{NUMI714}. However, since we require
charge identification to establish the $\nu_e$ ($\bar\nu_e$)
appearance against the $\bar{\nu}_e$ ($\nu_e$) disappearance from the
beam, we add an extra background from these disappearance neutrinos.
We assume the background after the CID selection to be 1\% of all
electron neutrino disappearance neutrinos.  We apply a lower energy
detection threshold at 0.5~GeV. Electron charge ID so far has been only
studied for a magnetized liquid Argon TPC and the numbers above
roughly match the ones indicated in~\cite{Rubbia:2001pk}. In the same
\Ref\ it was also pointed out that electron charge ID may have an
upper threshold beyond which it may no longer be possible to measure
the charge.  Electrons/positrons at higher energies tend to
shower early, which means that the track is too short and the
curvature is hardly measurable. Therefore, the CID of electrons and
positrons most likely is only  possible up to a certain energy threshold.

\begin{figure}[t!]
 \begin{center}
 \includegraphics[width=0.4\textwidth]{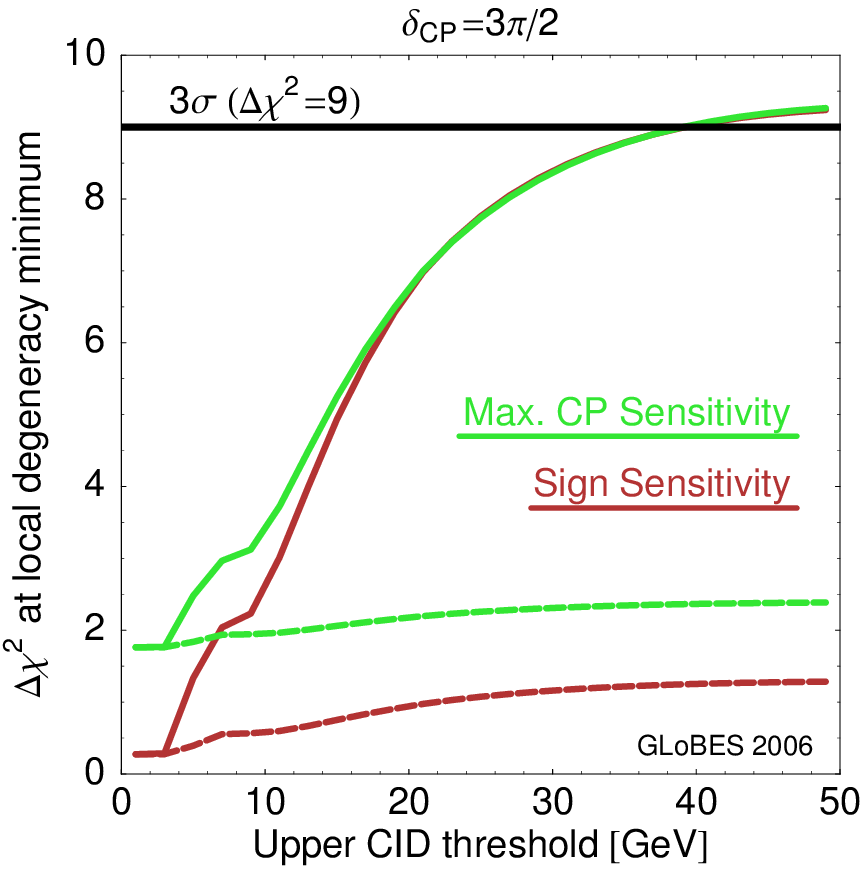} \hspace{0.1cm}
 \includegraphics[width=0.4\textwidth]{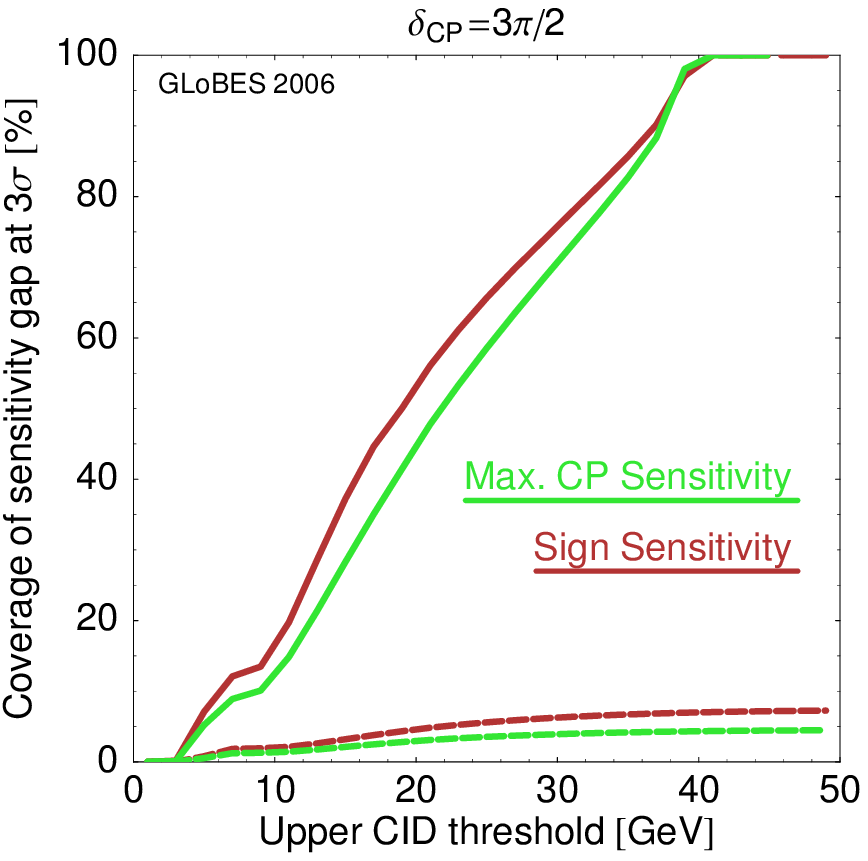} 
 \end{center}
 \mycaption{\label{fig:pltresh} The evolution of the height of the
   $\Delta \chi^2$ at the local degeneracy minimum (left) and the
   coverage of the sensitivity gap (right) as function of the upper
   electron/positron CID threshold for the sensitivity to CP
   violation.  The sensitivity gap (see \figu{SilverL}) and local
   degeneracy appear for about the true $\stheta\approx2.5\times
   10^{-3}$ if the true value $\deltacp=3\pi/2$ is assumed. The
   different curves refer to the sensitivity to maximal CP violation
   (light gray/green) and sensitivity to the normal mass hierarchy
   (dark grey/red).  The baseline is assumed to be 4000~km and the
   energy of the parent muons is 50~GeV. The dashed curves refer to
   the standard scenario with a lower detector mass (15~kt instead of
   50~kt) and lower efficiencies (20\% instead of 40\%).  }
 \end{figure}

 For the platinum channel, we will always assume the same baseline as
 for the golden channel, since it is at least in principle conceivable
 to use the same detector for both golden and platinum channel. We
 define two setups:
\begin{itemize}
\item Standard: {\bf Platinum} \\
  We assume a platinum channel detector with a fiducial mass
  of~15~kt, which may be the largest magnetizable volume for a liquid
  argon TPC. The signal efficiency is taken to be
  20\%~\cite{Rubbia:2001pk}, and the background rejection factors are
  summarized in \Tab~\ref{tab:platinumbckg}.  Furthermore, the energy
  resolution is assumed to be $15 \% \times E$. The upper threshold
  for the electron/positron CID is assumed to be 7.5~GeV. The CID
  background is assumed to be $1\%$~\cite{Rubbia:2001pk} and the other
  backgrounds are taken from~\Ref~\cite{NUMI714}.
\item Optimistic: {\bf Platinum$^*$} \\
  We assume a platinum channel detector with a fiducial mass of 50~kt.
  This choice is inspired by the possibility (at least in principle)
  to use the same, improved detector than for the golden channel. The
  signal efficiency is 40\%. The background rejection
  factors of \Ref~\cite{NUMI714} are extrapolated to higher energies.
  The CID background is the same than for the standard setup.
  Electron/positron CID is assumed to be possible to the highest
  energies and no upper threshold is imposed.
\end{itemize}

First,we will discuss the impact of the upper CID threshold and
discuss the performance of the additional platinum channel data
depending on the value of this threshold energy.  Again, as the first
case, we want to use the additional channel data to resolve the
degeneracies, which especially appear for the choice of true
$\deltacp=3\pi/2$. As indicated in \figu{cpdid}, the sensitivity gap
for maximal CP violation appears as a local minimum in the projected
$\Delta \chi^2$ at higher values of true $\stheta$, which is also true
for the mass hierarchy.  In the left-hand side of \figu{pltresh}, we
therefore show the height of these minima as function of the assumed
upper electron/positron CID threshold for the platinum channel. One
can easily see that the platinum channel data can help to resolve the
degeneracy and push the minimum above the $3\sigma$ confidence level
similar to the silver channel. However, it could only significantly
contribute, if the CID were possible up to high energies larger than
about 20 to 30~GeV. On the right-hand side of \figu{pltresh}, we show
how the width of the sensitivity gap (already discussed in
\figu{SilverL}) at $3\sigma$ evolves. For high CID thresholds, it can
be covered completely. The dashed curves show the same results but for
the reduced detector mass and efficiencies. One can see, that in this
case, the reduced statistics in the platinum channel data cannot help
to resolve the degeneracy. In order to show the maximal
contribution from platinum data and its usefulness for the physics
performance, we will only discuss platinum CID thresholds possible up
to $50\,\mathrm{GeV}$.
\begin{figure}[t!]
\begin{center}
\includegraphics[width=8cm]{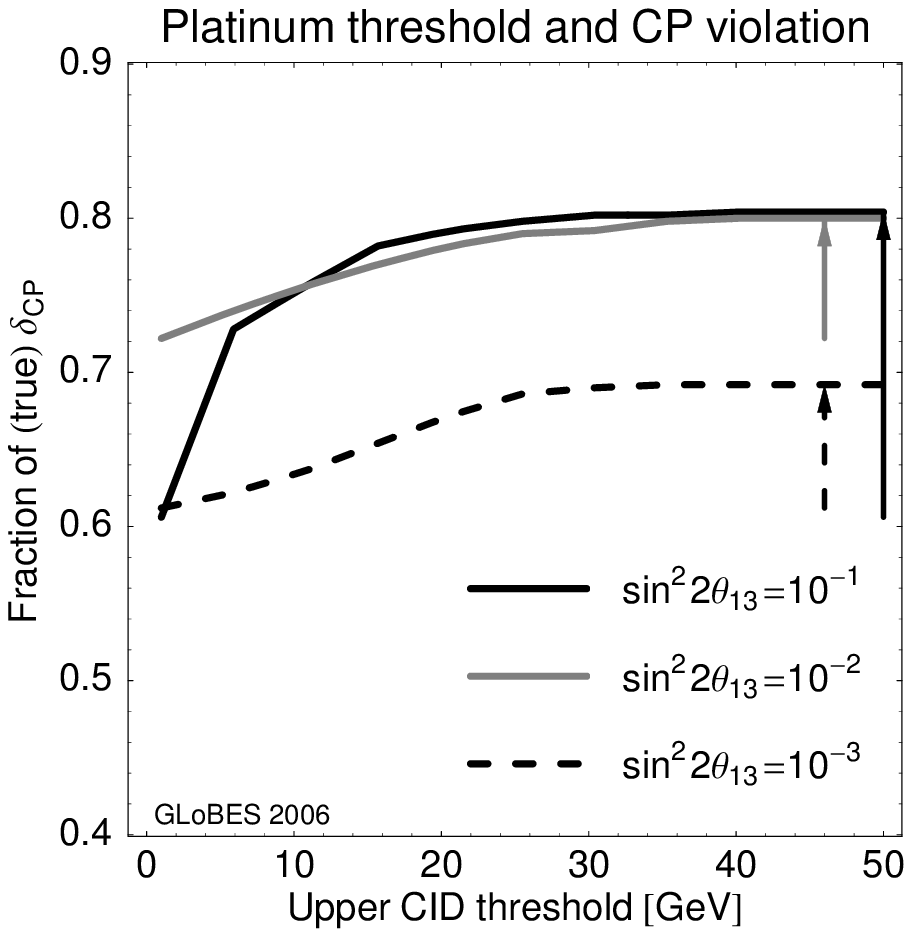}
\end{center}
\mycaption{\label{fig:platthresh} The fraction of (true) $\deltacp$
  for which CP violation can be discovered at the $3 \sigma$
  confidence level as function of the upper platinum CID threshold
  (for a normal mass hierarchy). As a setup, we use the golden and
  platinum channels (platinum with 50~kt mass, 40\% detection
  efficiency) at $4 \,000 \, \mathrm{km}$ for $E_\mu=50 \,
  \mathrm{GeV}$.  The different curves correspond to different values
  of $\stheta$ as given in the plot legend.  The arrows refer to the
  improvement in the physics potential by using the platinum channel.}
\end{figure}

As opposed to using the platinum channel to resolve degeneracies for
medium $\stheta$, we will also discuss its potential for large
$\stheta$. We show in \figu{platthresh} the fraction of (true)
$\deltacp$ for which CP violation can be discovered as function of the
upper platinum CID threshold, \ie, we discuss the performance for all
values of $\deltacp$ in this figure. As in \figu{pltresh}, one can
easily see the relatively shallow dependence on the threshold for
$\stheta \lesssim 10^{-2}$, whereas for large $\stheta$ already a $6
\, \mathrm{GeV}$ upper threshold can increase the fraction of
$\deltacp$ by about 10\%. This means that if it turns out that the
platinum channel is mainly useful for large $\stheta$, a relatively
low upper threshold will not harm. However, if we intend to use it for
medium $\stheta$ as a degeneracy resolver, the threshold will need to
be as high as 20 to 30~GeV.

\begin{table}[t]
\begin{center}
\begin{tabular}{lr}
\hline
Configuration & Fraction of $\deltacp$ \\
\hline
Golden only & 61\% \\
Golden+$\nu_e$/$\bar{\nu_e}$-disappearance (with CID) & 71\% \\
Golden+$\nu_e$/$\bar{\nu_e}$-disappearance (without CID) & 76\% \\
Golden+Platinum$^*$ ($\nu_e$/$\bar{\nu_e}$-appearance) & 80\% \\
Golden+Platinum$^*$+$\nu_e$/$\bar{\nu_e}$-disappearance (with CID) & 82\% \\
\hline
\end{tabular}
\end{center}
\mycaption{\label{tab:plat} The fraction of (true) $\deltacp$ in per
  cent for 
which CP violation can be discovered for $\stheta=0.1$ and a
normal mass hierarchy for several electron neutrino detection options. For the detector parameters (efficiencies, \etc ), the same values as for the
Platinum$^*$ channel have been used. For the option without CID, we have added the electron neutrino appearance and disappearance
signals. For all options, a 2.5\% normalization error has been used, but we have checked that the results do not change for a 1\% normalization error.}
\end{table}

While we have only considered electron neutrino (and anti-neutrino)
appearance in this section, one could also think about implementing
the electron neutrino disappearance channels. We have tested the
impact of these channels for $\stheta=0.1$ (where the effect on the
disappearance is largest), and we have found some improvement for
large $\stheta$, which is, however, not as good as the platinum
appearance potential. We show in \Tab~\ref{tab:plat} several options
with electron neutrino detection for large $\stheta=0.1$. Obviously,
for the platinum channel with CID the best potential can be achieved,
and an additional 2\% in the fraction of $\deltacp$ can be gained by
using the $\nu_e$/$\bar{\nu_e}$-disappearance channels as well.
However, if one cannot achieve CID to the anticipated level/upper
energies, the disappearance channel alone without CID can also provide
some additional information.  Surprisingly, the electron neutrino
disappearance channel with CID performs worse than the one without CID
(appearance and disappearance rates added), but note that the
combination without CID contains some information on $\deltacp$ as
well (as opposed to the one with CID) while the leading $\stheta$-term
is of the same order of magnitude.  We do not consider electron
neutrino disappearance for the rest of this paper anymore because we
expect the best results from the platinum channel as we have defined
it. Nevertheless, if electron neutrino detection is eventually
implemented, the disappearance information should be exploited as
well.

\subsection{Impact on physics reach}

In this section, we summarize the possible impact of the data from the
additional channels and the combination of golden, silver, and
platinum channels. Therefore, we discuss all three performance
indicators: Sensitivity to $\stheta$, maximal CP violation, and the
mass hierarchy.

\begin{table}[t!]
\begin{center}
\begin{tabular}{lrrr} \hline
 $\sin^22\theta_{13}=10^{-1}$ & Signal & Background & S/$\mathrm{\sqrt{B}}$ \\ \hline 
Golden & 31000 (6000) & 39 (73) & 5000 (700) \\ 
Silver & 210 (--) & 32 (--) & 37 (--) \\  
Silver@732km & 260 (--) & 110 (--) & 25 (--) \\
Silver$^*$ & 2100 (--) & 190 (--) & 150 (--) \\ 
Silver$^*$@732km & 2600 (--) & 670 (--) & 100 (--) \\
Platinum & 4 (120) & 140 (110) & 0.3 (11) \\ 
Platinum$^*$ & 6700 (27000) & 190000 (160000) & 15 (68) \\
$\mathrm{(Golden)_{\mathrm{MB}}}$ & 5100 (340) & 9 (17) & 1700 (83) \\ \hline
\end{tabular}

\vspace*{0.3cm}

\begin{tabular}{lrrr} \hline
 $\sin^22\theta_{13}=10^{-2.5}$ & Signal & Background & S/$\mathrm{\sqrt{B}}$ \\ \hline 
Golden & 1900 (450) & 39 (72) & 300 (53) \\ 
Silver & 3 (--) & 33 (--) & 0.5 (--) \\  
Silver@732km & 1.7 (--) & 110 (--) & 0.2 (--) \\
Silver$^*$ & 29 (--) & 200 (--) & 2.1 (--) \\ 
Silver$^*$@732km & 17 (--) & 680 (--) & 0.7 (--) \\
Platinum & 1 (5) & 170 (110) & 0.08 (0.5) \\ 
Platinum$^*$ & 500 (1600) & 190000 (160000) & 1.1 (4) \\
$\mathrm{(Golden)_{\mathrm{MB}}}$ & 200 (10) & 9 (17) & 67 (2.4) \\ \hline
\end{tabular}
\end{center}
\mycaption{\label{tab:channelevents} The (rounded) event rates in the $\mu^+$-stored phase 
($\mu^-$-stored phase) for the golden channel and the standard silver and platinum 
channels, as well as their optimized scenarios (indicated by the stars) at a baseline of 
4000~km and for $\mathrm{E_\mu= 50\,GeV}$. For reasons of comparison, the last row 
gives the event rates of the golden channel at the magic baseline of 7500~km and the silver channel
event rates are also given at a baseline of 732~km. 
The upper table is calculated for a large case $\sin^22\theta_{13}=10^{-1}$ and the lower 
table for a medium case $\sin^22\theta_{13}=10^{-2.5}$. For the other oscillation
parameters the true values are chosen as in \eq~\ref{equ:params} and $\delta_{CP}=0$.}
\end{table}
The relative contribution to the physics reach can be roughly
understood by looking at the statistical significance of the various
options. To this end we show the signal and background event rates for
two specific points in parameter space in
\Tab~\ref{tab:channelevents}.  In this table, the rounded signal and
background event rates, as well as signal over square root of the
background are given for either $\stheta=0.1$ or $\stheta=10^{-2.5}$.
Quite obviously the golden channel deserves its name, for both values
of $\stheta$ it by far has the most statistical significance. This is
due to the fact that muons are relatively straightforward to detect
and easy to distinguish from backgrounds. The platinum channel also
has very high statistics, but the background is very high as well.
Most importantly, the platinum channel has better statistics for the
$\mu^-$-stored phase when the golden channel is weaker because of the
matter effect suppression, and vice versa. Thus, it acts as an
anti-neutrino mode without matter effect suppression. The silver
channel, on the other hand, suffers from both very low statistics and
relatively high background.  The event rates for the silver cannel
scenarios are also given at a ECC detector baseline of~732~km, the
distance of the CERN to Gran Sasso baseline, where the OPERA detector
will be located. One can see, that the variation of the baseline has
not a big impact on the total rates here.  Note that the performance
of the golden channel can also be improved by a second detector at the
magic baseline and degeneracies can be effectively resolved.
Therefore, we also give the golden channel event rates at the magic
baseline for comparison.  Despite the almost doubled baseline, very
high statistics still remains with a much better signal to background
ratio than for the platinum channel. From this simple discussion we
expect that additional channels will be only useful in those regions
of the parameter space where the performance of a neutrino factory is
strongly comprised by either degeneracies or correlations.

 \begin{figure}[t!]
 \begin{center}
 \includegraphics[width=8cm]{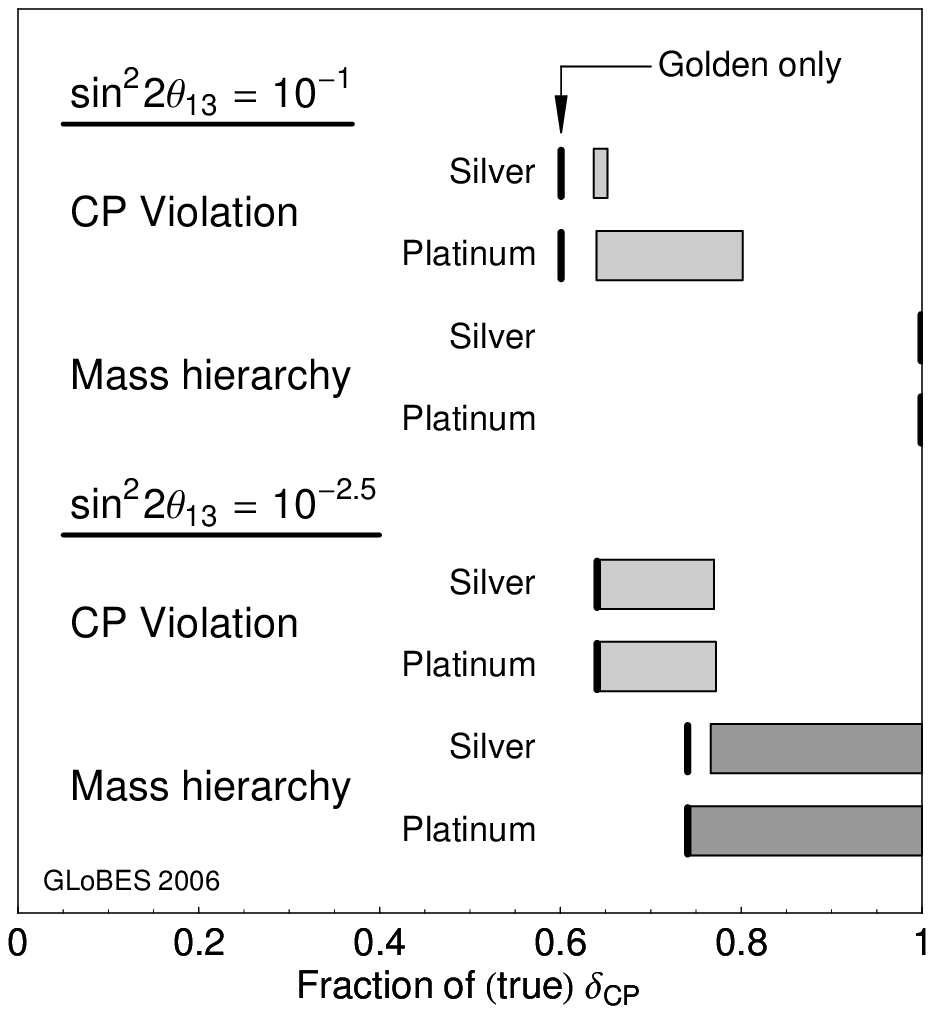}
 \end{center}
 \mycaption{\label{fig:chbars} Fraction of true values of
   $\delta_{CP}$ for which CP violation and a normal mass hierarchy
   can be established at the $3 \sigma$ confidence level for two
   different values of $\stheta$. The baselines are assumed to be
   4000~km for all channels and $\mathrm{E_\mu=50\,GeV}$. The black
   lines refer to the performance of the golden detector only, and the
   gray bars the potential of the additional channel data. The left
   edges of the bars are for the standard scenarios Silver and
   Platinum, whereas the right edges of the bars refer to the
   optimization potentials Silver$^*$ and Platinum$^*$.}
 \end{figure}

 In \figu{chbars}, the contribution from additional silver channel and
 platinum channel data is discussed for two true values of $\stheta$,
 which represent two conceptually different cases. For a medium
 $\stheta=10^{-2.5}$, the golden channel suffers from degeneracies,
 and the additional data could help resolve it. For a large
 $\stheta=10^{-1}$, the golden channel suffers from the uncertainty in
 the matter density and also there, additional channel data could
 improve the performance. For $\stheta \ll 10^{-2.5}$, however, we do
 not expect major contributions from any of the two channels because
 of a lack of statistics (silver) or the CID background (platinum).
 The black lines in \figu{chbars} refer to the golden channel only,
 and the improvement from the additional channels is visualized with
 the bars. The left edges of these bars represent the contribution
 from the standard scenarios Silver and Platinum, whereas the right
 edges represent the maximal contribution from the optimistic scenarios
 Silver$^*$ and Platinum$^*$. Thus, the finally achievable contribution
 most likely is within the bars.
 It can be read off from \figu{chbars} that the standard scenarios do
 not contribute in a sizable way, whereas there is a substantial
 contribution for the optimistic setups. We will therefore only discuss
 the scenarios Silver$^*$ and Platinum$^*$ in the following. In the case of
 a medium $\stheta=10^{-2.5}$, the impact of the silver channel and
 platinum channel is comparable. The sensitivity to the mass hierarchy
 is restored in the complete $\delta_{CP}$ range, and the fraction of
 true $\delta_{CP}$ where sensitivity to CP violation is given is
 significantly increased. However, in the case of the large value of
 $\stheta=10^{-1}$, the platinum channel performs noticeably better
 than the silver channel. One reason for this, lies in the tau
 production threshold for the silver channel which suppresses the most
 useful events around the first oscillation maximum. Note, that
 already the golden channel alone can distinguish the mass hierarchy
 over the whole $\delta_{CP}$ range and no improvement can come from
 the additional channel data.

 \begin{figure}[t!]
 \begin{center}
 \includegraphics[width=0.4\textwidth]{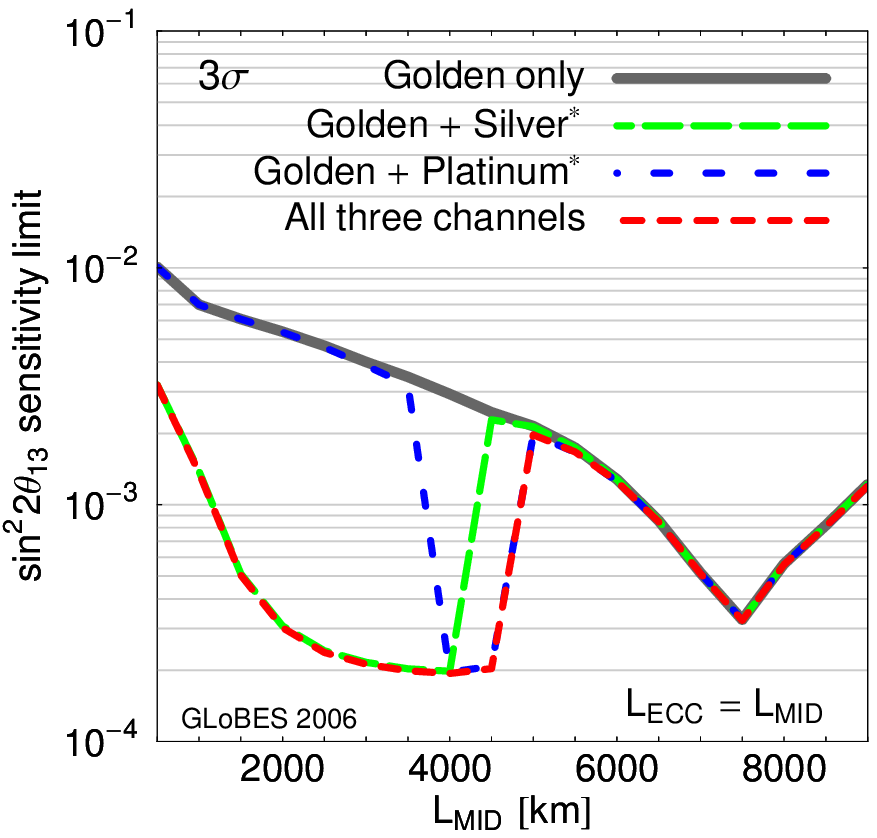} \hspace{0.1cm}
 \includegraphics[width=0.4\textwidth]{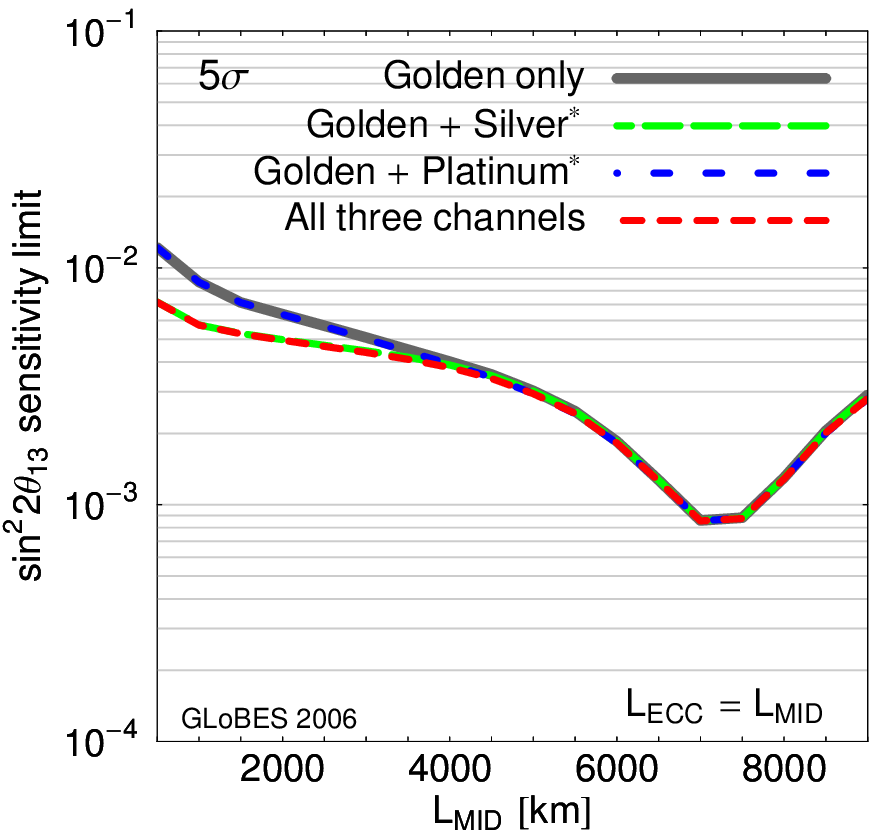}
 \end{center}
 \mycaption{\label{fig:ChannelsSens} The $\stheta$ sensitivity limit
   for the combination of different channels at $3 \sigma$ (left) and
   $5\sigma$ (right). The energy of the parent muons is fixed to 50
   GeV, and systematics, correlations, and degeneracies are taken into
   account.  }
 \end{figure}
 As it was shown in \figu{degdid}, the $\stheta$ sensitivity suffers
 from the intrinsic degeneracy, \ie, a second fit solution at
 $\stheta\approx 2.5 \times 10^{-3}$ which appears at the $3\sigma$
 confidence level. In \figu{ChannelsSens}, the sensitivity to
 $\stheta$ is therefore shown at the $3\sigma$ (left) and $5\sigma$
 (right) confidence level for the golden channel alone, the
 combination of golden and silver channel, the combination of golden
 and platinum channel, and the combination of all three channels as
 indicated by the plot legend. The baseline of all detectors at the
 same location is varied between 500 and 9000~km. Again, the effect of
 the magic baseline at $L \simeq 7500\, \mathrm{km}$ can be easily
 seen. At $5\sigma$, the degeneracy is still present in all
 combinations of channels, and the overall sensitivity to $\stheta$ is
 not affected. Only the silver channel improves the achievable limit
 to some extend at lower baselines. If, however, the $3\sigma$
 sensitivity is considered, the silver channel allows to resolve the
 degeneracy up to baselines of $L \simeq 4000\, \mathrm{km}$, and the
 platinum channel resolves the degeneracy at $L \simeq 4000 \,
 \mathrm{km}$. In both cases, the sensitivity makes a jump of one
 order of magnitude in $\stheta$, which comes from lifting the
 degenerate solution in \figu{degdid} above the $3 \sigma$ threshold.
 However, if one considers the depth of the local $\Delta \chi^2$
 minimum at $\stheta\approx 2.5 \times 10^{-3}$, it is only marginally
 above the $3\sigma$ confidence level. This effect could also be
 achieved by moderately increasing the detector mass of the golden
 detector.
 
\begin{figure}[t!]
 \begin{center}
 \includegraphics[width=0.4\textwidth]{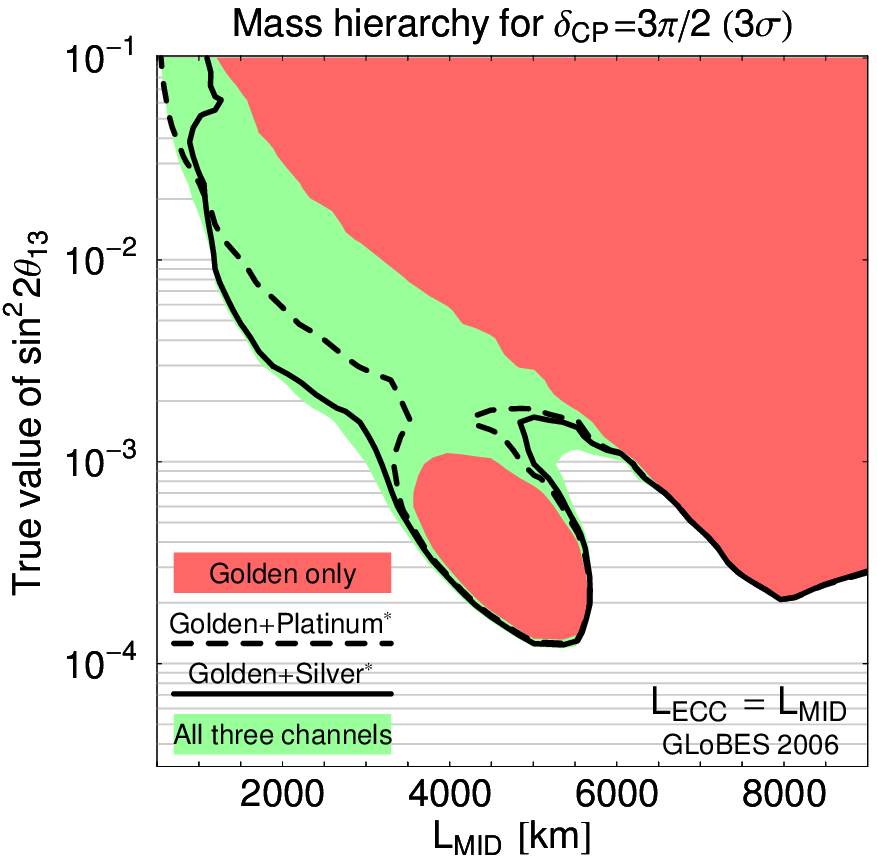} \hspace{0.1cm}
 \includegraphics[width=0.4\textwidth]{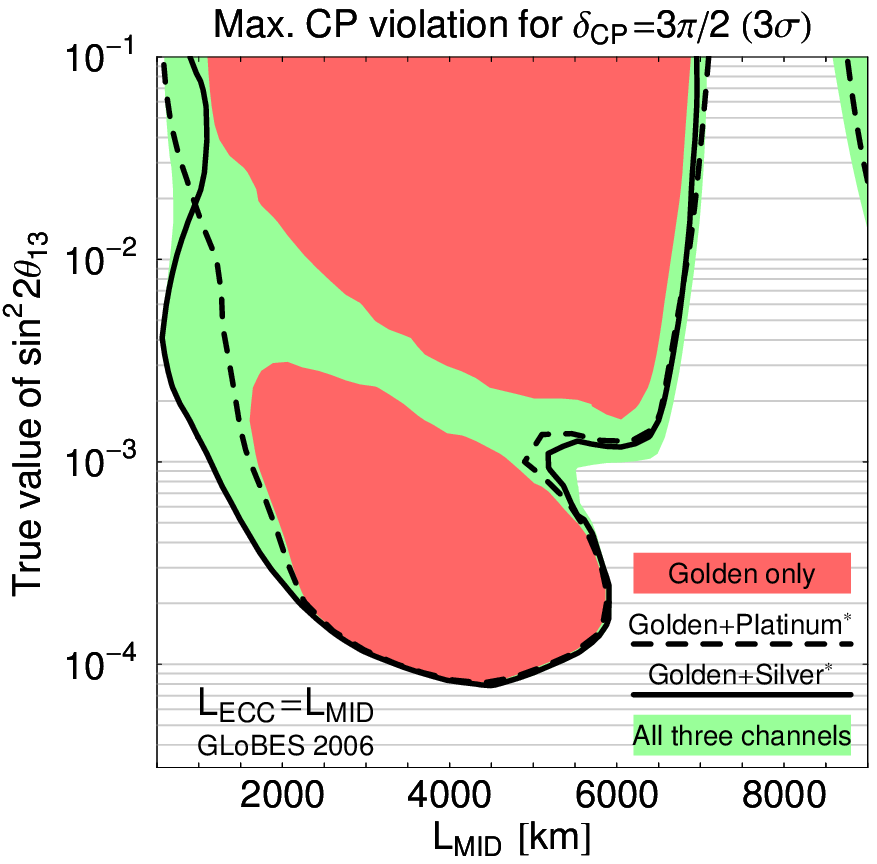}
 \end{center}
 \mycaption{\label{fig:ChannelsMHCP} The sensitivity to mass hierarchy
   (left) and sensitivity to maximal CP violation (right) at $3\sigma$
   for the combination of different channels as given in the plot
   legends. The true value for the CP phase is assumed to be
   $\deltacp=3\pi/2$, since for this value the golden channel suffers
   from the intrinsic degeneracy.  All correlations and degeneracies
   are taken into account.  }
 \end{figure}

 Since the silver and platinum channel appearance probabilities have a
 different dependence in $\deltacp$, the addition of the data from the
 two channels should help resolve degeneracies. In order to check the
 baseline optimization, we show in \figu{ChannelsMHCP} the sensitivity
 to the normal mass hierarchy (left) and maximal CP violation (right)
 for several combinations of channels and $\deltacp=3 \pi/2$, where
 the degeneracy problem is present.
 For the sensitivity to the mass hierarchy (left plot), the additional
 silver and platinum channel data can improve the sensitivity and
 close the sensitivity gap between the dark shaded regions in a large
 baseline window.\footnote{For $L=4 \, 000 \, \mathrm{km}$, the local
   minimum of the degenerate solution fitted with inverted hierarchy
   is found at $\Delta \chi^2 = 0.2$ for golden channel only, $\Delta
   \chi^2 = 13.0$ for golden and silver* channel, $\Delta \chi^2 =
   9.3$ for golden and platinum* channel, and $\Delta \chi^2 = 17.0$
   for the combination of all three channels.} This implies that the
 $4 \, 000 \, \mathrm{km}$ baseline alone is very good for the chosen
 $\deltacp \approx 3\pi/2$ compared to the magic baseline. We will
 test in \Sec~\ref{sec:comp} if this result holds for all values of
 $\deltacp$. We have also checked that the impact of the additional
 channels is small for $\deltacp=0$ and even negligible for
 $\deltacp=\pi/2$. At this point we would like to point out that the
 inclusion of the additional silver and platinum channel data does not
 affect the baseline choice of $4 \, 000 \, \mathrm{km}$.
 For the maximal CP violation sensitivity (right plot), the best reach
 of sensitivity to maximal CP violation in small $\stheta$ is still
 obtained at $\mathrm{L\approx4000\,km}$, while the degeneracy gap can
 be closed by either of the additional channels.\footnote{At
   $\mathrm{L=4000\,km}$ and $\mathrm{E_\mu=50\,GeV}$, the local
   minimum of the $\Delta \chi^2$ function at the degeneracy is found
   at $\Delta \chi^2 = 1.8$ for golden channel only, $\Delta \chi^2 =
   14.0$ for golden and silver* channel, $\Delta \chi^2 = 9.3$ for
   golden and platinum* channel, and $\Delta \chi^2 = 20.0$ for the
   combination of all three channels.} We have not shown the case of
 the true value $\deltacp=\pi/2$, but since there the effect of
 degeneracies is small, we have checked that the impact of the
 additional channels is negligible in that case for baselines around
 $4 \, 000 \, \mathrm{km}$.

 Besides the baseline optimization, we discuss the muon energy
 dependence in \App~\ref{app:ch_energy}. As the main result, the tau
 production threshold for the silver channel seems to point to muon
 energies higher than about $30 \, \mathrm{GeV}$.
 
 In this section, we have defined and tested several options for
 silver and platinum channels, and we have demonstrated their
 degeneracy resolving potential at individual points in the parameter
 space. We have shown that the golden channel baseline choice is not
 affected by the inclusion of the silver channel. In addition, we have
 found that the platinum channel could be especially useful for large
 $\stheta$, where the impact of the upper CID threshold is lowest.  In
 the next \Sec~\ref{sec:comp}, we will quantify the synergy of the
 different channels and other options in terms of the full relevant
 parameter space.  However, we will focus on the optimized setups
 found in this section (\ie, the ``star'' options).

%%%%%%%%%%%%%%%%%%%%%%%%%%%%%%%%%%%%%%%%%%%%%%%%%%%%%%%%%%%%%%%%%%%%%%%%%
\section{Comparison of optimized setups}
\label{sec:comp}

\begin{table}[t]
\begin{center}
\begin{tabular}{|l||l|l|}
\hline
Effort: Baselines $\rightarrow$ & One baseline & Two baselines \\
Detectors $\downarrow$ $~$ Overall $\searrow$ & (thin curves) & (thick curves) \\
\hline
\hline
Single detector & Golden &  not applicable \\
 & (Golden)$_{\mathrm{MB}}$ & \\
& Beta beam & \\
\hline
Double detector & (Golden)$_{\mathrm{2\mathcal{L}}}$ & Golden+(Golden)$_{\mathrm{MB}}$\\
 & Golden$^*$ &  \\
 & Golden+Silver$^*$ & \\
 & Golden+Platinum$^*$ & \\
\hline
Triple detector &   & Golden$^*$+(Golden$^*$)$_{\mathrm{MB}}$ \\
 &  & Golden+(Golden)$_{\mathrm{MB}}$+Platinum$^*$ \\
\hline
Quadruple detector &  & Golden$^*$+(Golden$^*$)$_{\mathrm{MB}}$+Platinum$^*$ \\
\hline
\end{tabular}
\end{center}
\mycaption{\label{tab:setupmatrix} 
Different (optimized) setups considered for
comparison. The column headings also contain the line styles used for \figu{mhcpcomp}
(for the convenience of comparing similar options). The detector effort is characterized
in terms of multiples of a conventional detector: Using an optimized detector,
a hybrid detector, additional (optimized) channel, or conventional detector with double luminosity increases the detector effort by one (simplified picture). Therefore, the different rows represent different levels of sophistication in terms of detector, whereas the different columns represent different
levels of sophistication in terms of number of baselines. 
For the muon energy, we use, unless
noted otherwise, $E_{\mu} = 50 \, \mathrm{GeV}$. For all neutrino factory channels, we use, unless noted otherwise, $L = 4 \, 000 \, \mathrm{km}$ and $m_{\mathrm{Det}} = 50 \, \mathrm{kt}$ (a number index refers to a different baseline, ``MB'' to the magic baseline, and the index $2 \mathcal{L}$ refers to double luminosity, \ie, $m_{\mathrm{Det}} = 100 \, \mathrm{kt}$). The stars refer to the optimized detectors: For the golden channel, a better threshold and energy resolution is implemented, as well as $E_\mu = 20 \, \mathrm{GeV}$ is used for all options including the optimized golden detector; for the silver channel, a $10 \, \mathrm{kt}$ ECC with a signal efficiency increased by a factor of five and a background increased by a factor of three is used, which could be achieved by the implementation of more decay channels of the tau lepton; for the platinum channel, the full golden detector mass of $50 \, \mathrm{kt}$ is used with an efficiency of 40\% in the whole analysis range. For the beta beam, we use the $\gamma=350$ option from \Ref~\cite{Burguet-Castell:2005pa} for reference. 
}
\end{table}

In order to compare different neutrino factory options and to discuss
where to focus the effort, we use a number of different setups and
classify them according to the sophistication of the detection system,
the total luminosity, and the number of baselines used. In
\Tab~\ref{tab:setupmatrix}, we list these setups in a matrix, where
the rows correspond to a similar effort to the detection system, and
the columns to an equal number of baselines. We define the ``detector
effort'' in terms of multiples of a conventional detector: Using an
optimized detector, a hybrid detector, additional (optimized) channel,
or conventional detector with double luminosity/mass increases the
detector effort by one. This picture may be a bit over-simplified,
since some approaches may be feasible from the current point of view
(such as the double mass detector), while others may even not be
possible to their full extent (such as the completely optimized
detector, or the silver or platinum channels). However, this
classification should somehow reflect the level of sophistication in
terms of the detection system.  For the number of baselines, we
restrict ourselves to one or two, \ie, additional detectors/channels
have to be placed such that this baseline constraint is not violated.
In summary, the effort in \Tab~\ref{tab:setupmatrix} increases from
the top to the bottom in terms of the detection system, from left to
right in terms of number of baselines, and diagonally from top left to
bottom right in total. Note, however, that an increased detector
effort and baseline effort may not be comparable at all, since a
second baseline depends on accelerator considerations (such as the
storage ring shape), while the increased detector effort is often
limited by technical feasibility.  Therefore, we visualize these
completely different degrees of freedom by the matrix choice in
\Tab~\ref{tab:setupmatrix}: The columns represent the {\em accelerator
  degree of freedom}, the rows the {\em detector degree of freedom}.
 
Here we compare optimal setups, \ie, the optimized choices from the
previous sections. We do not discuss the baseline and muon energy
optimization anymore, but we take the choices for these parameters
from the earlier discussion.  Let us now quickly explain the setups
and their labels as used in \Tab~\ref{tab:setupmatrix}.  For the muon
energy, we use, unless noted otherwise, $E_{\mu} = 50 \,
\mathrm{GeV}$. For all golden channels, we use $L = 4 \, 000 \,
\mathrm{km}$ and $m_{\mathrm{Det}} = 50 \, \mathrm{kt}$, where a
number index in the setup refers to a different baseline, and the
index ``MB'' refers to the magic baseline.  In addition, the index
``$2 \mathcal{L}$'' refers to double luminosity, \ie,
$m_{\mathrm{Det}} = 100 \, \mathrm{kt}$.  The stars refer to the
optimized, improved detectors. For the golden channel detector, a
better threshold and energy resolution is used, as well as $E_\mu = 20
\, \mathrm{GeV}$ for all options including the optimized golden
detector.  Therefore, although the setups Golden$^*$ represent a
detector with a high level of sophistication, the lower muon energy
may compensate for this effort. For the silver channel, a $10 \,
\mathrm{kt}$ ECC with a signal efficiency increased by a factor of
five and a background increased by a factor of three is used, which
could be achieved by the implementation of more decay channels of the
tau lepton. For the platinum channel, the full golden detector mass of
$50 \, \mathrm{kt}$ is used with an efficiency of 40\% in the whole
analysis range.  Therefore, we use for both silver and platinum
channels only the improved setups from the previous section, Silver$^*$ and
Platinum$^*$. Note that wherever the platinum channel is used, it is
used in (or at the location of) all golden detectors (such as for
Golden+(Golden)$_{\mathrm{MB}}$+Platinum$^*$). In addition, note that
the matter density error is assumed to be correlated among all
channels at the same baseline.  In order to compare the neutrino
factory with its possible alternative, a beta beam, we choose the the
$\gamma=350$ option from \Ref~\cite{Burguet-Castell:2005pa} for
reference.\footnote{This setup assumes eight years of simultaneous
  operation with $2.9 \cdot 10^{18}$ useful $^6$He and $1.1 \, \cdot
  10^{18}$ useful $^{18}$Ne decays per year and a $500 \, \mathrm{kt}$
  water Cherenkov detector. The gamma factor is $350$ for both
  isotopes, and the baseline is $L=730 \, \mathrm{km}$. The setups is
  simulated with the migration matrixes from
  \Ref~\cite{Burguet-Castell:2005pa}. In order to impose constraints
  to the atmospheric parameters, ten years of T2K disappearance
  information is added (such as in \Ref~\cite{Huber:2005jk}).}

The organization of the section is as follows: First, we will discuss
real synergies with respect to the physics potential only, \ie, we
compare options which use similar luminosities in terms of flux
$\times$ running time $\times$ total detector mass.  In the next part,
we will discuss specific physics scenarios and how the choice of
technology changes with respect to these.  And finally, we will focus
on the optimized physics potential, where we will demonstrate where to
focus the effort.

\subsection{Synergies in physics potential}

\begin{figure}[p]
\begin{center}
\vspace*{-0.5cm}
\includegraphics[height=0.9\textheight]{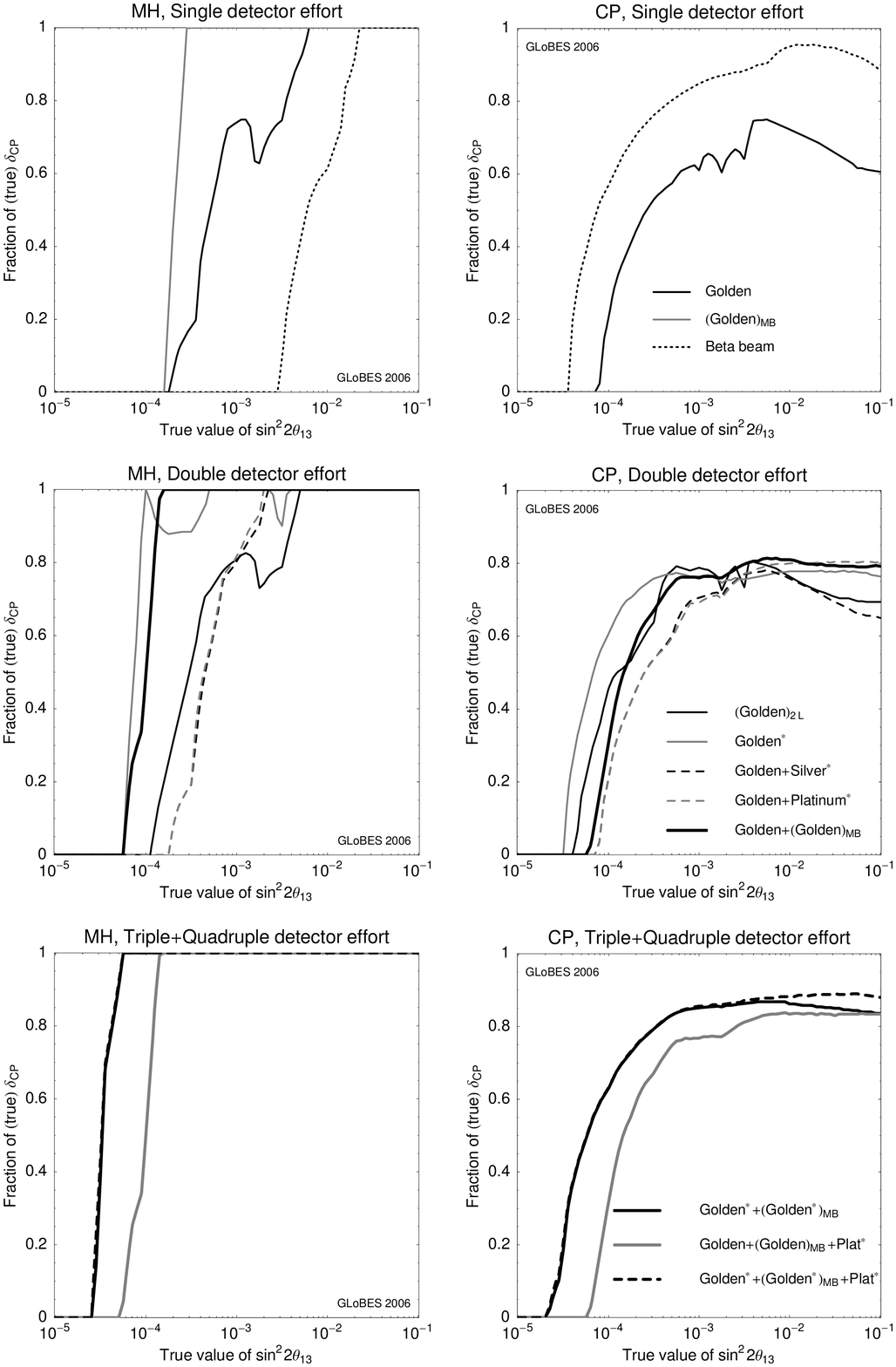}
\vspace*{-0.8cm}
\end{center}
\mycaption{\label{fig:mhcpcomp} Comparison of the different options
  from \Tab~\ref{tab:setupmatrix}, where we show the fraction of true
  $\deltacp$ as function of true $\stheta$ for the normal mass
  hierarchy sensitivity (left column, $3 \sigma$) and CP violation
  sensitivity (right column, $3 \sigma$). The different rows
  correspond to an increasing detector effort (from top to bottom)
  corresponding to the rows in \Tab~\ref{tab:setupmatrix}. Note that
  thin curves represent one-baseline options, and thick curves
  two-baseline options.}
\end{figure}

As in \Ref~\cite{Huber:2002rs}, we define the extra gain in the
combination of the experiments beyond a simple addition of statistics
as the synergy among two or more experiments. A reasonable definition
of synergy must therefore subtract, in a suitable way, the increase of
statistics of otherwise more or less equivalent experiments.
Therefore, from the physics point of view, we restrict the synergy
discussion to within the rows in \Tab~\ref{tab:setupmatrix}, since any
increase in the detector effort also increases statistics. In order to
compare our standard detector to more sophisticated detector
combinations (double detector effort), we have incorporated the setup
``(Golden)$_{2 \mathcal{L}}$'' (second row) with the doubled detector
mass.
In \figu{mhcpcomp}, a comparison of the different options from
\Tab~\ref{tab:setupmatrix} is performed, where we show the fraction of
true $\deltacp$ as function of true $\stheta$ for the normal mass
hierarchy sensitivity (left column) and CP violation sensitivity
(right column). The different rows correspond to an increasing
detector effort (from top to bottom) corresponding to the rows in
\Tab~\ref{tab:setupmatrix}. Note that thin curves represent
one-baseline options, and thick curves two-baseline options.  As far
as our synergy discussion is concerned, it is taking place within the
rows of \figu{mhcpcomp}.

The first row in \figu{mhcpcomp} (single detector effort) does not
contain any major surprises. For the mass hierarchy, the neutrino
factory at the ``magic baseline''~\cite{Huber:2003ak} has the best
potential because of the long baseline and reduced degeneracies,
which, however, does not have any CP violation sensitivity at all. The
neutrino factory faces severe problems with degenerate solutions for
intermediate values of $\stheta$ (making up the wiggles). The beta
beam is extremely good for CP violation measurements, but the
relatively short baseline makes the mass hierarchy determination
difficult. This property is common to higher gamma beta beams almost
independent of the gamma and baseline chosen, because its spectrum
easily covers at least one oscillation maximum, but the neutrino
energies are too low to go to long baselines with large
matter effects (see, \eg, \Ref~\cite{Huber:2005jk}).

There is a large number of observations from the second row in
\figu{mhcpcomp} (double detector effort). One can easily identify
synergies by the comparison of the simple luminosity upgrade (golden
channel with double luminosity) with the other shown options:
Wherever the alternative option is better than the luminosity
upgrade, we speak about ``synergy'' according to our definition, since
we gain complementary information to a detector upgrade. For the mass
hierarchy, we find such synergies for the silver and platinum channels
for intermediate $\stheta$, and for the improved detector (Golden$^*$)
and magic baseline (Golden+(Golden)$_{\mathrm{MB}}$) even for smaller
$\stheta$. Note that these latter two options are qualitatively
different: For the improved detector, correlations and degeneracies
are resolved by increased statistics. Therefore, some ``wiggles'' at
intermediate values of $\stheta$ remain. For the magic baseline,
statistics is lower because of the $1/L^2$ drop of the flux, but
correlations and degeneracies are intrinsically not present.
For CP violation, we identify the platinum channel and magic baseline
to be synergistic for large $\stheta$, and the improved detector for
small and large $\stheta$.  Note that the ability of the magic
baseline to help for large $\stheta$ is new, which comes from the
clean measurement of $\stheta$ and matter effects without correlation
with $\deltacp$.
As far as the silver channel is concerned, we only find very small
regions in parameter space where it could be useful compared to the
increased golden channel statistics for CP violation, and we have
identified much stronger alternatives for the mass hierarchy
determination. This means that we do not consider the silver channel
for further upgrades anymore. This also because it requires larger
muon energies above $E_\mu=30\,\mathrm{GeV}$. However, since the
platinum channel has the best potential for large $\stheta$, we will
include it in further discussions.

The last row in \figu{mhcpcomp} represents the most sophisticated
neutrino factory setups we could come up with, and all include a
second baseline. The option ``Golden~+ (Golden)$_{\mathrm{MB}}$~+
Platinum$^*$'' represents the best option for two baselines if the
golden channel detector cannot be improved (and the silver channel is
not used). In addition, we include two options ``Golden$^*$~+
(Golden$^*$)$_{\mathrm{MB}}$'' and ``Golden$^*$~+
(Golden$^*$)$_{\mathrm{MB}}$~+ Platinum$^*$'' with improved detector.
One can clearly read off this row that an improved golden detector
helps almost everywhere. The platinum channel is, in addition, useful
for large $\stheta$ (for small $\stheta$, the charge identification
error limits the potential), whereas it does not improve the mass
hierarchy reach for these sophisticated options at all.  As the last
observation, if one has the choice between R\&D for the platinum
channel and the improved golden channel detector, one can read off
that even for large $\stheta$ the improved golden channel is more
powerful in combination with the second baseline. In summary, the best
neutrino factory setup with two baselines, two improved golden channel
detectors, and electron neutrino detection at both detectors could,
for the chosen parameter values, measure the mass hierarchy down to
$\stheta \sim 10^{-4.5}$ and CP violation for most values of
$\deltacp$ down to $\stheta \sim 10^{-4}$ at the $3 \sigma$ confidence
level.

\subsection{Performance comparison for specific physics scenarios}

\begin{figure}[t!]
\vspace*{-1.5cm}
\begin{center}
\includegraphics[width=\textwidth]{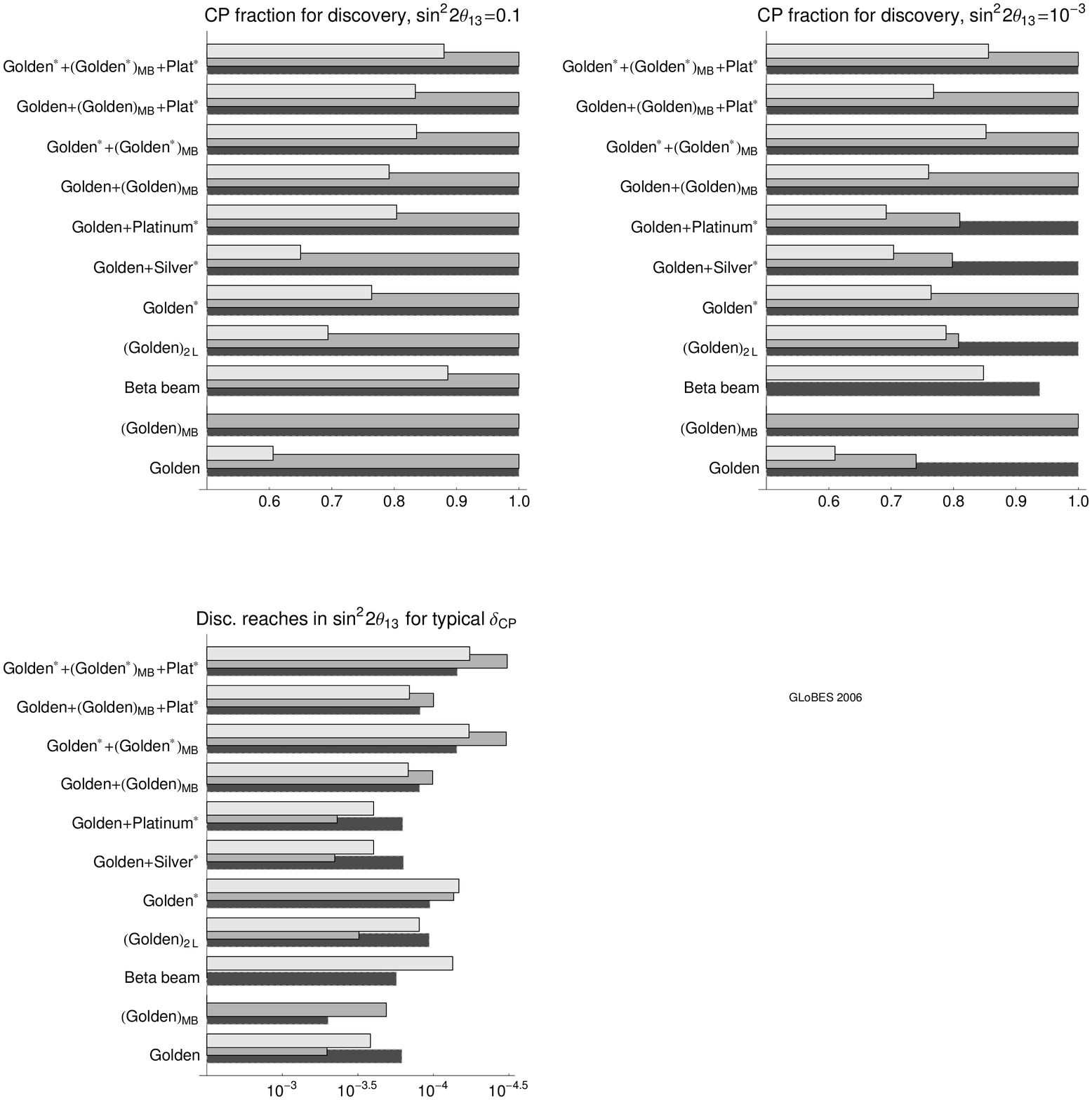}
\end{center}
\vspace*{-2cm} \mycaption{\label{fig:sumbars} Direct comparison of the
  $\stheta$ (dark bars, $5 \sigma$), mass hierarchy (medium bars, $3
  \sigma$), and CP violation (light bars, $3 \sigma$) discovery
  reaches for the setups from \Tab~\ref{tab:setupmatrix}. The upper
  left plot shows the fraction of $\deltacp$ for the physics scenario
  ``large $\stheta$'' ($\stheta=0.1$), the upper right plot the
  fraction of $\deltacp$ for the physics scenario ``medium $\stheta$''
  ($\stheta=10^{-3}$), and the lower plot the $\stheta$ discovery
  reach for the physics scenario ``small $\stheta$'' (for a CP
  fraction of $0.5$ corresponding to the ``typical value of
  $\deltacp$''). The detector effort (as defined in
  \Tab~\ref{tab:setupmatrix}) increases from bottom to top.}
\end{figure}

As far as the physics scenarios for a neutrino factory are concerned,
the main discriminator will certainly be $\stheta$. For example, one
may distinguish the following scenarios:
\begin{description}
\item[Large $\boldsymbol{\stheta}$] ($\stheta \gtrsim 10^{-2}$) In
  this case, the main question for a neutrino factory is the
  competitiveness to other options, such as superbeams or beta beams.
\item[Medium $\boldsymbol{\stheta}$] ($10^{-4} \lesssim \stheta
  \lesssim 10^{-2}$) This is the typical scenario for a neutrino
  factory. The main focus is on resolving degeneracies.
\item[Small $\boldsymbol{\stheta}$] ($\stheta \lesssim 10^{-4}$) For
  this range, statistics in combination with low backgrounds is
  important, which means that the selection of channels and baselines
  may look differently.
\end{description}
In order to discuss these physics scenarios further, we show in
\figu{sumbars} a direct comparison of the $\stheta$ (dark bars), mass
hierarchy (medium bars), and CP violation (light bars) discovery
reaches for the setups from \Tab~\ref{tab:setupmatrix}. The upper left
plot shows the fraction of $\deltacp$ for an example of the physics
scenario ``large $\stheta$'' ($\stheta=0.1$), the upper right plot the
fraction of $\deltacp$ for one representative example of the physics
scenario ``medium $\stheta$'' ($\stheta=10^{-3}$), and the lower plot
the $\stheta$ discovery reach for the physics scenario ``small
$\stheta$'' (for a CP fraction of $0.5$ corresponding to the ``typical
value of $\deltacp$''). This means that we use the $\stheta$ reach as
performance indicator for small $\stheta$ because it is more
discriminating than the CP fraction for a specific value of
$\stheta$\footnote{See \figu{mhcpcomp}: For a specific small
  $\stheta$, the CP fraction is easily either zero or very large for
  two very close curves.}

Let us first discuss the typical interesting range for a neutrino
factory (upper right plot). In this example, the $\stheta$ discovery
is not an issue for any of the neutrino factory options and any value
of $\deltacp$.  The mass hierarchy can be measured for all values of
$\deltacp$ by either improving the detector, or by adding the magic
baseline. For the measurement of CP violation, however, both of these
options may be necessary to increase to physics reach to the optimum
and make it competitive to the beta beam. Note that for CP violation
alone, increasing the mass of the golden channel detector may be
better than adding the second baseline because of the better
statistics. As far as the overall comparison with the beta beam is
concerned, a superior performance for the neutrino factory can be easily
established with respect to $\stheta$ and mass hierarchy discovery,
whereas the improved detector and second baseline will allow for the
best physics potential for all shown performance indicators.

For small values of $\stheta$, the $\stheta$ discovery reach is shown
in the lower panel of \figu{sumbars}. In this case, the optimal reach
for all indicators can clearly be obtained by an improved detector
operated at two baselines ``Golden$^*$+(Golden$^*$)$_{\mathrm{MB}}$'',
where the magic baseline is a key component to improve the mass
hierarchy reach. The final potential for each of the three performance
indicators is significantly better than the one of the beta beam.
However, note that also the golden channel alone performs already
fairly well, since correlations and degeneracies are less important
than statistics in this regime.

For large values of $\stheta$ (upper left plot in \figu{sumbars}), we
find that neither the mass hierarchy nor the $\stheta$ discovery are a
problem for any $\deltacp$ or option. Thus, the performance comparison
reduces to the CP violation potential. One can read off this figure
that the potential of the neutrino factory can be significantly
improved by adding the magic baseline, utilizing the platinum channel,
or improving the golden channel detector. In principle, one can also
think about a re-optimization of the final configuration in
two-baseline space using all available channels and improvements.
Although we find that this could increase the CP violation potential
slightly (such as using a $L=1 \, 500 \, \mathrm{km}$ baseline instead
of the very long baseline), it is hard to compete with the beta beam
in the entire range $\stheta \gtrsim 0.01$. Therefore, even after all
of these improvements, there is no clear advantage compared to the
beta beam, which serves as a representative for a number of
circulating beta beam and superbeam upgrade options at this place.

In summary, we have demonstrated that one might optimize a neutrino
factory for extremely good performances in $\stheta$, mass hierarchy,
and CP violation discovery reach below $\stheta \lesssim 10^{-2}$.
This means that we believe it to be difficult for alternative options
to compete in this range for {\em all} of these specific performance
indicators. However, we cannot establish the physics case for a
neutrino factory for $\stheta \gtrsim 0.01$ for sure. This means that
depending on systematics and achievable luminosities for alternative
options (beta beams, superbeams), as well as the utilization of the
platinum channel and the improvement of the golden detector at the
neutrino factory, alternative options could actually be better for CP
violation. Note that for large $\stheta$, the $\stheta$ and mass
hierarchy discoveries are very likely possible with many alternatives.

\subsection{Where to concentrate the efforts?}
\label{sec:efforts}

\begin{figure}[t]
\begin{center}
\includegraphics[width=\textwidth]{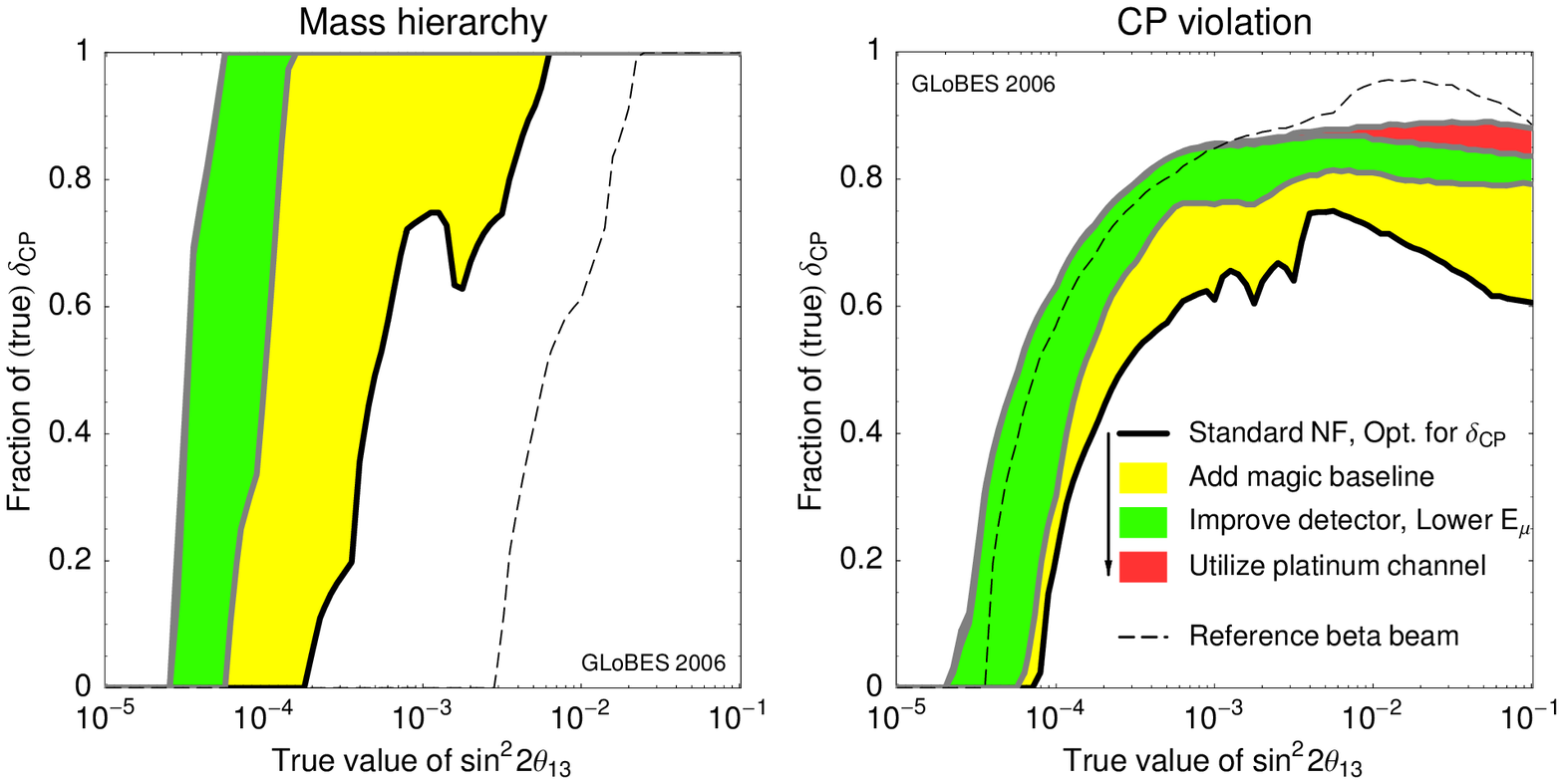}
\end{center}
\mycaption{\label{fig:optsummary} Summary of the optimization
  potential of a neutrino factory for mass hierarchy (left) and CP
  violation (right) at the $3\sigma$ confidence level.  The different
  shaded areas correspond to successively taking into account the
  additional optimizations as given in the plot legend. This means,
  for instance, that the best curves include both magic baseline and
  (improved) platinum channels.  For reference, we show the
  $\gamma=350$ beta beam from \Ref~\cite{Burguet-Castell:2005pa}.}
\end{figure}

Given the discussions in the last two subsections, let us summarize
the results of the optimization and where to concentrate the efforts.

For the {\em optimal baseline}, we find that CP violation measurements
favor a baseline around $4 \, 000 \, \mathrm{km}$, where baselines
between $3 \, 000 \mathrm{km}$ and $5 \, 000 \, \mathrm{km}$ do not
affect the sensitivity too much. For large values of $\stheta$,
shorter baselines $L \gtrsim 1 \, 500 \, \mathrm{km}$ are possible as
well. Note that the ``short'' baseline ($L \lesssim 5 \,000 \,
\mathrm{km}$) is affected by correlations and degeneracies for small
and intermediate values of $\stheta$, which means that it has moderate
$\stheta$ and mass hierarchy sensitivities. In addition, we have
tested this optimization result for larger values of $\ldm$, and it
does not change significantly (whereas the absolute physics potential
increases).

As far as {\em baseline upgrades} are concerned, a degeneracy
resolving baseline is necessary to improve the $\stheta$ sensitivity,
$\stheta$ discovery reach, and mass hierarchy discovery reach. The
``magic baseline'' at $L \sim 7 \, 000 \, \mathrm{km} - 7 \,500 \,
\mathrm{km}$ is a very robust such degeneracy resolver (independent of
the oscillation parameters, possibly over-estimated luminosities,
chosen confidence level, \etc) because the appearance probability does
not depend on $\deltacp$ there and the
($\deltacp,\theta_{13}$)-degeneracy can be unambiguously resolved.
Furthermore, matter effects are stronger than for the shorter
baseline, which means that the magic baseline measures a different
physics instead of being a pure luminosity upgrade. In addition, it
helps for CP violation measurements at large $\stheta$, and can
establish the MSW effect in Earth matter even for
$\stheta=0$~\cite{Winter:2004mt}. Since this baseline is useful in all
physics scenarios, one may want to choose a storage ring and neutrino
factory setup with two baselines already from the very beginning.

For {\em detector upgrades}, an improvement of the golden channel
detector is certainly the main objective.  Especially, improving the
detection threshold will greatly improve the physics potential in all
physics scenarios and for both mass hierarchy and CP violation
measurements. In particular, we have demonstrated that an improved
detector would allow to use a lower $E_{\mu} \sim 20 \, \mathrm{GeV}$
instead of $E_{\mu} \sim 50 \, \mathrm{GeV}$, which means that the
effort on the accelerator side would be much lower. However, note that
an improved detector will not be able to solve all degeneracy issues
on its own.

In addition, as an independent effort for {\em useful additional
  channels}, the platinum channel (electron neutrino detection) will
be very useful for large $\stheta \gtrsim 10^{-2}$ if the assumed
level of charge identification can be achieved up to large enough
energies (about $10$ to $15 \, \mathrm{GeV}$, \cf, \figu{platthresh}),
and enough statistics can be collected. This improvement would also be
complementary to the improved detector from the theoretical point of
view, since a different combination of CP violation and matter effects
would be measured (the channel behaves like an antineutrino channel
with neutrino matter effects, \ie, it is the T-conjugated channel to
the golden channel). However, it should be secondary objective after
improving the golden channel detection threshold. In addition, the
silver channel might be useful for a small fraction of the parameter
space for relatively large detectors and enough tau decay channels
implemented to improve statistics.  Note that the silver channel could
be interesting for different applications not tested here, such as new
physics tests or deviations from maximal mixing.

The {\em muon energy} for a neutrino factory should be around $40$ to
$50 \, \mathrm{GeV}$ in order to be optimized for all measurements,
where it may not have to be as high as $50 \, \mathrm{GeV}$ for
neutrino oscillation physics because of the matter resonance in the
Earth's mantle. An improvement of the detection threshold could reduce
the muon energy to $20 \, \mathrm{GeV}$ while having excellent physics
sensitivities, and the physics scenario ``large $\stheta$'' may even
allow for lower energies (while $50 \, \mathrm{GeV}$ do not harm).
Note that the use of the silver channel disfavors too low muon
energies, \ie, $E_\mu$ should in this case be larger than about $30 \,
\mathrm{GeV}$.

We show in \figu{optsummary} the summary of this optimization
discussion by successively switching on the second baseline, by
improving the detector performance, and by using the platinum channel
(Platinum$^*$). We have chosen this order because it we believe that
it somehow represents the order of the technical feasibility at this
point of time. The second baseline will be a major challenge from the
engineering point of view. However, the physics potential of this
baseline is well established and the technical feasibility should be
rather predictable. The improvement of the detector with respect to
energy resolution and threshold should be doable to a certain extent,
but it is not yet clear yet by how much exactly. The platinum channel
may be implemented in the golden detector, but the electron neutrino
detection may turn out to be technically not doable at this level and
might be limited to too low energies. In this figure, the beta beam
curves are once again given for reference. One can easily read off the
excellent combined potential for mass hierarchy and CP violation for
the neutrino factory below $\stheta \lesssim 10^{-2}$. Remember that
none of these suggested improvements could be achieved with a simple
luminosity upgrade, \ie, adding mass to the golden channel detector.
Therefore, we speak of real synergies.

\begin{figure}[t]
\begin{center}
\includegraphics[width=8cm]{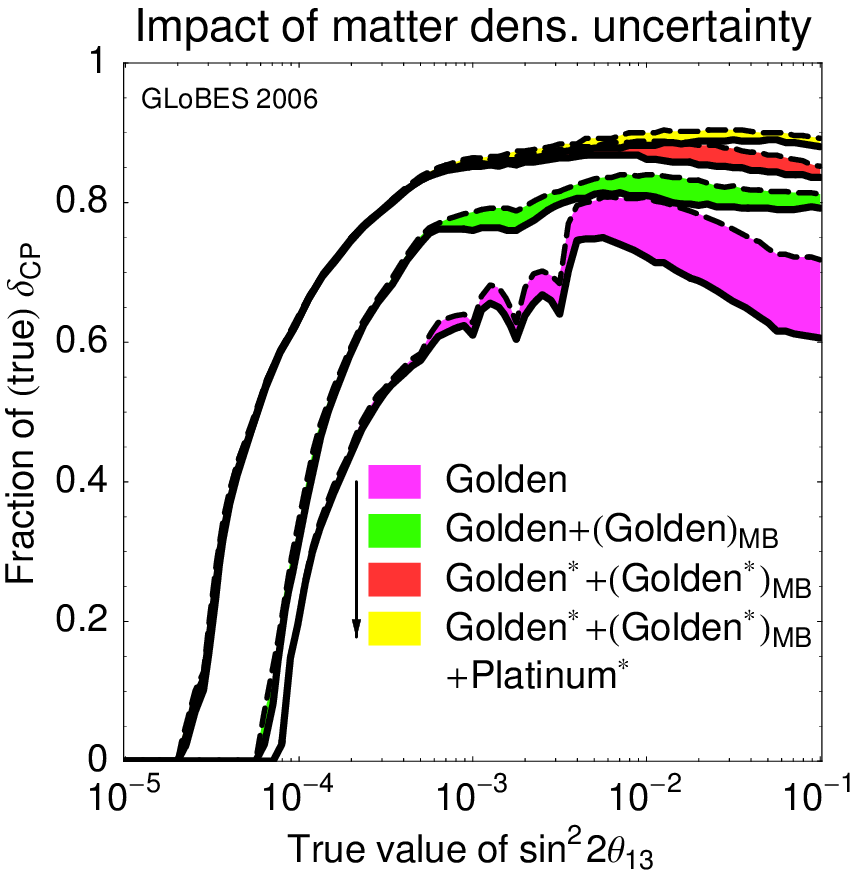}
\end{center}
\mycaption{\label{fig:matterdensity} Curves from \figu{optsummary}
  (right) for 5\% matter density uncertainty (solid) and 2\% matter
  density uncertainty (dashed), \ie, the shaded areas represent the
  improvement potential with respect to the unknown matter density
  profile. Note, that in going from Golden to Golden* the muon energy
  goes down from $E_\mu=50\,\mathrm{GeV}$ to $E_\mu=20\,\mathrm{GeV}$.}
\end{figure}

Finally, it is well known that the {\em matter density uncertainty} is
important for $\stheta$ and $\deltacp$ measurements at large $\stheta$
(see, \eg , \Refs~\cite{Huber:2002mx,Ohlsson:2003ip} for the relevant
regions in parameter space). Since magic baseline and platinum channel
extract the information on $\stheta$ (and $\deltacp$) in a different
way compared to the golden channel, one may suspect that the
correlation with the matter density can be partially eliminated. We
therefore test the impact of the matter density uncertainty on our
optimization summary in \figu{matterdensity}. Obviously, for the $L=4
\, 000 \mathrm{km}$ baseline alone, the impact of matter density
uncertainties is rather large (``Golden''). However, especially the
magic baseline and platinum channel reduce this dependency. This
result is very interesting, since it means that using magic baseline
and possibly platinum channel, an improvement of the knowledge on the
matter density profile may not be necessary anymore. Nevertheless,
note that a lower matter density uncertainty cannot replace the
detector, channel, and baseline improvements discussed in this
section.

%%%%%%%%%%%%%%%%%%%%%%%%%%%%%%%%%%%%%%%%%%%%%%%%%%%%%%%%%%%%%%%%%%%%%%%%%%%%%%%%%%
\section{Summary and conclusions}
\label{sec:summary}

In this study, we have optimized a neutrino factory for oscillation
parameter measurements.  Assuming the standard magnetized iron
calorimeter as detector we find the following results: For small
$\stheta$, we have found an excellent (maximal) CP violation
sensitivity for baselines $L \simeq 3 \,000 \, \mathrm{km}$ to $5 \,
000 \, \mathrm{km}$, mass hierarchy sensitivity for $L \gtrsim 6 \,
000 \, \mathrm{km}$, and $\stheta$ sensitivity at the ``magic
baseline'' $L \simeq 7 \, 500 \, \mathrm{km}$.  Thus, the optimal
baseline depends on the performance indicator one is optimizing for.
In summary, we have identified the combination of two baselines $L
\sim 4 \, 000 \, \mathrm{km}$ and $L \simeq 7 \, 500 \, \mathrm{km}$
as the optimal configuration given these different performance
indicators. For the muon energy, we have found that $E_\mu \gtrsim 40
\, \mathrm{GeV}$ should be sufficient.  Note that a longer baseline
(such as the $L \simeq 7 \, 500 \, \mathrm{km}$) is useful for the
$\pi/2 - \theta_{23}$ measurement as well as for seeing multiple
oscillation nodes.

For large $\stheta$, the determination of the mass hierarchy and $\stheta$
discovery are not a problem at all. Therefore, the optimization is
determined by the CP violation sensitivity. We have found that in this
case slightly shorter baselines $L \gtrsim 1 \, 500 \, \mathrm{km}$
and lower muon energies $E_\mu \gtrsim 20 \, \mathrm{GeV}$ are
sufficient, while the values identified for small $\stheta$ do not
harm.  In addition, the ``magic baseline'' improves the CP violation
measurement potential because of a ``clean'' ({\ie} unaffected by
$\deltacp$) measurement of $\stheta$.

In order to further improve the neutrino factory, we have demonstrated
that a lower energy threshold and higher energy resolution for the
muon neutrino (golden channel) detector would increase the physics
potential significantly. In that case, also, the muon energy can be
reduced to $E_\mu=20\,\mathrm{GeV}$. In particular, the lower energy
threshold increases the mass hierarchy and CP violation discovery
reaches. This statement is true even in the presence of a higher
background fraction from neutral currents and mis-identified events.
For large $\stheta$, we have found that all of the following help for
a single detector/baseline: lower threshold, better energy resolution,
and lower matter density uncertainties.  Therefore, improving the
golden channel detector will have an excellent potential to push the
neutrino factory physics reach independently of the value of
$\theta_{13}$.  Further studies are needed to demonstrate to what
extent the detector improvements, we have anticipated, can be
achieved.

As far as different channels are concerned, we have modeled electron
(platinum channel) and tau (silver channel) neutrino appearance. Due to
the assumption of standard oscillations of three active flavors, the
additional channels do not provide completely independent information.
From an experimental point of view, both channels represent a
considerably more difficult challenge than the golden channel. As a
consequence the number of events in either the silver or platinum
channel will typically be smaller than the one in the golden channel, and
the backgrounds always will be higher, which means that the statistical
significance is doubly suppressed. Therefore, additional channels can
only improve the physics reach of the golden channel in those regions
of parameter space where the performance of the golden channel alone
is sub-optimal. The two regions where this happens are either large
$\stheta\sim0.1$, where the matter density uncertainty spoils the
sensitivity to CP violation, or intermediate
$\stheta\sim3\times10^{-3}$, where the so-called $\pi$-transit of the
mass hierarchy degeneracy can
destroy CP violation and mass hierarchy sensitivities.  For the
standard configurations of both channels using currently accepted
parameter estimates, we found that the impact is small in comparison 
to the additional effort. Therefore, we have
considered improved versions of these
channels throughout this study, which we have introduced as Silver$^*$
 and Platinum$^*$. The improvements with
respect to the standard setups are mainly much larger statistics
(factor 5-10) and lower backgrounds (Silver$^*$), as well as a wider energy
range for charge identification (Platinum$^*$). These improvements
are, at current, hypothetical, and more detector studies are clearly
needed.
We have demonstrated that the improved platinum
channel is especially helpful for large $\stheta$, where the
impact of the upper CID threshold is moderate. For intermediate
values of $\stheta$, we find that both the improved silver and platinum
channel can resolve the effects of the $\pi$-transit. However, 
the platinum channel requires a much higher CID threshold in this range.
  Note that we
have not tested the use of the silver channel for significant
deviations from maximal mixing and for new physics test, which are
under discussion elsewhere (see, \eg , \Refs~\cite{Autiero:2003fu,Kitazawa:2006iq}). 
In summary, the addition of silver and platinum channels is most likely only justified
by their ability to provide crucial cross checks of the assumption of
standard three flavor oscillation. 

In the last part of this study, we have compared different options for
synergies and competitiveness including the option of multiple
baselines, as well as we have compared the neutrino factory to a
higher gamma beta beam representative. We have found that magic
baseline, improved golden detector, and platinum channels are
synergistic at a varying degree and competitive in different regions
of the parameter space.  Thus, the same physics potential cannot just
be achieved by increasing the luminosity. In addition, we have shown
that a neutrino factory can outperform the competing technologies,
such as beta beams, for $\stheta \lesssim 10^{-2}$.  This is especially
true with respect to the fact that a neutrino factory can address
{\em all} open issues in oscillation physics. For large $\stheta
\gtrsim 10^{-2}$, we have demonstrated that improving the golden
channel detector, adding the magic baseline, and using the platinum
channel would improve the physics potential. We have also found that
the use of magic baseline and platinum channel reduces the impact of
matter density uncertainties for large $\stheta$ significantly, \ie,
the more information is added, the less important the matter
density uncertainties become. However, it is yet unclear if not other
alternatives, such as higher gamma beta beams, can do the desired
measurements with a lower effort, which very much depends on the
systematics assumed for these experiments.

We conclude that the neutrino factory setup optimized for oscillation
parameter measurements has two baselines, one at $L \sim 4000 \,
\mathrm{km}$ and one at $L \simeq 7 \, 500 \, \mathrm{km}$, an
optimized golden channel detector with lower threshold and higher
energy resolution and a muon energy of $E_{\mu} \sim 20 \, \mathrm{GeV}$.
This set of improvements exhausts the optimization potential in the
majority of the parameter space. The only region where an additional
gain may be achieved is for large $\stheta\sim0.1$. Here the addition
of a high statistics platinum channel detector would decrease the
impact of the matter density uncertainty.

As far as future neutrino factory R\&D is concerned, we find that the
ability to operate two baselines as well as the lower detection
threshold of the golden detector are the most critical components to
the optimized physics potential. Furthermore, a better energy
resolution of the golden channel detector would improve the physics
potential further. 

\section*{Acknowledgments}

We would like to thank the conveners and contributors of the
International Scoping Study for a future Neutrino Factory and
Superbeam Facility for countless useful discussions and comments.  In
particular, we would like to thank Alain Blondel, Peter Dornan, Steve
King, Yorikiyo Nagashima, Lee Roberts, Osamu Yasuda, Mike Zisman, and
especially Ken Long for coordinating the work of the experimental
subgroup of the physics working group and numerous discussions and
input.

In addition, we are grateful for information and comments on various
aspects of this study from Scott Berg, Scott Menary, Mauro Mezzetto,
Pasquale Migliozzi, and Graham Rees.

Computing was performed on facilities supported by the US National
Science Foundation Grants EIA-032078 (GLOW), PHY-0516857 (CMS Research
Program subcontract from UCLA), PHY-0533280 (DISUN), and the
University of Wisconsin Graduate School/Wisconsin Alumni Research
Foundation, as well as on the Scheides Beowulf and Condor clusters at
the Institute for Advanced Study.

WW would like to acknowledge support from the W.~M.~Keck Foundation
through a grant-in-aid to the Institute for Advanced Study, and
through NSF grant PHY-0503584 to the Institute for Advanced Study.

PH would like to acknowledge the warm hospitality at the Institute for
Advanced Study and the Technische Universit\"at M\"unchen where parts
of this work were carried out.

\begin{appendix}

%%%%%%%%%%%%%%%%%%%%%%%%%%%%%%%%%%%%%%%%%%%%%%%%%%%%%%%%%%%%%%%%%
%\newpage

\section{Impact of muon energy on additional channels}
\label{app:ch_energy}

Besides the baseline, the muon energy (energy of the muons stored in
the storage ring) can be changed. In the former discussions of
additional channel data this parameter was set to the standard
neutrino factory value $E_\mu =50\,\mathrm{GeV}$. A reduction of the
muon energy may be especially interesting in connection with an
improved detector, as well as it may allow to use a larger fraction of
the platinum channel data if the upper platinum CID threshold is at
lower energies. On the other hand, the tau production threshold may
affect the usefulness of the silver channel for lower muon energies.
Therefore, we discuss in this appendix if the inclusion of additional
channel information changes the muon energy optimization.
 
\begin{figure}[t!]
 \begin{center}
 \includegraphics[width=0.4\textwidth]{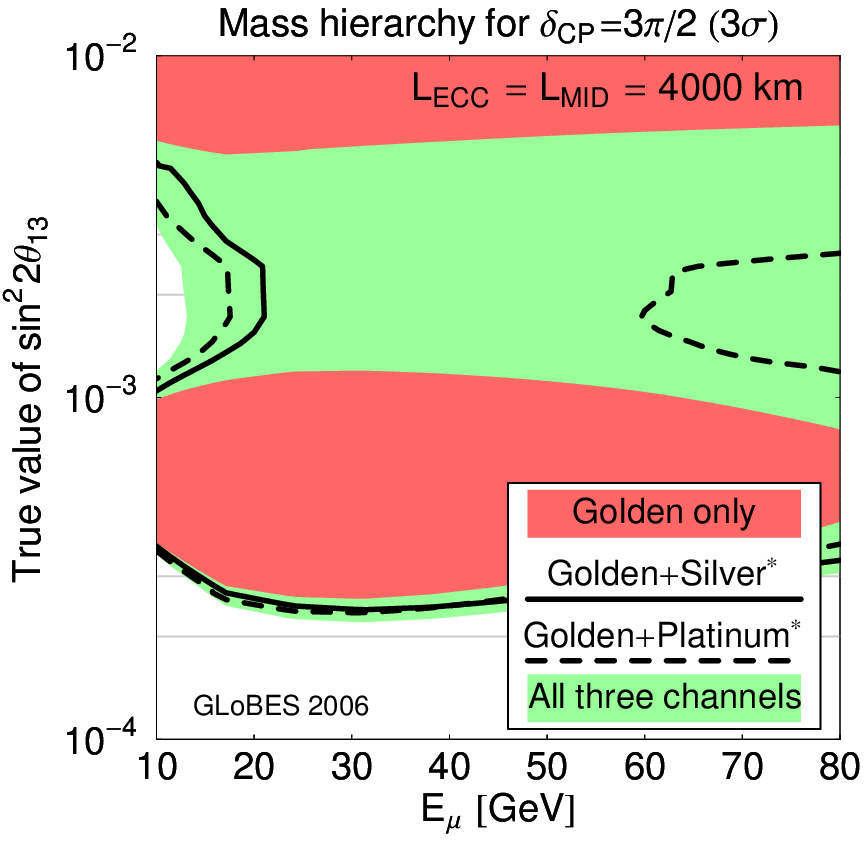} \hspace{0.1cm}
 \includegraphics[width=0.4\textwidth]{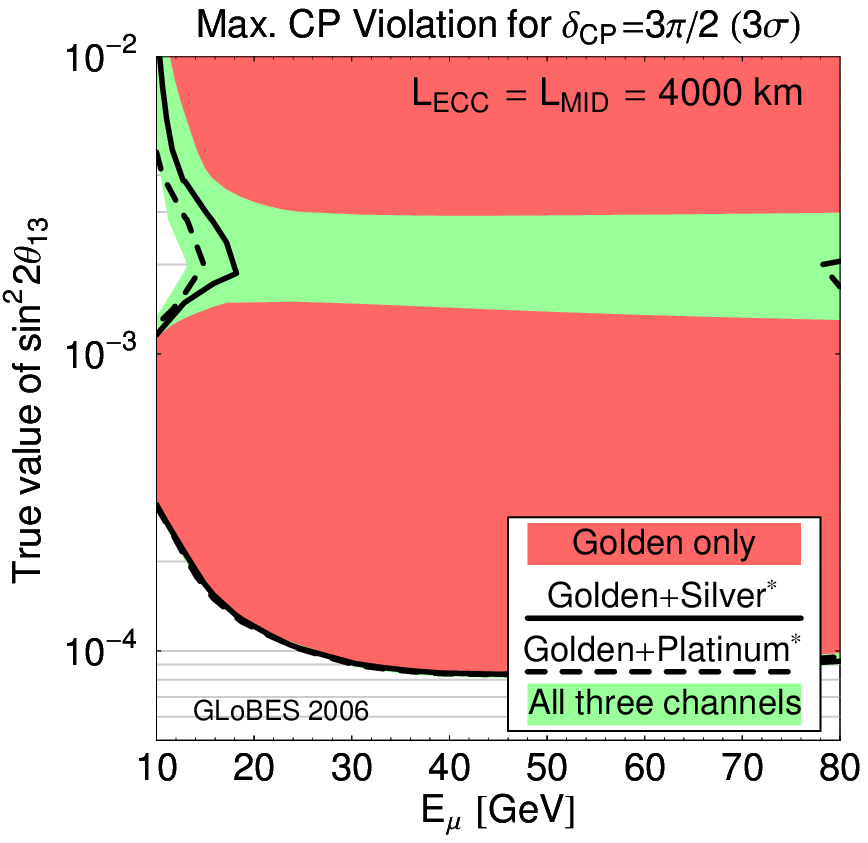}
 \end{center}
 \mycaption{\label{fig:EnergyMHCP} The sensitivity to the normal mass
   hierarchy (left) and to maximal CP violation (right) at $3\sigma$
   for the combination of different channels as given in the plot
   legend. The true value of the phase is assumed to be
   $\deltacp=3\pi/2$, and the hierarchy is assumed to be normal.  The
   parent energy of the stored muons is varied and the baseline is
   fixed to $L_{\mathrm{MID}}=L_{\mathrm{ECC}}=4000\, \mathrm{km}$ }
 \end{figure}

As far as the sensitivity to the normal mass hierarchy is concerned,
we show in \figu{EnergyMHCP} (left) the relevant $\stheta$ reach
$L_{\mathrm{MID}}=L_{\mathrm{ECC}}=4000\, \mathrm{km}$.  The variation
of the absolute reach in small values of true $\stheta$ is of minor
importance, and even improves slightly for the choice of smaller
parent energies. For the golden channel only, or golden and platinum
channels combined, the optimum is approximately reached for $E_\mu
\sim 30\,\mathrm{GeV}$. The lack of sensitivity to the mass hierarchy
in the gap between the dark gray-shaded regions cannot be resolved by
the golden channel alone independent of $\mathrm{E_\mu}$. However, if
combined with the silver or platinum channel, the sensitivity gap can
be closed for parent energies $E_\mu \gtrsim 20\,\mathrm{GeV}$
(platinum combinations) or larger than about $E_\mu \gtrsim
25\,\mathrm{GeV}$ (golden and silver combined). For the platinum
combinations (or all channels combined), the additional channel
information does not only allow to use a lower energy neutrino beam,
but also favors a lower parent energy of $E_\mu \sim
30\,\mathrm{GeV}$. If, on the other hand, only the silver channel data
is used, the tau production threshold disfavors too low muon energies.
 On the right-hand side of \figu{EnergyMHCP}, the sensitivity to
 maximal CP violation is shown as function of the muon energy. The
 qualitative observations are the same as for the mass hierarchy, but
 the silver channel favors $\mathrm{E_\mu\gtrsim 20\,GeV}$.

\end{appendix}

%%%%%%%%%%%%%%%%%%%%%%%%%%%%%%%%%%%%%%%%%%%%%%%%%%%%%%%%%%%%%%%%%%%%%%
%%%%%%%%%%             References                         %%%%%%%%%%%%
%%%%%%%%%%%%%%%%%%%%%%%%%%%%%%%%%%%%%%%%%%%%%%%%%%%%%%%%%%%%%%%%%%%%%%

%\clearpage

\end{document}